\documentclass[iop,numberedappendix]{emulateapj}

\bibpunct[; ]{(}{)}{;}{a}{,}{,}

\newcommand{\secpoint}{\mbox{$''\mskip-7.6mu.\,$}}
\newcommand{\secnopoint}{\mbox{$''\mskip-7.6mu\,$}}
\newcommand{\lya}{Ly$\alpha$}
\newcommand{\minpoint}{\mbox{$'\mskip-4.7mu.\mskip0.8mu$}}

\shorttitle{Narrowband Lyman-Continuum Imaging of Galaxies at $z \sim 2.85$}
\shortauthors{Mostardi, R.~E., et al.}

\begin{document}

\title{Narrowband Lyman-Continuum Imaging of Galaxies at $z \sim 2.85$}

\author{R.~E. Mostardi\altaffilmark{1}} \altaffiltext{1}{Department of Physics \& Astronomy, University of California,
Los Angeles, 430 Portola Plaza, Los Angeles, CA 90095}

\author{A.~E. Shapley\altaffilmark{1}}

\author{D.~B. Nestor\altaffilmark{1}}

\author{C.~C. Steidel\altaffilmark{2}} \altaffiltext{2}{Cahill Center for Astrophysics, California Institute of
Technology, MS 249-17, Pasadena, CA 91125}

\author{N.~A. Reddy\altaffilmark{3}} \altaffiltext{3}{Department of Physics and Astronomy, University of California,
Riverside, 900 University Avenue, Riverside, CA 92521}

\author{R.~F. Trainor\altaffilmark{2}}

\begin{abstract}

We present results from a survey for $z\sim2.85$ Lyman-Continuum (LyC) emission in the HS1549+1933 field and place
constraints on the amount of ionizing radiation escaping from star-forming galaxies. Using a custom narrowband filter
(NB3420) tuned to wavelengths just below the Lyman limit at $z\geq2.82$, we probe the LyC spectral region of 49 Lyman
break galaxies (LBGs) and 91 \lya-emitters (LAEs) spectroscopically confirmed at $z\geq2.82$. Four LBGs and seven LAEs
are detected in NB3420. Using $V$-band data probing the rest-frame non-ionizing UV, we observe that many
NB3420-detected galaxies exhibit spatial offsets between their LyC and non-ionizing UV emission and are characterized
by extremely blue NB3420$-V$ colors, corresponding to low ratios of non-ionizing to ionizing radiation
$(F_{UV}/F_{LyC})$ that are in tension with current stellar population synthesis models. We measure average values of
$(F_{UV}/F_{LyC})$ for our LBG and LAE samples, correcting for foreground galaxy contamination and HI absorption in
the IGM. We find $(F_{UV}/F_{LyC})_{corr}^{LBG}=82\pm45$ and $(F_{UV}/F_{LyC})_{corr}^{LAE}=7.4\pm3.6$. These
flux-density ratios correspond respectively to relative LyC escape fractions of $f_{esc,\,rel}^{LBG}=5-8$\% and
$f_{esc,\,rel}^{LAE}=18-49$\%, absolute LyC escape fractions of $f_{esc}^{LBG}=1-2$\% and $f_{esc}^{LAE}=5-15$\%, and a
comoving LyC emissivity from star-forming galaxies of
$8.8-15.0\times10^{24}\;\mathrm{ergs}\;\mathrm{s}^{-1}\;\mathrm{Hz}^{-1}\;\mathrm{Mpc}^{-3}$. In order to study the
differential properties of galaxies with and without LyC detections, we analyze narrowband \lya\ imaging and
rest-frame near-infrared imaging, finding that while LAEs with LyC detections have lower \lya\ equivalent widths on
average, there is no substantial difference in the rest-frame near-infrared colors of LBGs or LAEs with and without
LyC detections. These preliminary results are consistent with an orientation-dependent model where LyC emission escapes
through cleared paths in a patchy ISM.

\end{abstract}


\section{Introduction}    \label{sec_intro}

One of the foremost goals in the study of cosmological reionization is determining the sources of the ionizing
photons. Quasars (QSOs), while able to maintain an ionized universe from $z \sim 0 - 2$ \citep{cowie09}, rapidly fall
in number density at redshifts greater than $z \sim 2$ \citep{hopkins07}. Recent studies of the faint-end slope of the
QSO luminosity function indicate that while the QSO contribution to the overall ionizing budget at $z \gtrsim 4$ may
not be negligible, QSOs alone are still unable to sufficiently account for all of the ionizing radiation during the
epoch of reionization \citep{fontanot12,glikman11,siana08}. It is generally assumed that star-forming galaxies fill in
the remainder of the gap in the ionizing budget. Therefore, studying the ionizing Lyman-continuum (LyC) properties of
high-redshift star-forming galaxies can provide vital information about the evolution of the intergalactic medium
(IGM). Because the IGM at $z \gtrsim 6$ is opaque to LyC photons, we cannot directly observe the ionizing radiation at
the redshifts corresponding to reionization. In order to directly measure the ionizing radiation escaping galaxies and
study the galactic properties that give rise to significant LyC escape fractions, we must locate and study
lower-redshift analogs to the star-forming galaxies that reionized the universe.

There have been several recent studies of LyC emission at both low and intermediate redshifts. To probe $0<z<2$,
space-based far-UV observing facilities such as $HST$/STIS, $GALEX$, and $FUSE$ have searched for LyC radiation from
star-forming galaxies \citep{leitherer95,malkan03,siana07,siana10,grimes07,grimes09,cowie09,bridge10} only to obtain null
results \citep[but see, e.g.,][]{leitet13}. At $z \sim 3$, ground-based optical studies have shown that roughly 10\%
of star-forming galaxies have a moderately high escape fraction of ionizing radiation
\citep[$f_{esc}>0.2$;][]{shapley06,iwata09,nestor11,nestor13}. At even higher redshift ($z\sim4$), \citet{vanzella12}
find only one LyC-emitter out of 102 LBGs, although this small number of LyC detections might reasonably be attributed
to the rapidly increasing IGM opacity rather than to processes internal to galaxies. \citet{siana10} investigate the
conspicuous lack of LyC-emitting galaxies at low redshift, where LyC transmission through the IGM is high. Given the
similar stellar populations for UV-luminous galaxies at $z \sim 1.3$ and $z \sim 3$, \citet{siana10} infer that LyC
production does not change with redshift but the mechanism governing LyC escape must vary.
 
\begin{figure}
\epsscale{1.0}
\plotone{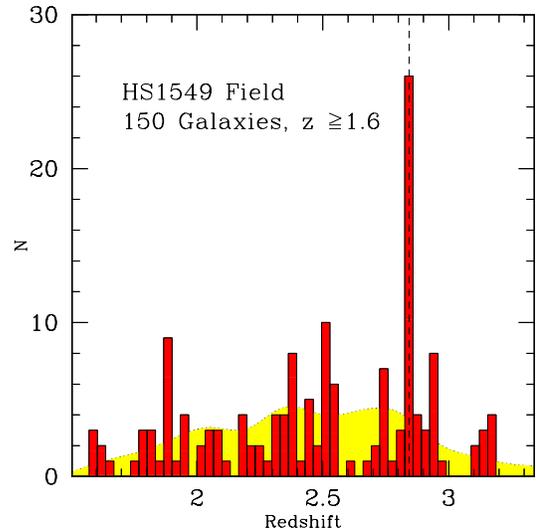}
\caption{
\small
Redshift histogram of galaxies in the HS1549 field. The yellow curve represents the overall redshift selection
function of the LBG survey \citep{steidel03,steidel04}, normalized to the observed number of galaxies in HS1549. The
overdensity of objects at $z=2.85$ is indicated by the dashed line, representing the redshift of the HS1549
protocluster.
\label{fig_zhist}
}
\end{figure}

One key observational method used to measure LyC flux is deep imaging through a narrowband filter tuned to wavelengths
just bluewards of the Lyman limit. Narrowband imaging provides a very effective way to simultaneously probe the LyC of
large samples of galaxies at the same redshift. We have designed a narrowband imaging program to study the LyC
properties of galaxies in the HS1549+1919 field. This field was observed as part of a larger survey of UV-selected
star-forming galaxies at $z \sim 2 - 3$ \citep{steidel03,steidel04,steidel11,reddy08} and contains a galaxy
protocluster with a redshift-space overdensity of $\delta_{gal} \sim 5$ at $z = 2.85 \pm 0.03$ (Figure
\ref{fig_zhist}). The ``spike" redshift coincides with that of a hyperluminous QSO \citep{trainor12}. More than 350
UV-selected galaxies have been identified in the HS1549 field, $\sim160$ of which have been spectroscopically
confirmed at $1.5 \leq z \leq 3.5$. Additionally, narrowband imaging with a 4670\AA\ filter tuned to the wavelength of
\lya\ at the redshift spike has revealed $\sim$300 potential \lya\ Emitters (LAEs) and several \lya\ ``blobs"
\citep{steidel00,steidel11}. Such a large sample of star-forming galaxies at approximately the same redshift greatly
facilitates the systematic narrowband imaging search for leaking LyC emission.

Our work in the HS1549 field parallels that of \citet{nestor11,nestor13} and \citet{iwata09}, who investigated another
high-redshift protocluster (SSA22a; $z = 3.09$). Several questions emerged from these initial narrowband LyC studies,
including the nature of galaxies with large offsets (occasionally reaching several kpc) between the centroids of their
non-ionizing UV-continuum and LyC emission. While significant offsets have been predicted in some simulations modeling
the escape of ionizing photons \citep[e.g.,][]{gnedin08}, in practice it is difficult to determine whether the
observed offsets provide information about the interstellar medium of a LyC-leaking galaxy, or are simply the result
of a foreground contaminant \citep{vanzella10,vanzella12}. Another key result presented in the SSA22a studies is the
high apparent ratio of escaping ionizing to non-ionizing UV radiation measured for several LAEs. For some objects,
this ratio exceeded unity \citep{nestor11,inoue10}. If such measurements are free of foreground contamination, they
are at odds with standard models for the intrinsic spectral energy distribution of star-forming galaxies \citep{bc03}.
Consequently, as discussed by \citet{vanzella12}, the most critical goal for LyC studies is to minimize the
possibility that candidate LyC-leaking galaxies are contaminated by low-redshift interlopers. While previous $z\sim3$
LyC studies have been plagued by small samples and lack of spectroscopic redshifts, in this work we present a sample
of 131 spectroscopically confirmed galaxies (49 LBGs and 91 LAEs, 9 of which are constituents of both samples). While
our seeing-limited, ground-based imaging makes it difficult to distinguish individual cases of foreground
contamination, we have performed simulations to model the amount of expected contamination in our samples as a whole
\citep[as in][]{nestor11,nestor13}.

\begin{figure*}
\epsscale{1.0}
\plotone{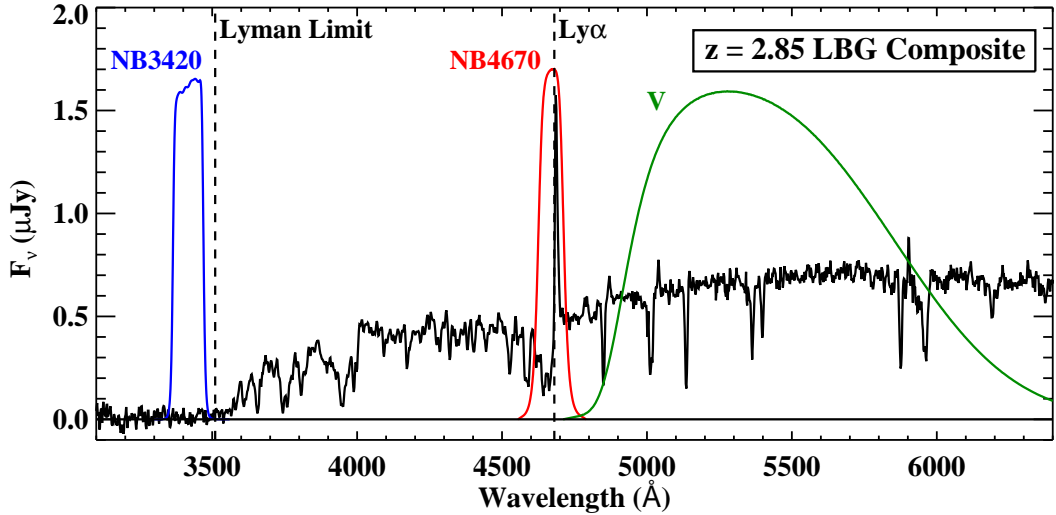}
\caption{
\small
Narrowband filter transmission curves overlaid on a composite LBG spectrum \citep{shapley06} redshifted to $z=2.85$.
The NB3420 filter (shown in blue) is located just bluewards of the Lyman limit at $z=2.85$, and probes LyC emission
for galaxies at $z \geq 2.82$. The $V$ filter (shown in green) probes the non-ionizing UV continuum (rest-frame
$\sim1400$\AA) of galaxies at $z \sim 2.85$. We note that the $V$ image was taken with the d500 dichroic, which blocks
light bluewards of 5000\AA, slightly truncating the transmission curve shown in this figure. The NB4670 filter (shown
in red) is centered on the \lya\ emission line for objects at $z \sim 2.85$ and was used along with the $V$ band to
select the sample of LAEs. Dashed lines indicate the locations of the Lyman limit and \lya\ feature at $z=2.85$.
\label{fig_spectrum}
}
\end{figure*}

In this work, we build upon previous high-redshift LyC studies by considering a large spectroscopic sample of galaxies
in an independent field from the SSA22a observations. We present imaging for 12 new $z\sim2.85$ galaxies with putative
LyC detections and correct for foreground contamination and IGM absorption. For our LBG and LAE samples, we calculate
the escape fraction of ionizing photons both in an absolute sense and relative to the escape fraction of non-ionizing
UV photons. We also explore the differential multiwavelength properties of objects with and without leaking LyC
radiation with regard to their \lya\ equivalent widths, rest-frame near-infrared photometry, and stellar populations.
The paper is organized as follows. In Section \ref{sec_obsred}, we describe our observations, data reduction, and
galaxy sample. In Section \ref{sec_ana}, we present the LyC and broadband photometric measurements and error analysis.
Section \ref{sec_contam} contains a discussion of the complexities involved in identifying foreground contaminants and
our methods for correcting measured LyC magnitudes for both foreground contamination and IGM absorption. In Section
\ref{sec_res}, we discuss the colors of individual LyC-detected objects, the average properties of the LBG and LAE
samples, and the implied LyC escape fraction and comoving emissivity. We present the multiwavelength properties of our
targets in Section \ref{sec_prop} and summarize our results in Section \ref{sec_summary}. Throughout the paper we
employ the AB magnitude system and assume a cosmology with $\Omega_m=0.3$, $\Omega_{\Lambda}=0.7$, and $H_0=
70$~km~s$^{-1}$~Mpc$^{-1}$. At $z=2.85$, 1\secnopoint\ corresponds to 7.8 proper kpc.

\section{Observations and Data Reduction} \label{sec_obsred}
\subsection{Photometric Observations} \label{ssec_obs}

For observations of LyC emission, we used a custom narrowband filter manufactured by Barr Associates with central
wavelength 3420\AA\ and FWHM 105\AA. This filter (hereafter referred to as NB3420) is designed to probe the LyC region
blueward of 912\AA\ for galaxies at $z \geq 2.82$ such that no contaminating flux ($\leq 0.4\%$) from the galaxy's
non-ionizing UV continuum is transmitted. The filter lies well within one LyC mean free path for galaxies at $z=2.85$,
minimizing the effect of intervening Lyman Limit systems and allowing for a more direct probe of the LyC absorption
properties internal to galaxies. At a redshift of 2.85, current estimates place the LyC mean free path at $\sim$ 100
proper Mpc \citep{rudie13,fauchergiguere08,songaila10} which corresponds to a rest-frame wavelength interval of $\sim
830 - 912$\AA. At $z=2.85$, the range of rest-frame wavelengths probed by the NB3420 filter is $872 - 904$\AA. Figure
\ref{fig_spectrum} shows the wavelengths probed by the NB3420 filter with respect to a typical LBG spectrum at $z =
2.85$, along with the locations of the NB4670 and $V$ filters relevant to the identification of LAEs and the
photometry of the non-ionizing UV spectral region.

Our imaging was centered on the HS1549 field, at ($\alpha$, $\delta$) = (15:51:53.7, +19:10:42.3). Observations were
taken through the NB3420 filter using the blue side of the Low Resolution Imaging Spectrometer
\citep[LRIS,][]{oke95,steidel04} on the Keck I telescope. During observing runs on 23 $-$ 24 June 2009, 11 $-$ 12 July
2010, and 9 August 2010, we acquired a total of 19.2 hours of data comprising thirty-seven 1800 second exposures, one
1500 second exposure, and one 900 second exposure. We dithered the telescope between exposures to minimize the effects
of bad pixels and to cover the gap between the two LRIS-B CCDs. Because these two 2K $\times$ 4K detectors have
slightly different quantum efficiencies below 4000\AA, we obtained half of the exposures at a sky position angle of
$0^{\circ}$ and half at $180^{\circ}$ in order to even out systematics between the two chips. Conditions were
photometric during all observing runs, and the effective seeing FWHM in the final stacked NB3420 image is 0\secpoint7
with a $1 \sigma$ surface brightness limit of $29.4 \mbox{ mag}/\mbox{arcsec}^{2}$.\footnote{The surface brightness
limit quoted is a rough estimation made by evaluating sky counts in 1\secpoint0 apertures placed randomly on blank
areas of the image. We conduct a more detailed study of the NB3420 image properties and our photometric accuracy using
Monte Carlo simulations, as described in Section \ref{ssec_uncert}}

Additional data in the HS1549 field includes broadband ground-based optical ($U$, $G$, $V$, $\cal{R}$) and near-IR
($J$, $K$) imaging, \emph{Spitzer} IRAC and MIPS ($24 \mu \mbox{m}$) photometry, along with $HST$/WFC3 $F160W$ and
$F475W$ imaging and morphologies for a small portion of the field. At $z=2.85$, the ground-based $V$ and $\cal{R}$
images both probe the non-ionizing UV continuum in the vincinity of 1500\AA\ and do not suffer from contamination by
the \lya\ emission line or \lya\ forest line blanketing. Given the significant redshift spike in the HS1549 field,
narrowband 4670\AA\ imaging (NB4670; $\lambda_{eff}=4667$\AA, $\mbox{FWHM}=88$\AA) probing \lya\ at $z=2.85$ was also
obtained. We used the combination of NB4670 and broadband $V$ imaging to select LAEs near the redshift of the HS1549
protocluster \citep{steidel11} and the NB4670$-V$ continuum subtracted image to examine the morphology of the \lya\
emission. Replacing the $V-$band image with $G$ also yields information about LAEs\footnote{While the $G$-band image
is contaminated by emission from the \lya\ feature, the $V-$band image is offset from \lya\ in wavelength and thus may
include a color term if the continuum is not flat in $f_{\nu}$.}, so we experimented with selecting LAEs using
NB4670$-G$ colors and examined the \lya\ morphology in the NB4670$-G$ image as well. Table \ref{tab_data} gives a
summary of the imaging in the HS1549 field most relevant to the LyC observations. Detailed photometry and error
analysis are performed on the NB3420, $V$, and $\cal{R}$ images, as described in Section \ref{sec_ana}.

\begin{deluxetable}{cccc} 
\tablewidth{0pt} 
\footnotesize
\tablecaption{Keck/LRIS Imaging Observations \label{tab_data}}
\tablehead{
\colhead{Filter\tablenotemark{a}} &
\colhead{$\lambda_{eff}$} &
\colhead{Seeing FWHM} &
\colhead{Exposure} 
\cr \colhead{} & \colhead{(\AA)} & \colhead{($''$)} & \colhead{(s)}  
}
\startdata 
NB3420   & 3420 & 0.7 & 69000  \\
NB4670   & 4670 & 0.7 & 18000  \\
$V$         & 5506 & 1.0 & 10800  \\
$\cal{R}$   & 6830 & 1.3 & 4800
\enddata
\tablenotetext{a}{The NB3420 and NB4670 imaging were taken on LRIS-B, while the $V$ and $\cal{R}$ imaging were
taken on LRIS-R. Additionally, the $V-$band transmission is slightly affected by the use of the d500 dichroic.}
\end{deluxetable}

\subsection{NB3420 Imaging Reduction} \label{ssec_red}

We used standard IRAF image reduction procedures to reduce individual narrowband exposures and create the final NB3420
stacked image. For each exposure, we divided the image by a flatfield constructed from images of the twilight sky,
subtracted the background, and masked out cosmic rays and satellite trails. In order to stack the 39 NB3420 frames, we
registered $>$500 objects in the NB3420 frames with their counterparts in the astrometrically corrected $\cal{R}$-band
image and resampled the pixels to the $\cal{R}$-band plate scale of 0\secpoint2119/pixel. Accurate image registration
is necessary for creating a spatial map of the relative positions and morphology of escaping LyC and non-ionizing UV
continuum radiation and identifying likely low-redshift contaminants.

In the process of combining the individual exposures, which ranged in airmass from 1.00 to 1.71, we scaled the flux in
each frame such that each exposure was effectively observed at the minimum airmass. Narrowband images were calibrated
onto the AB magnitude system using observations of spectrophotometric standard stars from the list of
\citet{massey88}. These AB magnitudes were also corrected for a Galactic extinction of $E(B-V) = 0.045$, based on IRAS
$100\mu\mbox{m}$ cirrus emission maps from \citet{schlegel98} and the extinction law from \citet{odonnell94}. The
final stacked image has an area of 35.5 $\mbox{arcmin}^{2}$.

\subsection{Spectroscopic Observations, Data Reduction, and Analysis} \label{ssec_spec}

Previous spectroscopy has already been performed in the HS1549 field, resulting in spectroscopic redshifts for a
significant number of LBGs and LAEs (Section \ref{ssec_samp}). In order to augment the existing spectroscopic sample
and confirm the redshifts of potential LyC-leaking galaxies, we obtained additional spectra, favoring objects with
NB3420 detections. Another of our original intentions was to acquire deep spectroscopy in the LyC region of galaxies
with NB3420 detections, but we were limited by poor weather conditions.

We performed multi-object spectroscopy in May 2011 on the Keck 1 telescope, using the blue side of LRIS. We observed
four slitmasks with exposure times of 16560, 9000, 8400, and 8100 seconds, respectively. For all masks, we used the
400 line mm$^{-1}$ grism blazed at 3400\AA, acheiving a spectral resolution of R = 800 for 1\secpoint2 slits. The
``d500'' dichroic beam splitter was used for the first mask (originally designed for deep LyC spectroscopy) and the
``d560'' dichroic was used for the three additional masks (designed to acquire redshifts). The conditions during the
observing run were suboptimal, with intermittent clouds and a seeing FWHM of 0\secpoint7 $-$ 1\secpoint0 during clear
spells.

When designing the slitmasks, we targeted both LBGs and LAEs with NB3420 detections. Slits were centered on the
coordinates of the $V$ (NB4670) centroid for LBGs (LAEs). While most LAEs were selected using the $V-$band image as
the continuum band, a small fraction (20\%) were selected using $G-$band (henceforth referred to as GNBs). Overall, we
observed 46 objects on the four slitmasks, 29 of which had repeat observations.

Standard IRAF tasks were used to cut up the multi-object slitmask images into individual slitlets, flatfield the
spectra using twilight sky flats, mask out cosmic rays, subtract the sky background, and average individual exposures
to make stacked 2D spectra. These spectra were then extracted to one dimension, wavelength calibrated, and shifted
into the vaccuum frame. Details of these spectroscopic reduction procedures are discussed in \citet{steidel03}. The
centroid of the \lya\ emission feature ($\lambda$ = 1215.67) was used to estimate redshifts for LAEs. Both \lya\
emission and interstellar absorption lines (when detected) were used for LBGs, yielding separate emission and
absorption redshifts when both types of features were detected. For objects with spectra taken on multiple masks, we
averaged these spectra in order to determine redshifts in cases when doing so increased the S/N.

We successfully measured redshifts for 9 of the 12 LBG candidates, confirming three objects (MD5, M16, and D24) to be
at $z > 2.82$ and identifying one as a star. We also measured redshifts for 11 out of 27 LAEs, thus providing us with
five new LAE candidate LyC emitters. Of the remaining six LAE candidates for which we measured redshifts, two already
had redshifts determined from previous spectroscopic studies of Steidel et al. ($lae4680$, $lae7832$; these objects
were on the mask that was designed to directly detect LyC emission spectroscopically), one object had an NB3420
detection that was flagged as contamination after the spectroscopy was completed ($lae3208$), one object ($lae6856$)
was at $z = 2.807$ (slightly too low redshift for NB3420 filter to probe uncontaminated LyC emission), and two objects
($lae4152$, $z=2.447$; $lae5165$, $z=1.873$) were at too low a redshift to be members of the HS1549 protocluster. Out
of the seven GNBs, we acquired three redshifts (GNB2861, GNB4769, GNB5270), all of which placed the objects at $z \sim
2.85$, i.e., in the correct redshift range for our study. We note that in cases where no redshift could be measured,
we did not draw conclusions about the quality of the object; the poor weather conditions, combined with the faintness
of our targets, made it impossible to remove objects from our sample on the basis of a non-detection.

In summary, the analysis of the spectra allowed us to confirm $z \geq 2.82$ redshifts for 3 LBGs, 5 LAEs, and 3 GNBs
with NB3420 detections. With these new redshifts, we were able to include the three additional LBGs in our LyC
analysis. However, all five of the LAEs for which we confirmed redshifts have $m_{4670} > 26.0$. Because we do not
have a complete and unbiased spectroscopic sample of LAEs with $m_{4670} > 26.0$ or GNBs (see Section
\ref{ssec_samp}), we did not include these objects in the LyC analysis. In the appendix, we present their postage
stamp images and uncorrected NB3420 magnitudes in Figure \ref{fig_GNB_faint_LAE} and Table \ref{tab_GNB_faint_LAE},
respectively.


\subsection{Sample}    \label{ssec_samp}

Our initial sample consisted of 363 UV-selected galaxies and 289 narrowband-color selected LAEs. The UV-selected
galaxies were identified using the $UG\cal{R}$ color selection criteria discussed in \citet{steidel03,steidel04}.
Spectroscopic redshifts of 145 of these UV-selected galaxies were previously measured in this field
\citep{reddy08,steidel11} and our follow-up spectroscopy (described in Section \ref{ssec_spec}) yielded an additional
eight redshifts. For the purposes of studying LyC emission, we kept only galaxies that were spectroscopically
confirmed to be at $z \geq 2.82$ with no AGN signatures in their spectra. This sample consists of 49 LBGs with
$z_{spec} \geq 2.82$.\footnote{We note that one of the 49 spectroscopically confirmed LBGs (D27) has an absorption
redshift of 2.814 and a \lya\ emission redshift of 2.816. Although it is at a lower redshift than our conservative
cutoff of $z=2.82$, less than 1\% of the flux redwards of the Lyman limit is transmitted through the NB3420 filter at
$z=2.814$. Accordingly, we include it in the sample of spectroscopically confirmed LBGs.} We note that while the
spectroscopic studies of \citet{reddy08} and \citet{steidel11} were conducted without reference to the LyC properties
of the LBGs, our follow-up spectroscopy was aimed at confirming redshifts of LBGs with NB3420 detections. While the
addition of these objects may introduce a slight bias in the average non-ionizing to ionizing UV flux-density ratio
for the full LBG sampe, it allows us to study the individual flux-density ratios of a larger number of LyC-emitting
LBGs and perform a more useful differential analysis of the stellar populations of LBGs with and without LyC emission.

LAEs were selected via a broadband filter ($V$) and a narrowband filter (NB4670) designed to probe the \lya\ emission
line for galaxies in the range of $2.803 < z < 2.876$. In previous work \citep[e.g.,][]{steidel00,nestor11}, a
broadband minus narrowband color excess of 0.7 magnitudes was used to identify LAEs, corresponding to an equivalent
width threshold of 80~\AA\ (corresponding to rest-frame 20~\AA). Our HS1549 photometric LAE sample, however, is
comprised of objects with $V -$ NB4670 $>$ 0.6, a slightly lower threshold designed to increase the sample size by
including LAEs with observed \lya\ equivalent widths slightly less than 80~\AA.\footnote{Three of the LAE candidates
in the HS1549 field identified on the basis of an early generation of NB4670 and $V$-band photometry by Steidel et al.
have been shown by subsequent deeper NB4670 data to have colors not quite red enough to satisfy $V - $NB4670 $>$ 0.6.
However, as these objects have already been confirmed spectroscopically to be at $z \sim 2.85$, we include them in the
LAE sample regardless of $V-$NB4670 color.} Spectroscopic follow-up of 116 of these narrowband excess objects
confirmed 99 to be LAEs at the redshift of the protocluster (Trainor et al., in prep). Of the 17 objects not confirmed
to be at $z \sim 2.85$, two were found to be low-redshift galaxies (at $z=1.983$ and $z=2.7773$) and the other 15 did
not yield spectroscopic redshifts. With the exclusion of \lya\ blobs, galaxies with evidence for AGN emission in their
spectra, and two LAEs that lie just below the redshift limit of the NB3420 filter ($z < 2.82$), the
spectroscopically-confirmed LAE sample consists of 91 galaxies. The spectroscopic follow-up of LAEs was conducted
independently of LyC observations; thus, the sample of LAEs with spectroscopic redshifts is unbiased with respect to
LyC properties.

We also created an additional photometric LAE sample that includes 33 photometric candidates whose NB4670 magnitudes
are in the same range as the LAEs with spectroscopy ($m_{4670} \leq 26$), and include photometry and postage-stamp
images for these LAE photometric candidates in Appendix B. The lack of spectroscopic confirmation of these photometric
LAE candidates creates a small likelihood of contamination due to lower-redshift galaxies. The first source of
contamination arises from the fact that the NB4670 filter selects for \lya\ emission at a range that extends down to
$z \sim 2.80$, while the NB3420 filter only measures uncontaminated LyC emission for galaxies at $z \geq 2.82$.
However, only 2 out of 99 LAEs confirmed by spectroscopy has $z_{spec} < 2.82$; these objects (at $z=2.811$ and
$z=2.801$) lie in the tail of the redshift overdensity centered at $z=2.85$. With the assumption that the redshift
distribution of the LAEs without spectroscopy matches the distribution of those with spectroscopy, contamination from
additional galaxies in the low-redshift tail should be negligible. The second potential source of contamination arises
from [O{\sc II}]-emitters at $z \sim 0.24-0.26$ whose emission lines fall within the NB4670 filter bandpass, but this
type of contamination is also unlikely. Not only is the volume probed at $z \sim 2.80-2.88$ forty times larger than
that probed at $z \sim 0.24-0.26$, but the photometrically measured equivalent widths of the LAE candidates, while
typical for LAEs, would be considered exceptionally large if the objects were in fact low-redshift [O{\sc
II}]-emitters \citep[see, e.g.,][]{hogg98}. Additionally, as described in Section \ref{ssec_det}, the relative spatial
positions of the NB3420, $V$, and NB4670-$V$ emission point toward the LAE candidates being high-redshift objects.
However, because of the increased likelihood of contaminated NB3420 detections within the photometric LAE sample, we
do not include these objects in the LAE analysis.

In summary, the samples of LBGs and LAEs with $z_{spec} \geq 2.82$ consist of 49 and 91 galaxies, respectively. There
is some overlap in the final LBG and LAE samples: 9 LBGs are also LAEs. We exclude one of these overlap objects
(MD12/\emph{lae3540}) from our analysis of global LyC properties because of its complex morphology, but discuss its
photometric properties in detail in Section \ref{ssec_nbmorph}. With the exclusion of MD12/\emph{lae3540}, the final
LBG and LAE samples consist of 48 and 90 galaxies, respectively.

\begin{deluxetable*}{cccccccccccccccc}
\tablecolumns{12}
\tablewidth{0pt}
\footnotesize
\tabletypesize{\footnotesize}
\tablecaption{Uncertainties in Simulated Photometry for $\cal{R}$, $V$, and NB3420 \label{tab_sim}}
\tablehead{
\colhead{Magnitude Bin\tablenotemark{a}} & \colhead{$\Delta \cal{R}$\tablenotemark{b}} &
\colhead{$\sigma_{\cal{R}}^{+}$\tablenotemark{c}} & \colhead{$\sigma_{\cal{R}}^{-}$\tablenotemark{c}} & &
\colhead{$\Delta V$\tablenotemark{b}} &\colhead{$\sigma_{V}^{+}$\tablenotemark{c}} &
\colhead{$\sigma_{V}^{-}$\tablenotemark{c}} & & \colhead{$\Delta$NB\tablenotemark{b}} &
\colhead{$\sigma_{\mathrm{NB}}^{+}$\tablenotemark{c}} & \colhead{$\sigma_{\mathrm{NB}}^{-}$\tablenotemark{c}}
}
\startdata
22.5 $-$ 23.0 & 0.00 & 0.03 & 0.03 & & 0.00 & 0.02 & 0.02 & & 0.00 & 0.02 & 0.02  \\
23.0 $-$ 23.5 & 0.00 & 0.05 & 0.05 & & 0.00 & 0.03 & 0.03 & & 0.00 & 0.03 & 0.03  \\
23.5 $-$ 24.0 & 0.00 & 0.08 & 0.07 & & 0.00 & 0.05 & 0.05 & & 0.00 & 0.05 & 0.04  \\
24.0 $-$ 24.5 & 0.00 & 0.12 & 0.11 & & 0.00 & 0.07 & 0.06 & & 0.00 & 0.07 & 0.07 \\
24.5 $-$ 25.0 & 0.00 & 0.17 & 0.15 & & 0.00 & 0.10 & 0.09 & & 0.00 & 0.11 & 0.10  \\
25.0 $-$ 25.5 & 0.00 & 0.25 & 0.21 & & 0.00 & 0.14 & 0.13 & & 0.00 & 0.17 & 0.14 \\
25.5 $-$ 26.0 & 0.01 & 0.34 & 0.26 & & 0.00 & 0.22 & 0.18 & & 0.01 & 0.25 & 0.20 \\
26.0 $-$ 26.5 & 0.06 & 0.45 & 0.31 & & 0.01 & 0.32 & 0.25 & & 0.03 & 0.34 & 0.26 \\
26.5 $-$ 27.0 & $0.17$\tablenotemark{d} & 0.54 & 0.36 & & 0.04 & 0.38 & 0.28 & & 0.09 & 0.44 & 0.31 \\
27.0 $-$ 27.5 & $0.37$\tablenotemark{d} & 0.71 & 0.43 & & 0.13 & 0.51 & 0.34 & & $0.24$\tablenotemark{d} & 0.60 & 0.38
\\
27.5 $-$ 28.0 & $0.70$\tablenotemark{d} & 1.22 & 0.56 & & $0.25$\tablenotemark{d} & 0.60 & 0.38 & &
$0.40$\tablenotemark{d} & 0.79 & 0.45
\enddata
\tablenotetext{a}{Recovered magnitude range of simulated galaxies. The statistics in this table are calculated for an
object whose flux lies in the center of the bin; for example, this flux corresponds to a magnitude of 22.72 for the
brightest bin.}
\tablenotetext{b}{Average value of the difference between recovered and input magnitudes. In the fainter bins,
significant departures from zero imply systematic biases in the photometry.}
\tablenotetext{c}{The standard deviation in the difference between recovered and input magnitudes, used to calculate
photometric uncertainties. Note that uncertainties in magnitudes are asymmetric because the standard deviation is
calculated from the simulated flux distribution (Section \ref{ssec_uncert}).}
\tablenotetext{d}{Magnitudes of objects with systematic biases greater than one third of the object's uncertainty are
adjusted to reflect the systematic bias, as discussed in Section \ref{ssec_uncert}.}
\end{deluxetable*}

\section{Detecting LyC Emission} \label{sec_ana}
\subsection{Photometric Measurements} \label{ssec_phot}

Source identification and photometry were performed using SExtractor \citep{bertin96}. Objects were detected
independently in all images to allow for different spatial distributions of light at different wavelengths. For
$\cal{R}$-band and $V$-band images, we ran SExtractor in single-image mode using a Gaussian smoothing kernel with a
FWHM of 2 pixels and a detection threshold of 1.0 standard deviation above the local smoothed background. For the
NB3420 image, we ran SExtractor in dual-image mode with separate detection and measurement images, following the
methodology of \citet{nestor11}. For detection, we smoothed the NB3420 image by a Gaussian kernel with FWHM = 2.35
pixels and used a detection threshold of 2.25 standard deviations. For measurement, we used the unsmoothed NB3420
image. In the case of the NB3420 photometry, SExtractor parameters were chosen to reflect the higher signal to noise
of the NB3420 image, to produce object number counts similar to those in \citet{vanzella10}, and to be complete in the
faint magnitudes of interest for studying LyC emission. In all images, magnitudes were computed using ``Kron-like"
elliptical apertures (i.e., MAG\_AUTO in SExtractor).

Because there has been some contention in this field concerning the most appropriate way to photometrically measure
the flux-density ratio between ionizing (NB3420) and non-ionizing ($V$) UV emission
\citep{vanzella12,nestor11,nestor13}, we explain our methods here. Galaxies at $z \sim 3$ are known to exhibit clumpy
morphologies \citep[see e.g.,][]{law07}. As LyC photons may not escape isotropically from all portions of the galaxy,
it is possible that LyC emission will be observed emanating from only one clump of a given galaxy. With the
seeing-limited resolution of our images, it is often impossible to distinguish individual star-forming clumps; thus
LyC emission may appear to be offset from the centroid of the $V-$band emission. The idea of offset LyC emission is
also supported by results from galaxy formation simulations. Simulations by \citet{gnedin08} claim
that increased star-formation efficiencies in the more luminous LBGs 
result in higher scale heights of young stars relative to the HI in 
the disk. Therefore, minor interactions can disrupt the HI 
sufficiently to reveal the young stars and allow significant 
emission of ionizing photons to escape the galaxy.  With the proper 
orientation, this configuration can be recognized as LyC emission 
offset from the primary UV-continuum emission.  Alternatively, 
\citet{ricotti02} suggests that the formation of globular clusters 
may have reionized the universe, since luminous OB associations in 
the outer haloes may have escape fractions approaching unity. 
Globular clusters are still forming at $z \sim 3$ \citep{stetson96}, 
and would be observable in our data as LyC emission offset spatially 
from the primary galaxy. Because these models support the 
possibility that the spatial distribution of LyC emission may be 
offset from $V-$band emission or have different morphology (e.g., 
compact vs. diffuse), we must choose photometric methods that can 
account for such scenarios.

While \citet{vanzella12} suggest that flux-density ratios should be measured using both LyC and non-ionizing UV
emission \emph{only} from regions of escaping LyC emission, we argue here that the entire non-ionizing and ionizing UV
flux should be measured. Measuring the flux-density ratio within an individual LyC-emitting region (defined by the
isophote of the LyC emission) may provide useful information about the stellar populations of that region, but it does
not provide information about the average LyC escape fraction among star-forming galaxies or the global LyC emissivity
produced by these galaxies. The average escape fraction must be computed from the non-ionizing to ionizing
flux-density ratio of an ensemble of $z \sim 3$ galaxies, including both galaxies without observed LyC emission and
the non-LyC-emitting regions from galaxies with LyC emission. The calculation of the ionizing emissivity, moreover,
relies entirely on the $z \sim 3$ UV luminosity function constructed by measuring non-ionizing UV light from
\emph{entire} galaxies, not from their constituent clumps. Alternatively, we find that a method in which LyC is
measured solely
within an isophote defined by the galaxy's non-ionizing UV also does not 
accomplish our photometric goals.  As this ``isophotoal" method does 
not allow for LyC emission to have a different spatial distribution 
from that of the non-ionizing UV, it will result in the systematic 
loss of LyC flux.  A differential analysis of LAE magnitudes using 
this isophotal photometric method versus the ``Kron-like" apertures to 
measure $V-$band and NB3420 magnitudes separately (i.e., the method 
presented here) corroborates the idea that LyC emission is 
systematically missed in the isophotal method; the average LAE 
non-ionizing to ionizing flux-density ratio using the isophotal method 
is a factor of $\sim1.3$ larger than the methods we use.  While a 
factor of 1.3 is well within the 1$\sigma$ uncertainty\footnote{The 
two photometric methods agree so well because the NB3420 detections in 
the LAE sample have very small offsets from the corresponding $V-$band 
detections.}, it is an unnecessary loss of NB3420 flux. Therefore, we maintain that using the large ``Kron-like"
apertures to measure the full $V-$band and NB3420 magnitudes is the correct photometric method for the measurement of
the global LyC escape fraction and emissivity.

\subsection{Characterizing Photometric Uncertainties} \label{ssec_uncert}

We characterized both the statistical and systematic photometric uncertainties by running Monte Carlo simulations
designed to reproduce our photometric measurement procedures. In the simulations, we added to each image fake galaxies
of known magnitude and a range of radial profiles representative of the observed galaxies. In each iteration, one
hundred fake galaxies were placed in random, empty positions chosen to avoid image edges and photometric blending with
existing objects or previously placed fake galaxies. We ran SExtractor on each filter ($\cal{R}$, $V$, NB3420) using
the same parameters as for the actual photometry. The process of adding fake galaxies was repeated until we recovered
50,000 in each simulation, enough to populate all magnitude bins of interest with a statistically significant number
of objects. The bins of recovered magnitude are 0.5 magnitudes wide and span a range of $22.5 \leq m \leq 28$ in each
band to encompass the magnitude range of the observed LBGs and LAEs.

Table \ref{tab_sim} lists the systematic bias and photometric uncertainty associated with each magnitude bin, as
derived from the results of the simulations. In each bin, the systematic bias is defined to be the average difference
between input and recovered fluxes of fake galaxies, while the photometric uncertainty is defined to be the standard
deviation of the distribution of differences between input and recovered fluxes. The uncertainty associated with an
object of a given flux is a local quadratic interpolation between the uncertainties of neighboring bins. Because the
errors in flux are Gaussian, all errors quoted in magnitudes are double-sided. Using the uncertainties estimated by
our simulations, we define a 2$\sigma$ photometric limit that corresponds to a magnitude limit of 27.33 in $\cal{R}$,
27.58 in $V$, and 27.30 in NB3420. In all filters, the simulations show that SExtractor misses a higher percentage of
flux from fainter objects, resulting in a systematic offset in magnitude that is larger for fainter objects. In order
to correct for this bias, we added the systematic offset determined from the simulation analysis to the flux of an
object if the offset was greater than one third of the object's associated uncertainty. This threshold was chosen in
order to avoid the unnecessary addition of noise into our measurements by adding systematic offsets much smaller than
the 1$\sigma$ error. In practice, this correction only affected a few of the fainter bins (see Table \ref{tab_sim}).

\subsection{Object Matching} \label{ssec_objmat}

As we have two broadband filters ($V$ and $\cal{R}$) that probe the non-ionizing rest-frame UV ($\sim$1500\AA) for
galaxies at $z=2.85$, it is necessary to choose one to represent the non-ionizing UV flux. The spatial distributions
of the $V$ and $\cal{R}$ detections are well matched and their centroids agree within 0\secpoint2. The deeper $V-$band
image has 1\secpoint0 seeing, while the $\cal{R}$ image has 1\secpoint3 seeing, larger photometric errors (see Table
\ref{tab_sim}), and an elongated PSF. Although the $\cal{R}-$band image was originally used to identify LBGs, we adopt
the $V-$band image for our non-ionizing UV photometry based on its superior quality and the fact that more LAEs are
detected in $V$ than in $\cal{R}$. We note that all but one of the LBGs with LyC detections is detected in $V$; M4
does not have a $V$ detection because it falls off the edge of the $V$-band image. For this object, we calculate a $V$
magnitude by linearly interpolating between its measured $\cal{R}$ magnitude and a $G$ magnitude computed from its
$G-\cal{R}$ color obtained from the parent LBG survey \citep{reddy08,steidel04}.

The coordinates of LBGs are defined by $V-$band centroids, while the coordinates of LAEs are defined by NB4670
centroids. In order to find non-ionizing rest-frame UV counterparts for LAEs, we compiled the positions and
photometric measurements for any SExtractor detection in the $V-$band image within a 5 pixel (1\secpoint1 = 8.6 kpc at $z=2.85$) offset from the NB4670 coordinates of the target and successfully found counterparts for 77 out of 91 LAEs. Given that the largest galaxies reach sizes of $\sim10$ kpc at $z\sim2-3$ \citep{fs09,law12nature} and that we may be looking for LyC emission in the outskirts of these galaxies, we used a similar matching technique with a larger radius (9 pixel = 1\secpoint9 = 14.8 kpc at $z=2.85$) to find SExtractor detections in the NB3420 image near our LBGs and LAEs.  This generous matching radius guarantees the inclusion of all NB3420 detections potentially associated with the galaxies in the sample.  Our candidate with the largest offset between the galaxy centroid and the associated NB3420 detection (M16) has $\Delta_{Ly\alpha,LyC}=1\secpoint26=9.8$ kpc, but the majority of 
our NB3420 detections (especially around LAEs) are at much smaller offset ($<3$~kpc).

\begin{figure}
\epsscale{0.75}
\plotone{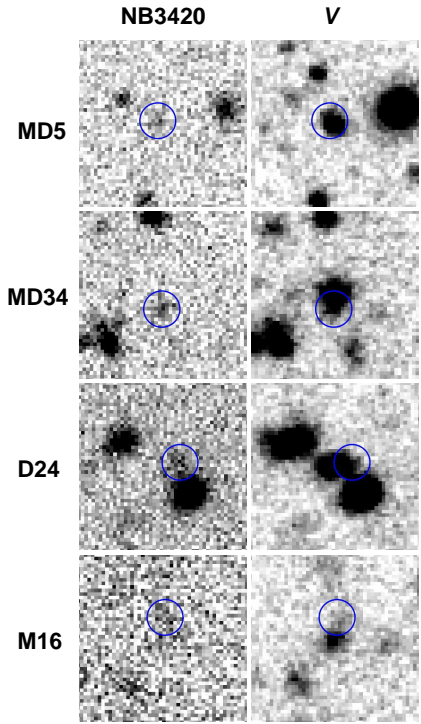}
\caption{
\small
10\secnopoint\ $\times$ 10\secnopoint\ postage stamp images of the 4 LBGs with NB3420 detections. Each object is
displayed in two bands: NB3420 (indicating the LyC) and $V$ (indicating the non-ionizing UV continuum). All postage
stamps are centered on the $V$-band centroid and blue circles (1\secnopoint\ radius) indicate the centroid of the
NB3420 emission. All postage stamps follow the conventional orientation, with north up and east to the left.
\label{fig_LBG}
}
\end{figure}

In order to remove any obvious false matches corresponding to neighboring sources, we visually inspected each NB3420
source matched with a known LBG or LAE in the $\cal{R}$, $V$, and $G$ images (probing rest-frame non-ionizing UV), and
the $F160W$ image (probing rest-frame optical), if available. We also inspected LAEs with NB3420 matches in the
continuum-subtracted NB4670$-V$ image, which probes the spatial distribution of \lya\ emission. NB3420 matches were
removed if they corresponded spatially to a visible counterpart in another image. We also removed matches where the
LyC emission was further than 1\secpoint9 away from the non-ionizing UV-continuum emission, which occasionally
occurred because the LAE matching radius was centered around the \lya\ emission, not the non-ionizing UV. Altogether,
we retained matches to 4 LBGs and 7 LAEs, labeling the NB3420 detections corresponding to 4 LBGs and 16 LAEs as
invalid.

\subsection{Targets with NB3420 Detections} \label{ssec_det}

\begin{figure}
\epsscale{1.125}
\plotone{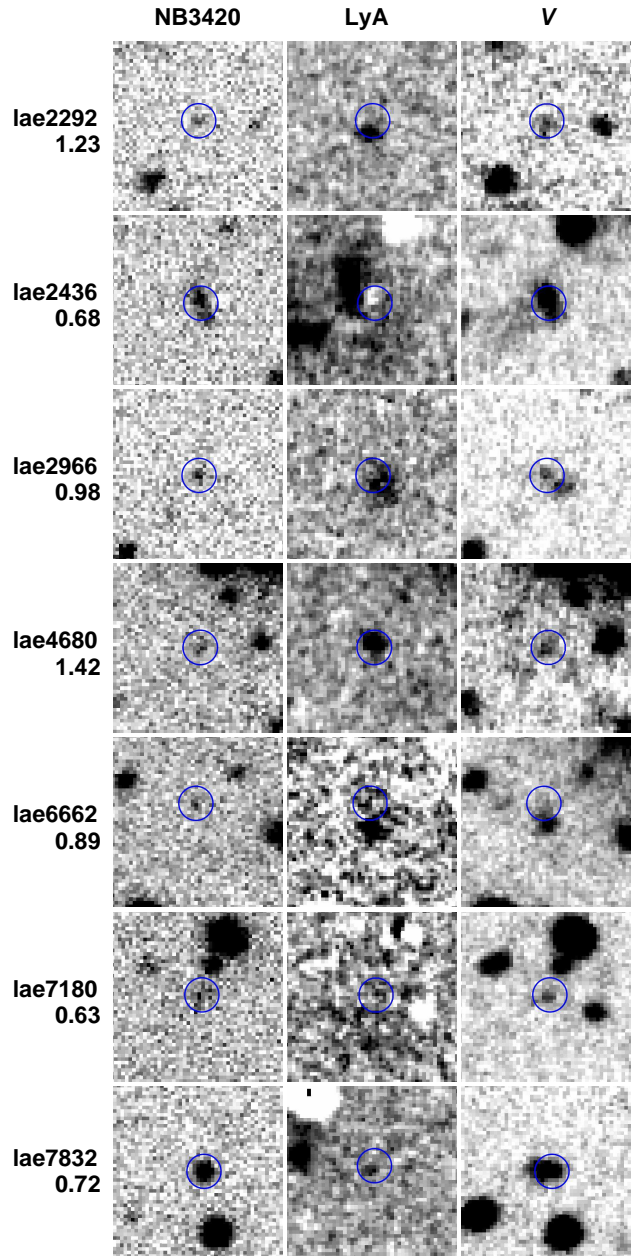}
\caption{
\small
10\secnopoint\ $\times$ 10\secnopoint\ postage stamp images of the 7 LAEs with NB3420 detections. Each object is
displayed in three bands: NB3420 (indicating the LyC), NB4670$-V$ (indicating \lya\ emission and labeled LyA), and $V$
(indicating the non-ionizing UV continuum). The $V-$NB4670 color of each LAE is indicated below the object name. All
postage stamps are centered on the $V$-band centroid and blue circles (1\secnopoint\ radius) indicate the centroid of
the NB3420 emission. All postage stamps follow the conventional orientation, with north up and east to the left. We
note that LAEs with more diffuse \lya\ emission may be difficult to distinguish in the NB4670$-V$ image, even though
their $V-$NB4670 colors identify them as LAEs and their redshifts have been confirmed by spectroscopy; for such
objects, we have increased the stretch of the NB4670$-V$ image to make the diffuse emission more easily visible.
\label{fig_LAE}
}
\end{figure}

\begin{figure}
\epsscale{1.0}
\plotone{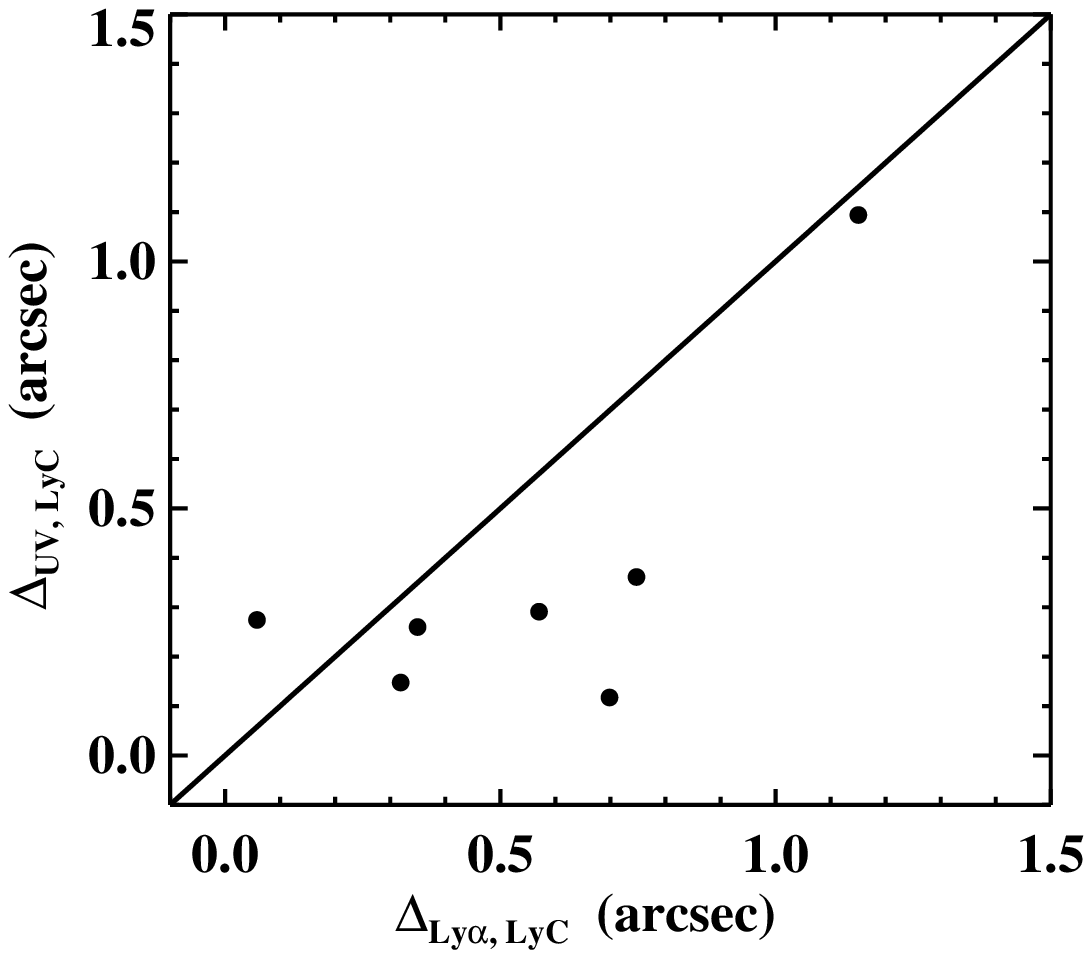}
\caption{
\small
Comparison of the offsets between UV-continuum and LyC emission ($\Delta_{UV,\,LyC}$) and between \lya\ and LyC
emission ($\Delta_{Ly\alpha,\,LyC}$) for LAEs with NB3420 detections. On average, $\Delta_{UV,\,LyC} <
\Delta_{Ly\alpha,\,LyC}$, reflecting the fact that \lya\ photons are resonantly scattered; therefore, \lya\ emission
is unlikely to coincide exactly with the location of LyC emission. This observed trend supports the claim that the
LAEs are not low-redshift [OII]-emitters (see Section \ref{ssec_det}).
\label{fig_offset_scatter_plot}
}
\end{figure}

We report NB3420 detections in 4 out of 48 LBGs and 7 out of 90 LAEs. MD12/\emph{lae3540} also has an NB3420 detection
and is discussed in Section \ref{ssec_nbmorph}. Although they are not the focus of our study, we note that 2 of the 4
AGNs in the field at $z \geq 2.82$ have NB3420 detections. Tables \ref{tab_LBG} and \ref{tab_LAE} display photometric
information for the NB3420-detected LBGs and LAEs. Postage-stamp images of targets with NB3420 detections are shown in
Figures \ref{fig_LBG} and \ref{fig_LAE}, indicating the NB3420 (LyC), NB4670$-V$ (\lya), and $V$ (non-ionizing UV
continuum) morphologies.

\begin{deluxetable*}{lcccccccc}
\tablewidth{0pt}
\tabletypesize{\scriptsize}
\tablecaption{LBG Photometry\label{tab_LBG}}
\tablehead{
\colhead{ID} & \colhead{RA} & \colhead{Dec} & \colhead{$z_{em}$\tablenotemark{a}} &
\colhead{$z_{abs}$\tablenotemark{b}} & \colhead{$V$} & \colhead{NB3420} &
\colhead{$\Delta_{UV,\,LyC}$\tablenotemark{c}} & \colhead{$\frac{F_{UV}}{F_{LyC}}_{obs}$\tablenotemark{d}} \\
 & \colhead{(J2000)} & \colhead{(J2000)} & & & & & &
}
\startdata
BX84 & 15:51:53.696 & 19:12:24.64 & \nodata & 2.823 & 24.41 & $>$27.30 & \nodata & $>$14.3 \\
MD5  & 15:51:45.211 & 19:11:05.13 & 3.146 & 3.139   & 24.96 & 26.89 & 0\secpoint3 & 5.9 $\pm$  2.0 \\
MD7 & 15:51:47.521 & 19:10:13.56 & 2.857 & 2.851 & 25.11 & $>$27.30 & \nodata & $>$7.5 \\
MD9 & 15:51:50.711 & 19:09:38.20 & 2.852 & 2.843 & 24.13 & $>$27.30 & \nodata & $>$18.6 \\
MD12\tablenotemark{e} & 15:51:51.882 & 19:10:41.16 & 2.856 & 2.849 & 24.56 & 26.74 & 1\secpoint0 & 7.5 $\pm$ 2.5 \\
MD34 & 15:52:06.315 & 19:12:48.60 & \nodata & 2.849 & 24.23 & 26.56 & 0\secpoint8 & 8.6 $\pm$  2.7 \\
C1 & 15:51:39.653 & 19:10:40.67 & 2.845 & \nodata & 24.77 & $>$27.30 & \nodata & $>$10.3 \\
C2 & 15:51:40.960 & 19:13:12.77 & 3.100 & \nodata & 23.76 & $>$27.30 & \nodata & $>$26.1 \\
C4 & 15:51:44.413 & 19:11:24.75 & 2.863 & 2.857 & 25.01 & $>$27.30 & \nodata & $>$8.2 \\
C5 & 15:51:44.627 & 19:10:59.73 & 3.173 & \nodata & 25.14 & $>$27.30 & \nodata & $>$7.3 \\
C6 & 15:51:44.760 & 19:10:32.84 & \nodata & 2.828 & 24.36 & $>$27.30 & \nodata & $>$15.0 \\
C7 & 15:51:45.247 & 19:12:13.21 & \nodata & 2.841 & 24.53 & $>$27.30 & \nodata & $>$12.8 \\
C8 & 15:51:45.386 & 19:08:49.84 & \nodata & 2.935 & 25.03 & $>$27.30 & \nodata & $>$8.1 \\
C9 & 15:51:46.707 & 19:11:52.39 & 2.925 & 2.919 & 23.67 & $>$27.30 & \nodata & $>$28.2 \\
C10 & 15:51:48.424 & 19:09:24.93 & 3.193 & 3.183 & 24.56 & $>$27.30 & \nodata & $>$12.4 \\
C12 & 15:51:49.360 & 19:09:52.55 & \nodata & 2.835 & 24.93 & $>$27.30 & \nodata & $>$8.8 \\
C13 & 15:51:49.685 & 19:10:58.10 & 2.843 & \nodata & 24.92 & $>$27.30 & \nodata & $>$9.0 \\
C14 & 15:51:50.616 & 19:09:18.46 & \nodata & 2.841 & 24.87 & $>$27.30 & \nodata & $>$9.4 \\
C15 & 15:51:51.352 & 19:10:19.50 & 2.849 & 2.849 & 25.39 & $>$27.30 & \nodata & $>$5.8 \\
C17 & 15:51:55.283 & 19:12:19.48 & 2.941 & 2.934 & 24.92 & $>$27.30 & \nodata & $>$9.0 \\
C19 & 15:52:00.196 & 19:10:08.73 & 3.166 & 3.158 & 24.41 & $>$27.30 & \nodata & $>$14.3 \\
C20 & 15:52:00.402 & 19:08:40.75 & 3.115 & \nodata & 25.09 & $>$27.30 & \nodata & $>$7.6 \\
C22 & 15:52:03.833 & 19:09:43.21 & \nodata & 2.960 & 24.65 & $>$27.30 & \nodata & $>$11.5 \\
C24 & 15:52:05.618 & 19:13:11.73 & 2.834 & 2.826 & 25.24 & $>$27.30 & \nodata & $>$6.7 \\
C25 & 15:52:06.069 & 19:11:28.37 & 3.159 & \nodata & 24.60 & $>$27.30 & \nodata & $>$12.0 \\
C27 & 15:52:07.041 & 19:12:19.29 & 2.931 & 2.922 & 24.32 & $>$27.30 & \nodata & $>$15.5 \\
D3 & 15:51:43.712 & 19:09:12.37 & 2.942 & 2.934 & 24.06 & $>$27.30 & \nodata & $>$19.7 \\
D4 & 15:51:43.976 & 19:11:39.67 & 2.863 & 2.856 & 24.43 & $>$27.30 & \nodata & $>$14.1 \\
D6 & 15:51:45.191 & 19:09:05.31 & 2.849 & 2.840 & 24.64 & $>$27.30 & \nodata & $>$11.5 \\
D7 & 15:51:46.246 & 19:09:50.11 & 2.943 & 2.932 & 24.51 & $>$27.30 & \nodata & $>$13.0 \\
D11 & 15:51:49.764 & 19:09:02.94 & \nodata & 2.837 & 23.72 & $>$27.30 & \nodata & $>$27.1 \\
D13 & 15:51:51.724 & 19:10:15.89 & 2.852 & 2.842 & 24.39 & $>$27.30 & \nodata & $>$14.5 \\
D14 & 15:51:53.262 & 19:11:01.00 & 2.851 & 2.851 & 24.87 & $>$27.30 & \nodata & $>$9.3 \\
D16 & 15:51:54.848 & 19:11:31.18 & 3.139 & 3.130 & 24.00 & $>$27.30 & \nodata & $>$20.8 \\
D17 & 15:51:57.435 & 19:11:02.56 & 2.841 & 2.825 & 25.02 & $>$27.30 & \nodata & $>$8.1 \\
D18 & 15:51:59.695 & 19:09:39.25 & 2.850 & \nodata & 24.60 & $>$27.30 & \nodata & $>$12.1 \\
D19 & 15:52:00.270 & 19:09:40.75 & 2.847 & 2.844 & 25.28 & $>$27.30 & \nodata & $>$6.4 \\
D20 & 15:52:00.484 & 19:10:27.55 & \nodata & 2.825 & 24.11 & $>$27.30 & \nodata & $>$18.8 \\
D23 & 15:52:03.743 & 19:09:24.47 & 2.902 & 2.893 & 24.49 & $>$27.30 & \nodata & $>$13.3 \\
D24  & 15:52:05.278 & 19:09:45.17 & 2.951 & 2.942   & 24.24 & 27.01 & 1\secpoint1 & 12.8 $\pm$ 4.0 \\
D25 & 15:52:07.999 & 19:08:55.80 & \nodata & 2.825 & 24.79 & $>$27.30 & \nodata & $>$10.1 \\
D27 & 15:52:08.314 & 19:09:48.82 & 2.816 & 2.814 & 23.95 & $>$27.30 & \nodata & $>$21.8 \\
M2 & 15:51:41.355 & 19:10:06.11 & 2.875 & 2.867 & 25.32 & $>$27.30 & \nodata & $>$6.2 \\
M5 & 15:51:41.970 & 19:08:22.21 & 2.936 & 2.931 & 25.00 & $>$27.30 & \nodata & $>$8.3 \\
M6 & 15:51:43.678 & 19:09:44.58 & 2.891 & \nodata & 23.77 & $>$27.30 & \nodata & $>$25.9 \\
M16  & 15:51:53.636 & 19:09:29.49 & 2.955 & 2.953   & 25.28 & 26.56 & 1\secpoint3 & 3.2 $\pm$  1.4 \\
M21 & 15:52:01.356 & 19:13:00.78 & 2.834 & \nodata & 25.14 & $>$27.30 & \nodata & $>$7.3 \\
M22 & 15:52:02.705 & 19:09:40.06 & 3.159 & 3.149 & 24.77 & $>$27.30 & \nodata & $>$10.3 \\
M23 & 15:52:05.748 & 19:12:08.72 & \nodata & 3.409 & 24.85 & $>$27.30 & \nodata & $>$9.6 
\enddata
\tablenotetext{a}{Emission redshift of \lya.}
\tablenotetext{b}{Interstellar absorption redshift.}
\tablenotetext{c}{Spatial offset between the centroids of $V$-band and NB3420 flux-densities.}
\tablenotetext{d}{Observed ratio and uncertainty in the ratio of non-ionizing UV to LyC emission, inferred from the
NB3420$-V$ color. This value has not been corrected for either contamination by foreground sources or IGM absorption.}
\tablenotetext{e}{MD12 is not included in the LBG sample.}
\end{deluxetable*}

\begin{deluxetable*}{lccccccccc}
\tablewidth{0pt}
\tabletypesize{\scriptsize}
\tablecaption{LAE Photometry \label{tab_LAE}}
\tablehead{
\colhead{ID} & \colhead{RA}\tablenotemark{a} & \colhead{Dec}\tablenotemark{a} & \colhead{z} & \colhead{NB4670} &
\colhead{$V$} & \colhead{NB3420} & \colhead{$\Delta_{UV,\,LyC}$\tablenotemark{b}} &
\colhead{$\Delta_{Ly\alpha,\,LyC}$\tablenotemark{c}} & \colhead{$\frac{F_{UV}}{F_{LyC}}_{obs}$\tablenotemark{d}} \\
 & \colhead{(J2000)} & \colhead{(J2000)} & & & & & & &
}
\startdata
lae32 & 15:51:38.692 & 19:10:04.89 & 2.846 & 25.32 & 26.78 & $>$27.30 & \nodata & \nodata & $>$1.6 \\
lae274 & 15:51:39.406 & 19:09:42.66 & 2.851 & 24.41 & 25.47 & $>$27.30 & \nodata & \nodata & $>$5.4 \\
lae367 & 15:51:39.993 & 19:09:00.13 & 2.882 & 25.45 & 25.96 & $>$27.30 & \nodata & \nodata & $>$3.4 \\
lae413 & 15:51:39.673 & 19:13:15.43 & 2.846 & 24.47 & 27.21 & $>$27.30 & \nodata & \nodata & $>$1.1 \\
lae447 & 15:51:41.908 & 19:10:16.77 & 2.844 & 25.93 & 27.24 & $>$27.30 & \nodata & \nodata & $>$1.1 \\
lae576 & 15:51:40.601 & 19:10:58.09 & 2.846 & 25.17 & 26.24 & $>$27.30 & \nodata & \nodata & $>$2.6 \\
lae599 & 15:51:40.758 & 19:11:00.26 & 2.834 & 25.07 & $>$27.58 & $>$27.30 & \nodata & \nodata & \nodata \\
lae618 & 15:51:39.812 & 19:10:53.58 & 2.847 & 24.07 & 27.58 & $>$27.30 & \nodata & \nodata & $>$0.8 \\
lae633 & 15:51:39.624 & 19:11:04.99 & 2.852 & 23.91 & 26.78 & $>$27.30 & \nodata & \nodata & $>$1.6 \\
lae661 & 15:51:41.061 & 19:11:54.91 & 2.836 & 25.43 & 26.31 & $>$27.30 & \nodata & \nodata & $>$2.5 \\
lae1012 & 15:51:42.495 & 19:10:34.26 & 2.839 & 24.21 & 25.69 & $>$27.30 & \nodata & \nodata & $>$4.4 \\
lae1206 & 15:51:43.025 & 19:10:39.98 & 2.863 & 24.78 & 26.61 & $>$27.30 & \nodata & \nodata & $>$1.9 \\
lae1261 & 15:51:43.596 & 19:11:57.22 & 2.851 & 25.70 & 26.99 & $>$27.30 & \nodata & \nodata & $>$1.3 \\
lae1359 & 15:51:43.780 & 19:11:40.20 & 2.861 & 25.02 & 26.49 & $>$27.30 & \nodata & \nodata & $>$2.1 \\
lae1375 & 15:51:43.969 & 19:11:02.19 & 2.864 & 25.83 & 27.20 & $>$27.30 & \nodata & \nodata & $>$1.1 \\
lae1500\tablenotemark{e} & 15:51:43.955 & 19:11:39.63 & 2.863 & 23.93 & 24.43 & $>$27.30 & \nodata & \nodata & $>$14.1
\\
lae1528 & 15:51:44.682 & 19:09:24.59 & 2.846 & 25.66 & $>$27.58 & $>$27.30 & \nodata & \nodata & \nodata \\
lae1540 & 15:51:44.490 & 19:11:47.22 & 2.845 & 24.90 & 26.88 & $>$27.30 & \nodata & \nodata & $>$1.5 \\
lae1552 & 15:51:44.709 & 19:08:44.92 & 2.838 & 25.09 & $>$27.58 & $>$27.30 & \nodata & \nodata & \nodata \\
lae1599\tablenotemark{f} & 15:51:44.424 & 19:11:24.64 & 2.863 & 23.96 & 25.01 & $>$27.30 & \nodata & \nodata & $>$8.2
\\
lae1610 & 15:51:44.873 & 19:11:02.12 & 2.840 & 24.77 & 26.54 & $>$27.30 & \nodata & \nodata & $>$2.0 \\
lae1679 & 15:51:45.164 & 19:12:34.40 & 2.842 & 25.40 & 26.34 & $>$27.30 & \nodata & \nodata & $>$2.4 \\
lae1751 & 15:51:45.449 & 19:11:11.71 & 2.840 & 25.50 & 27.12 & $>$27.30 & \nodata & \nodata & $>$1.2 \\
lae1765 & 15:51:45.323 & 19:09:29.49 & 2.847 & 24.04 & 26.20 & $>$27.30 & \nodata & \nodata & $>$2.7 \\
lae1774 & 15:51:53.129 & 19:10:57.03 & 2.852 & 23.20 & 25.04 & $>$27.30 & \nodata & \nodata & $>$8.0 \\
lae1787 & 15:51:45.646 & 19:11:28.60 & 2.836 & 25.98 & 26.63 & $>$27.30 & \nodata & \nodata & $>$1.8 \\
lae1803 & 15:51:45.697 & 19:11:53.03 & 2.852 & 25.55 & $>$27.58 & $>$27.30 & \nodata & \nodata & \nodata \\
lae1835 & 15:51:45.882 & 19:09:45.11 & 2.836 & 25.45 & 27.50 & $>$27.30 & \nodata & \nodata & $>$0.8 \\
lae1843 & 15:51:45.676 & 19:09:58.21 & 2.847 & 25.81 & 27.18 & $>$27.30 & \nodata & \nodata & $>$1.1 \\
lae2015 & 15:51:46.504 & 19:12:38.62 & 2.845 & 25.81 & $>$27.58 & $>$27.30 & \nodata & \nodata & \nodata \\
lae2063 & 15:51:45.966 & 19:08:22.13 & 2.849 & 24.49 & 25.91 & $>$27.30 & \nodata & \nodata & $>$3.6 \\
lae2158 & 15:51:47.005 & 19:11:02.98 & 2.847 & 25.34 & 26.11 & $>$27.30 & \nodata & \nodata & $>$3.0 \\
lae2174 & 15:51:47.001 & 19:08:22.06 & 2.885 & 25.89 & 26.69 & $>$27.30 & \nodata & \nodata & $>$1.8 \\
lae2183 & 15:51:47.235 & 19:09:53.38 & 2.869 & 25.76 & 27.41 & $>$27.30 & \nodata & \nodata & $>$0.9 \\
lae2292 & 15:51:47.635 & 19:10:00.50 & 2.851   & 25.86 & 27.19 & 27.11 & 0\secpoint3 & 0\secpoint6 & 0.9 $\pm$ 0.5 \\
lae2306\tablenotemark{g} & 15:51:47.509 & 19:10:13.34 & 2.857 & 24.60 & 25.11 & $>$27.30 & \nodata & \nodata & $>$7.5
\\
lae2358 & 15:51:47.945 & 19:09:03.21 & 2.841 & 25.31 & $>$27.58 & $>$27.30 & \nodata & \nodata & \nodata \\
lae2436 & 15:52:03.231 & 19:12:52.43 & 2.832   & 23.44 & 24.11 & 25.21 & 0\secpoint3 & 0\secpoint3 & 2.7 $\pm$ 0.4 \\
lae2489 & 15:51:48.187 & 19:08:30.15 & 2.845 & 25.24 & 27.05 & $>$27.30 & \nodata & \nodata & $>$1.3 \\
lae2551 & 15:51:48.539 & 19:12:03.11 & 2.849 & 25.59 & 27.08 & $>$27.30 & \nodata & \nodata & $>$1.2 \\
lae2561 & 15:51:48.346 & 19:13:13.88 & 2.828 & 24.71 & 26.13 & $>$27.30 & \nodata & \nodata & $>$2.9 \\
lae2668 & 15:51:49.100 & 19:11:22.54 & 2.841 & 24.73 & 27.19 & $>$27.30 & \nodata & \nodata & $>$1.1 \\
lae2747 & 15:51:49.311 & 19:08:44.33 & 2.842 & 24.77 & 26.84 & $>$27.30 & \nodata & \nodata & $>$1.5 \\
lae2796 & 15:51:49.579 & 19:10:41.28 & 2.844 & 25.66 & 26.69 & $>$27.30 & \nodata & \nodata & $>$1.8 \\
lae2854 & 15:51:49.793 & 19:12:47.89 & 2.836 & 25.84 & $>$27.58 & $>$27.30 & \nodata & \nodata & \nodata \\
lae2856 & 15:51:49.718 & 19:10:49.05 & 2.843 & 25.15 & 25.85 & $>$27.30 & \nodata & \nodata & $>$3.8 \\
lae2949 & 15:51:49.870 & 19:10:53.62 & 2.840 & 25.14 & 27.45 & $>$27.30 & \nodata & \nodata & $>$0.9 \\
lae2966 & 15:51:49.995 & 19:10:41.59 & 2.841   & 24.61 & 26.87 & 26.82 & 0\secpoint1 & 0\secpoint7 & 1.0 $\pm$ 0.4 \\
lae2984\tablenotemark{h} & 15:51:49.695 & 19:10:57.98 & 2.843 & 23.83 & 24.92 & $>$27.30 & \nodata & \nodata & $>$9.0
\\
lae3167 & 15:51:50.781 & 19:10:34.29 & 2.846 & 25.62 & 26.89 & $>$27.30 & \nodata & \nodata & $>$1.5 \\
lae3208 & 15:51:50.908 & 19:11:16.18 & 2.844 & 25.28 & 27.32 & $>$27.30 & \nodata & \nodata & $>$1.0 \\
lae3339\tablenotemark{i} & 15:51:51.354 & 19:10:19.71 & 2.849 & 24.69 & 25.39 & $>$27.30 & \nodata & \nodata & $>$5.8
\\
lae3354 & 15:51:51.525 & 19:10:47.02 & 2.843 & 24.70 & 27.29 & $>$27.30 & \nodata & \nodata & $>$1.0 \\
lae3540\tablenotemark{j} & 15:51:51.878 & 19:10:41.09 & 2.856 & 23.93 & 24.56 & 26.74 & 1\secpoint0 & 0\secpoint9 &
7.5 $\pm$ 2.5 \\
lae3763 & 15:51:51.530 & 19:10:58.22 & 2.842 & 23.27 & 26.60 & $>$27.30 & \nodata & \nodata & $>$1.9 \\
lae3798 & 15:51:53.127 & 19:10:34.46 & 2.852 & 25.47 & 27.01 & $>$27.30 & \nodata & \nodata & $>$1.3 \\
lae3808 & 15:51:52.858 & 19:11:41.05 & 2.839 & 24.03 & 25.75 & $>$27.30 & \nodata & \nodata & $>$4.2 \\
lae3866 & 15:51:52.742 & 19:11:39.09 & 2.839 & 23.32 & 25.42 & $>$27.30 & \nodata & \nodata & $>$5.7 \\
lae3922 & 15:51:53.457 & 19:11:43.89 & 2.856 & 24.63 & 25.97 & $>$27.30 & \nodata & \nodata & $>$3.4 \\
lae4147 & 15:51:54.158 & 19:11:05.14 & 2.841 & 24.70 & 27.12 & $>$27.30 & \nodata & \nodata & $>$1.2  \\
lae4366 & 15:51:54.775 & 19:11:06.42 & 2.844 & 24.44 & 25.75 & $>$27.30 & \nodata & \nodata & $>$4.2 \\
lae4680 & 15:51:56.206 & 19:09:56.72 & 2.848   & 25.24 & 27.07 & 27.05 & 0\secpoint1 & 0\secpoint3 & 1.0 $\pm$ 0.4 \\
lae4684 & 15:51:56.167 & 19:11:56.65 & 2.849 & 25.50 & 26.15 & $>$27.30 & \nodata & \nodata & $>$2.9 \\
lae4730 & 15:51:56.413 & 19:10:42.03 & 2.843 & 25.46 & 26.12 & $>$27.30 & \nodata & \nodata & $>$3.0 \\
lae4796 & 15:51:56.367 & 19:11:06.15 & 2.868 & 24.68 & 26.30 & $>$27.30 & \nodata & \nodata & $>$2.5 \\
lae4804 & 15:51:56.621 & 19:12:26.86 & 2.860 & 25.49 & 26.99 & $>$27.30 & \nodata & \nodata & $>$1.3 \\
lae4882 & 15:51:56.873 & 19:13:06.72 & 2.835 & 25.23 & 27.35 & $>$27.30 & \nodata & \nodata & $>$1.0 \\
lae4947 & 15:51:57.132 & 19:08:48.24 & 2.857 & 24.95 & 25.49 & $>$27.30 & \nodata & \nodata & $>$5.3 \\
lae5132 & 15:52:01.057 & 19:09:47.89 & 2.849 & 24.76 & 27.18 & $>$27.30 & \nodata & \nodata & $>$1.1 \\
lae5193 & 15:52:00.760 & 19:10:53.36 & 2.853 & 25.26 & 26.21 & $>$27.30 & \nodata & \nodata & $>$2.7 
\enddata
\end{deluxetable*}

\setcounter{table}{3}

\begin{deluxetable*}{lccccccccc}
\tablewidth{0pt}
\tabletypesize{\scriptsize}
\tablecaption{LAE Photometry \label{tab_LAE}}
\tablehead{
\colhead{ID} & \colhead{RA}\tablenotemark{a} & \colhead{Dec}\tablenotemark{a} & \colhead{z} & \colhead{NB4670} &
\colhead{$V$} & \colhead{NB3420} & \colhead{$\Delta_{UV,\,LyC}$\tablenotemark{b}} &
\colhead{$\Delta_{Ly\alpha,\,LyC}$\tablenotemark{c}} & \colhead{$\frac{F_{UV}}{F_{LyC}}_{obs}$\tablenotemark{d}} \\
 & \colhead{(J2000)} & \colhead{(J2000)} & & & & & & &
}
\startdata
lae5322 & 15:52:00.102 & 19:10:17.14 & 2.832 & 23.53 & 25.44 & $>$27.30 & \nodata & \nodata & $>$5.5 \\
lae5371\tablenotemark{k} & 15:52:00.265 & 19:09:40.41 & 2.848 & 24.54 & 25.28 & $>$27.30 & \nodata & \nodata & $>$6.4
\\
lae5458\tablenotemark{l} & 15:51:59.696 & 19:09:39.30 & 2.849 & 24.04 & 24.60 & $>$27.30 & \nodata & \nodata & $>$12.1
\\
lae5470 & 15:51:59.582 & 19:11:40.17 & 2.843 & 25.85 & $>$27.58 & $>$27.30 & \nodata & \nodata & \nodata \\
lae5713 & 15:51:58.999 & 19:09:20.60 & 2.857 & 25.57 & $>$27.58 & $>$27.30 & \nodata & \nodata & \nodata \\
lae5720 & 15:51:58.755 & 19:13:00.43 & 2.833 & 25.56 & $>$27.58 & $>$27.30 & \nodata & \nodata & \nodata \\
lae5740 & 15:51:59.315 & 19:10:35.10 & 2.831 & 24.96 & 25.90 & $>$27.30 & \nodata & \nodata & $>$3.6 \\
lae5900 & 15:51:58.057 & 19:11:21.84 & 2.845 & 24.24 & 25.42 & $>$27.30 & \nodata & \nodata & $>$5.6 \\
lae5995\tablenotemark{m} & 15:51:57.456 & 19:11:02.63 & 2.833 & 24.54 & 25.02 & $>$27.30 & \nodata & \nodata & $>$8.1
\\
lae6193 & 15:52:07.992 & 19:11:23.26 & 2.836 & 25.79 & 26.10 & $>$27.30 & \nodata & \nodata & $>$3.0 \\
lae6312 & 15:52:07.610 & 19:08:47.48 & 2.827 & 25.72 & 26.43 & $>$27.30 & \nodata & \nodata & $>$2.2 \\
lae6662 & 15:52:06.357 & 19:10:42.74 & 2.833   & 25.16 & 25.94 & 27.23 & 1\secpoint1 & 1\secpoint2 & 3.3 $\pm$ 1.3 \\
lae6774 & 15:52:06.090 & 19:11:36.81 & 2.845 & 25.80 & $>$27.58 & $>$27.30 & \nodata & \nodata & \nodata \\
lae6979 & 15:52:04.940 & 19:09:53.42 & 2.863 & 25.63 & 26.25 & $>$27.30 & \nodata & \nodata & $>$2.6 \\
lae7180 & 15:52:04.662 & 19:11:42.19 & 2.930 & 25.86 & 26.38 & 25.85 & 0\secpoint3 & 0\secpoint1 & 0.6 $\pm$ 0.2 \\
lae7542 & 15:52:03.575 & 19:12:50.73 & 2.841 & 24.28 & 25.48 & $>$27.30 & \nodata & \nodata & $>$5.4 \\
lae7577 & 15:52:03.189 & 19:09:09.08 & 2.828 & 25.35 & $>$27.58 & $>$27.30 & \nodata & \nodata & \nodata \\
lae7803 & 15:52:02.304 & 19:09:03.58 & 2.865 & 25.97 & 26.89 & $>$27.30 & \nodata & \nodata & $>$1.4 \\
lae7830 & 15:52:02.146 & 19:09:55.06 & 2.826 & 23.77 & $>$27.58 & $>$27.30 & \nodata & \nodata & \nodata \\
lae7832 & 15:52:02.201 & 19:10:48.59 & 	2.829   & 24.23 & 24.83 & 25.11 & 0\secpoint4 & 0\secpoint7 & 1.3 $\pm$ 0.2 \\lae7893 & 15:52:01.861 & 19:12:49.92 & 2.858 & 25.32 & $>$27.58 & $>$27.30 & \nodata & \nodata &
 \nodata 
\enddata
\tablenotetext{a}{Coordinates of LAEs are based on NB4670 centroids.}
\tablenotetext{b}{Spatial offset between the centroids of $V$ and NB3420 emission.}
\tablenotetext{c}{Spatial offset between the centroids of NB4760 and NB3420 emission.}
\tablenotetext{d}{Observed ratio and uncertainty in the ratio of non-ionizing UV to LyC flux-densities, inferred from
the NB3420$-V$ color. This value has not been corrected for either contamination by foreground sources or IGM
absorption.}
\tablenotetext{e}{D4}
\tablenotetext{f}{C4}
\tablenotetext{g}{D7}
\tablenotetext{h}{C13}
\tablenotetext{i}{C15}
\tablenotetext{j}{MD12.  This object is not included in either the LAE or LBG samples.}
\tablenotetext{k}{D19}
\tablenotetext{l}{D18}
\tablenotetext{m}{D17}
\end{deluxetable*}

\begin{deluxetable*}{lccccc}
\tablecolumns{16}
\tablewidth{0pt}
\tablecaption{Photometry in Stacked Images \label{tab_stacks}}
\tablehead{Sample & N$_{gal}$ & $V$\tablenotemark{a} & NB3420\tablenotemark{a} & NB3420$-V$ & $\langle F_{UV}/F_{LyC}
\rangle_{obs}$\tablenotemark{b}}
\startdata

LBG, all & 48 & $24.59^{+0.09}_{-0.08}$ & $>28.67$ & $>4.08$ & $>42.7$  \\
LBG, detect & 4 & $24.44^{+0.73}_{-0.43}$ & $27.05^{+0.63}_{-0.39}$ & $2.60^{+0.55}_{-1.16}$ & $11.0 \pm 7.2$  \\
LBG, non-detect & 44 & $24.60^{+0.09}_{-0.08}$ & $>28.62$ & $>4.03$ & $>40.8$  \\
\tableline
LAE, all & 90 & $ 26.03^{+0.14}_{-0.13}$ & $>29.01$ & $>2.98$ & $>15.6$ \\
LAE, detect & 7 & $ 25.49^{+0.40}_{-0.29}$ & $ 25.97^{+0.35}_{-0.26}$ & $ 0.48^{+0.38}_{-0.58}$ & $ 1.6 \pm 0.6 $ \\
LAE, non-detect & 83 & $ 26.09^{+0.15}_{-0.13}$ & $>28.97$ & $>2.88$ & $>14.2$ \\
\tableline
LAE, $24<V< 25$ & 5 & $ 24.67^{+0.09}_{-0.08}$ & $ 26.51^{+2.01}_{-0.66}$ & $ 1.84^{+0.67}_{-2.04}$ & $ 5.5 \pm 4.6 $
\\
LAE, $25<V< 26$ & 21 & $ 25.53^{+0.08}_{-0.07}$ & $>28.22$ & $>2.69$ & $>11.9$ \\
LAE, $26<V< 27$ & 30 & $ 26.31^{+0.07}_{-0.06}$ & $>28.41$ & $>2.10$ & $>6.9$ \\
LAE, $V> 27$ & 34 & $ 27.31^{+0.18}_{-0.16}$ & $>28.48$ & $>1.17$ & $>2.9$ \\ 
\enddata
\tablenotetext{a}{Uncertainties listed are 1$\sigma$ and include both photometric error and sample variance. All
photometric lower limits are 3$\sigma$.}
\tablenotetext{b}{Observed non-ionizing UV to LyC flux-density ratios and uncertainties, inferred from the NB3420$-V$
color of each stacked subsample. These values have not been corrected for either contamination by foreground sources
or IGM absorption.}
\end{deluxetable*}

As indicated by studies of the SSA22a field \citep{nestor13,nestor11,inoue10,iwata09} and implied by the simulations
of \citet{gnedin08} and \citet{ricotti02} (see Section \ref{ssec_phot}), the centroids of the UV-continuum and LyC
emission may not always coincide. The NB3420 versus $V-$band offsets for the four LBGs with NB3420 detections span a
large range: 0\secpoint34, 0\secpoint85, 1\secpoint06, and 1\secpoint26. The offsets for the 7 LAEs with NB3420
detections, however, are $\leq 0\secpoint36=2.8$ kpc for all but one LAE (\emph{lae6662}), which has an offset of
1\secpoint09. The smaller offsets of the NB3420 detections around LAEs strengthen the argument that these detections
are truly associated with the LAEs in question, and also may indicate that LAEs are more compact galaxies than LBGs.
For LAEs, we may also examine the offset between \lya\ emission and LyC emission ($\Delta_{Ly\alpha,\,LyC}$) which is
on average larger than both the offset between the UV-continuum and LyC ($\Delta_{UV,\,LyC}$) and the offset between
UV-continuum and \lya\ ($\Delta_{UV,Ly\alpha}$). While some spatial discrepancy between \lya\ and LyC emission is to
be expected due to the resonant scattering of \lya\ photons, large offsets point to an increased probability of
contamination and weaken the argument that the NB3420 detection is LyC emission associated with the high-redshift
galaxy \citep{nestor11,nestor13,vanzella10,vanzella12}. However, while it is a good rule of thumb to assume that
objects with $\Delta_{UV,\,LyC}\gtrsim1$\arcsec\ are likely contaminants, \citet{nestor13} demonstrated with
high-resolution spectroscopy in the SSA22a field individual cases where putative LyC detections at large offset (for
example, $\Delta_{UV,\,LyC}=$1\secpoint0) were found to be associated with the high-redshift galaxy, and detections at
small offset (for example, $\Delta_{UV,\,LyC}=$0\secpoint3) were found to be contaminants.

The fact that $\Delta_{Ly\alpha,\,LyC}$ is often greater than $\Delta_{UV,\,LyC}$ provides additional evidence against
significant contamination from low-redshift [OII]-emitters in the LAE sample, corroborating the conclusion drawn in
Section \ref{ssec_samp}. Figure \ref{fig_offset_scatter_plot} compares the offsets $\Delta_{Ly\alpha,\,LyC}$ and
$\Delta_{UV,\,LyC}$ for LAEs, demonstrating the tendency for NB3420 detections to be more closely associated with $V$
rather than NB4670 detections. If the LAEs were low-redshift [OII]-emitters rather than high-redshift LAEs, the NB3420
image would probe the rest-frame UV, the NB4670-$V$ image would probe [OII] emission, and the $V$ image would probe
the rest-frame optical. In this case, both the NB3420 and NB4670-$V$ images would probe active star formation, so a
smaller offset would be expected between the centroids of the detections in these two images. However, larger values
of $\Delta_{Ly\alpha,\,LyC}$ support the opposite interpretation; namely, the LAE candidates are truly high-redshift
objects whose LyC and \lya\ emission correlate more strongly with the non-ionizing UV-continuum than with each other.
In Tables \ref{tab_LBG} and \ref{tab_LAE}, we list the offsets $\Delta_{UV,\,LyC}$ and $\Delta_{Ly\alpha,\,LyC}$ for
galaxies with NB3420 detections.

In addition to exploring the properties of individual objects leaking LyC radiation, we created stacked NB3420 and $V$
images of subsamples of our targets to examine the average LyC emission properties and attempt a LyC measurement in
the stacks of objects without NB3420 detections. These subsamples include LBGs, LAEs, and LAEs in bins of $V$
magnitude. For each subsample, we also created two additional stacks comprising objects with and without NB3420
detections, respectively. Stacked images were made by averaging postage stamps of individual galaxies centered on the
coordinates of either their $V$ (LBGs) or NB4670 (LAEs) detections. Pixels contaminated by nearby objects were
excluded from the stacks by creating a mask from the object detection isophotes in the SExtractor segmentation image.
Stacked image photometry was performed using the IRAF PHOT routine using a 1\secpoint9 aperture, corresponding to the
matching radius used to find detections. Although the postage stamps for individual galaxies were previously
background subtracted as part of the data reduction process, we found that a second pass of background subtraction on
the stacked image was necessary to remove remaining sky subtraction systematics. Bootstrap resampling was employed for
each stack in order to include both sample variance and photometric error in the calculation of stack uncertainties.
The results of the stacking analysis are presented in Table \ref{tab_stacks}. None of the NB3420 stacks of individual
LBGs and LAEs without LyC detections exhibited any significant flux; the NB3420 stack of LBG nondetections reached a
3$\sigma$ limiting magnitude of 28.62 and the NB3420 stack of LAE nondetections reached a 3$\sigma$ limiting magnitude
of 28.97. These limits reflect only photometric errors. We note that for the NB3420 stacks of all LBGs and all LAEs,
the addition of noise from a large number of galaxies undetected in NB3420 overpowered the signal from the few
detected galaxies. As the resulting signal in the stack was detected at less than 3$\sigma$, we quote a lower limit in
magnitude in Table \ref{tab_stacks}.

\section{Accounting for Contamination} \label{sec_contam}

In narrowband imaging studies of $z \sim 3$ LyC emission, it is critical to determine whether the detection in the
narrowband filter is actually high-redshift LyC emission or contaminating radiation from a lower-redshift object. In
this section, we analyze the morphology of several candidate LyC-emitters for which we have high-resolution $HST$
imaging and discuss the complexities of identifying contaminants. Because we do not have high-resolution, multi-band
imaging for all of our candidate LyC-emitters, we also discuss statistical corrections applied to our samples in order
to account for foreground contamination by low-redshift galaxies (which artificially boosts the LyC signal) and LyC
absorption by the IGM (which decreases the observed LyC emission).

\begin{figure*}
\epsscale{1.0}
\plotone{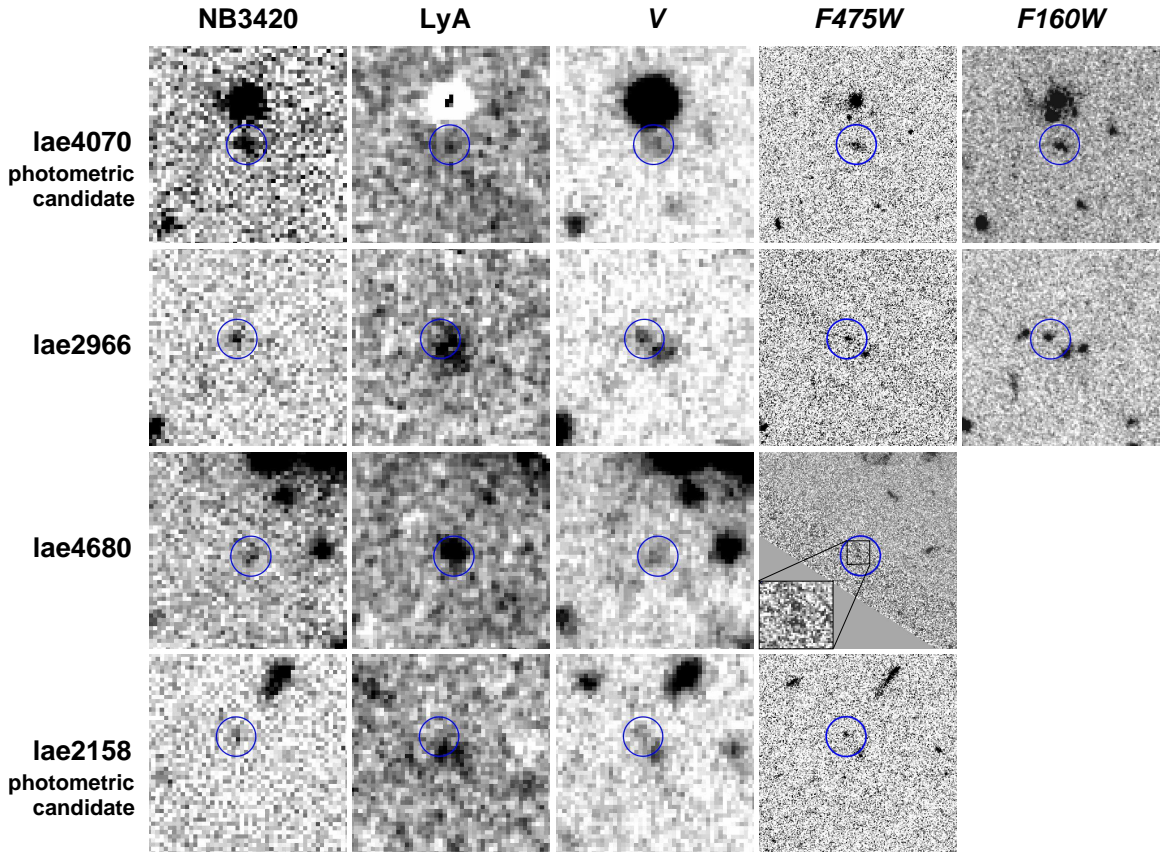}
\caption{
\small
10\secnopoint\ $\times$ 10\secnopoint\ postage stamp images of four LAEs with HST imaging and NB3420 detections (the
LAE photometric candidates \emph{lae4070} and \emph{lae2158}, along with spectroscopically confirmed LAEs
\emph{lae2966} and \emph{lae4680}). The postage stamps highlight the difficulties associated with interpreting the
morphologies of these objects with ground-based resolution. Objects are displayed in up to 5 filters: NB3420
(indicating the LyC), NB4670$-V$ (indicating \lya\ emission and labeled LyA), $V$ (indicating the non-ionizing UV
continuum), $F475W$ (rest-frame $\sim$1200\AA), and $F160W$ (rest-frame $\sim$4000\AA). All postage stamps are
centered on the $V$-band centroid and blue circles (1\secnopoint\ radius) indicate the centroid of the NB3420
emission. All postage stamps follow the conventional orientation, with north up and east to the left.
\label{fig_HSTstamps}
}
\end{figure*}

\begin{figure*}
\epsscale{1.0}
\plotone{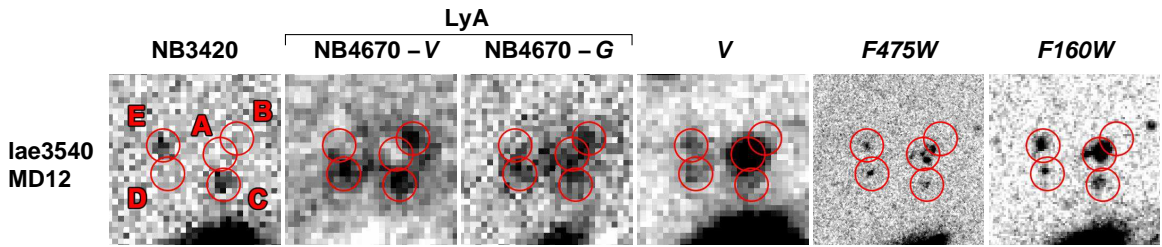}
\caption{
\small
3\secpoint2 $\times$ 3\secpoint2 postage stamp images of the MD12/\emph{lae3540}, an additional object with a
potential LyC detection. Originally identified as both an LBG and LAE, MD12/\emph{lae3540} was removed from both
samples due to its complex morphology (see Section \ref{ssec_nbmorph}). The object is displayed in 6 filters: NB3420
(indicating the LyC), NB4670$-V$ and NB4670$-G$ (both indicating \lya\ emission), $V$ (indicating the non-ionizing UV
continuum), $F475W$ (rest-frame $\sim$1200\AA), and $F160W$ (rest-frame $\sim$4000\AA). The clumpy morphology for
MD12/\emph{lae3540} is indicated by five red circles labeled A$-$E. Region A corresponds to the LBG and LAE centroids
for MD12/\emph{lae3540}, while regions B, C, and D correspond to areas of \lya\ emission indicated by the NB4670$-V$
image. Region C also corresponds to the location of the NB3420 emission. Region E likely indicates a lower-redshift
galaxy along the line of sight. All postage stamps follow the conventional orientation, with north up and east to the
left.
\label{fig_MD12}
}
\end{figure*}

\subsection{Morphology of NB3420 Detections} \label{ssec_nbmorph}

In order to cull potential LyC detections from obvious neighbors, we have visually inspected each NB3420 detection in
all available bands (as discussed in Section \ref{ssec_objmat}). This process of visual inspection, however, is
limited by the depth and resolution of the data. Within our dataset, the $HST$/WFC3 imaging has the best resolution,
followed by the ground-based optical data from Keck. The WFC3/UVIS \emph{F475W} and WFC3/IR \emph{F160W} images have
PSF FWHMs of 0\secpoint08 and 0\secpoint18, respectively \citep{law12}, while the seeing FWHMs of the Keck data are
larger (0\secpoint7 $-$ 1\secpoint0). Unfortunately, only 1 (2) of the targets in our samples with NB3420 matches fall
within the small 2\minpoint3 $\times$ 2\minpoint1 (2\minpoint9 $\times$ 2\minpoint7) field of view of the WFC3 IR
(UVIS) imaging. Three additional objects with NB3420 detections (two in the photometric LAE sample and one LAE/LBG
that we removed from both samples for its complicated morphology) also fall in the field of view of the WFC3 IR and
UVIS images. Examining the morphologies of these objects across the ground-based and HST images highlights the
difficulties associated with object matching.

Figure \ref{fig_HSTstamps} shows four LAEs with HST imaging and NB3420 detections (the LAE photometric candidates
\emph{lae4070} and \emph{lae2158}, along with spectroscopically confirmed LAEs \emph{lae2966} and \emph{lae4680})
displayed in five filters: NB3420 (LyC), NB4670$-V$ (\lya), $V$ (UV continuum), F475W (rest-frame $1240$\AA), and
F160W (rest-frame $4000$\AA). \emph{Lae4070} is an example of an object with a simple morphology. In all images, a
single detection is visible, and the offsets between the centroids of the detection in each image are very small.
Similarly, \emph{lae4680} can be indentified with a single source in all images except $F475W$, where the object
breaks into two clumps located 0\secpoint3 apart. As this offset is very small, these clumps are likely part of the
same system, although foreground contamination is still a possibility. For \emph{lae4070} and \emph{lae4680}, the only
possibilities of contamination arise from either the small probability of [OII] emission being misidentified as \lya\
emission or a lower-redshift foreground galaxy coincident with the LAE along the line of sight. We discussed the first
possibility in Section \ref{ssec_samp} and concluded that it is small enough to ignore. The second possibility is
quantified in Section \ref{sssec_foreground_contam} in the discussion of the contamination simulations. The analysis
of the remaining two objects is more complex. In the case of \emph{lae2966}, multiple clumps are visible in the
\emph{V}-band, $F475W$, and $F160W$ imaging, and the NB3420 emission is associated with only one clump. The
spectroscopically-confirmed \lya\ emission is extended and not distinctly associated with any single clump. Therefore,
it is unclear whether the NB3420 detection is LyC emission from one clump of a $z \sim 2.85$ galaxy or a low-redshift
interloper along the line of sight. In the case of \emph{lae2158}, the HST imaging can be used to identify the NB3420
detection as contamination. While it is not possible to determine in the $V$-band image whether the UV-continuum light
associated with the LAE belongs to multiple objects or whether it is merely extended, there are two distinct galaxies
visible in the $F475W$ image. Since the northeastern galaxy is associated with the NB3420 detection while the
southwestern galaxy is associated with the \lya\ emission, we classify this system as a case of contamination.

We also highlight one object whose complex morphology led us to remove it from our analysis completely. This object
(shown in Figure \ref{fig_MD12}, with multiple clumps indicated) was originally identified in our catalogs as both an
LBG (MD12) and an LAE (\emph{lae3540}), with both the LBG and LAE centroids coincident with the bright $V-$band
detection (indicated by region A). Analysis of the \lya\ morphology in both the NB4670$-V$ and NB4670$-G$ images
reveals extended emission in both \lya\ images. Although both \lya\ images show nearly identical morphology for all
other LAEs with NB3420 detections, the extended \lya\ emission around MD12 appears clumpy in the NB4670$-V$ image
(associated with regions B, C, and D) and more diffuse in the NB4670$-G$ image. Also, while the \lya\ emission appears
to extend down to region C where the NB3420 detection is located (1\secpoint0 to the south of the LBG centroid), this
location is also coincident with detections in the $F475W$ and $F160W$ images that are not visibly connected to the
LBG centroid in region A. Unfortunately, due to the slit position of $74^{\circ}$, the LRIS spectrum of this object
only provides us with information about region A (where we observe double-peaked \lya\ emission in the spectrum) and
misses the region associated with the NB3420 detection. We note that region E likely indicates a lower-redshift
interloper along the line of sight, as its position does not coincide with \lya\ emission and it is located at the
large offset of 2\secpoint1 from region A (the centroid of MD12).

The interpretation of the suite of imaging for MD12 is not straightforward, and it is impossible to fully understand
the nature of these clumps with the current data. The clumps may comprise an extended, perhaps interacting, system,
with all clumps at the same redshift. Alternatively, as region A has \lya\ emission in its spectrum and regions B, C,
and D are all associated with fairly compact \lya\ emission in the NB4670$-V$ image, these four clumps may simply be
several protocluster members along the line of sight located at slightly different redshifts; in this case, they
should be treated as separate LAEs. A final possibility (supported by the fact that the \lya\ emission appears diffuse
in the NB4670$-G$ image) is that the extended \lya\ emission actually originates in the central part of the MD12
system and is being projected over a large area; in this case, the clump with the NB3420 emission in region C may be
an interloper along the line of sight. Because of the ambiguity regarding the nature of MD12 as a either single,
complex system or the superposition of multiple galaxies, we exclude both MD12 and \emph{lae3540} from our LBG and LAE
analysis. Nevertheless, this object represents another possible LyC-emitter.

Determining conclusively whether the NB3420 detections are contamination or true LyC emission requires imaging of each
candidate galaxy with high enough resolution to discern the individual clumps and spectroscopy at high enough
resolution that distinct spectra are obtained (and redshifts calculated) for each emitting region. This method has
been recently implemented in the SSA22a field using $HST$/WFC3/$F336W$ to acquire high-resolution imaging below the
Lyman limit for three $z \sim 3.1$ LBGs. Additionally, near-infrared spectroscopy of rest-frame optical [OIII] nebular
emission was obtained using NIRSPEC on Keck II \citep[][Siana et al., in prep]{nestor13}. The seeing of
$\sim$0\secpoint5 for these near-infrared observations enabled the spatial separation of clumps. One of three LBGs was
confirmed to have escaping LyC radiation, and two others showed evidence of low-redshift contamination. Unfortunately,
a similar study for the HS1549 field is not possible using ground-based or current space-based instrumentation. There
is no appropriate imaging filter on $HST$ that probes the region just below the Lyman limit at $z=2.85$ without some
contamination redwards of the limit. Furthermore, spectroscopy from the ground is unfeasible because the rest-frame
optical features do not fall within windows of atmospheric transmission at $z=2.85$. The next best option is to
estimate photometric redshifts of each clump by acquiring high-resolution multi-band imaging. Such a technique has
been successfully employed by \citet{vanzella12} for a sample of 19 LBGs with potential LyC emission at $3.4 \le z \le
4.5$ using multi-band HST imaging from the GOODS and CANDELS surveys.

\subsection{Contamination from Foreground Galaxies} \label{sssec_foreground_contam}

Although we currently do not have the high-resolution imaging or spectroscopic data required to prove the validity of
each individual LyC detection, we can characterize the probability of foreground contamination statistically. With the
assumption that all of our targets have been correctly identified as high-redshift galaxies, the NB3420 flux we
measured can either be associated with LyC emission from the high-redshift galaxy itself or with contaminating
radiation from a spatially coincident lower-redshift interloper. In order to statistically characterize the fraction
of contaminated NB3420 detections in a given sample, we performed a simulation to calculate the expected number of
uncontaminated LyC detections and the contamination-corrected average NB3420 magnitude for that sample. We summarize
the simulation below, and further details are described in Section 5.1 of \citet{nestor13}.

\begin{figure}
\epsscale{1.0}
\plotone{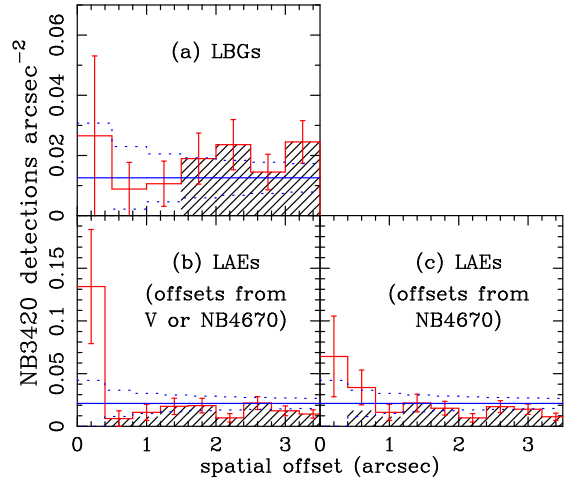}
\caption{
\small
Radial surface density of NB3420 detections within 3\secpoint5 of galaxies in our LBG and LAE samples (solid red
histograms). The radial bin size for the LBG histogram is 0\secpoint5, and the bin size for the LAE histograms is
0\secpoint4. Red histogram error bars represent 1$\sigma$ Poisson uncertainties in the number of NB3420 detections in
each bin. The blue solid line represents the average surface density of NB3420 detections within the magnitude range
($\pm 1\sigma$) of the LAEs/LBGs, corresponding to the expected amount of contamination. Blue dotted lines indicate
the 1$\sigma$ uncertainties in the average surface density in each bin. Black hashed regions on the red histogram
represent NB3420 detections that have been rejected by eye as contaminants and are not considered candidates for LyC
emission. As described in Section \ref{sssec_foreground_contam}, the surface density of NB3420 detections around LAEs
is plotted using the NB3420 offsets calculated from LAE position defined by either a combination of NB4670 and
$V-$band centroids (panel b) or NB4670 centroids only (panel c). In both cases, we observe a significant excess in
surface density of NB3420 detections at small offset from LAEs, which indicates that many of the LAE NB3420 detections
are likely uncontaminated LyC emission. We also note that panels (b) and (c) confirm that LyC emission is on average
more closely associated with $V-$band emission than with NB4670, as the average offset in panel (b) is less than that
in panel (c). As there are only 4 LBGs with NB3420 detections, the LBG excess over the average surface density at low
spatial offset is much smaller.
\label{fig_contam_hist}
}
\end{figure}

The contamination simulation was run separately for the LBGs and LAEs. Within the simulation, we considered as
possible LyC detections all NB3420 SExtractor detections within 3\secpoint5 of our targets that fall within 1$\sigma$
of the magnitude range of our reported LyC detections, corresponding to $26.19 \leq m_{3420} \leq 27.59$ for LBGs and
$24.98 \leq m_{3420} \leq 27.70$ for LAEs. These SExtractor detections included both the detections considered to be
candidates for LyC emission and detections previously identified by eye as belonging to interlopers visible in other
wavelength bands. We computed the probability of each NB3420 detection being an interloper, which depends on both the
global surface density of objects in the NB3420 image and the local surface density at the offset of the NB3420
detection from each LBG and LAE. In each iteration of the simulation, we used these probabilities to randomly
determine whether or not each NB3420 detection was flagged as an interloper. Detections at large radial offset $-$
where the local surface density approaches the global surface density $-$ are more likely to be flagged as
interlopers, as demonstrated in Figure \ref{fig_contam_hist}. We removed the NB3420 magnitudes of objects flagged as
interlopers before the average NB3420 magnitude was calculated, resulting in the contamination-corrected average
NB3420 magnitude. The simulation was repeated 1000 times, and we recorded the average number of uncontaminated NB3420
detections and the average contamination-corrected NB3420 magnitude for both samples. Uncertainties in these average
values were computed from the distribution of simulation results.

In Figure \ref{fig_contam_hist}, we present histograms of the radial surface density of NB3420 detections around
galaxies in the LBG and LAE samples, compared to the global surface density. The surface density of NB3420 detections
is plotted using NB3420 offsets computed relative to $V-$band (non-ionizing UV continuum) for LBGs. For LAEs, offsets
were computed relative to both $V-$band and NB4670 (a combination of \lya\ and continuum emission in the vicinity of
the \lya\ wavelength). As discussed in Section \ref{ssec_det}, LyC emission is spatially more closely associated with
the UV-continuum than with \lya\ emission. Therefore, we adopt $V$-band coordinates to represent the LAE centroids,
except in cases where the LAE is undetected in $V$ and we must use NB4670 coordinates. Panels (b) and (c) of Figure
\ref{fig_contam_hist}, displaying $V-$band and NB4670 offsets, respectively, confirm that LyC emission is on average
more closely associated with $V-$band emission than with NB4670, as the average offset in panel (b) is less than that
in panel (c). All panels show an excess surface density at small offsets (although this excess is only statistically
significant for the LAEs), suggesting that a large number of the NB3420 detections are physically associated with the
LBG and LAE targets; the contamination simulation statistically quantifies this number. For the 4 LBGs detected in
NB3420, the simulations predict on average 1.5 $\pm$ 1.0 uncontaminated detections. For the 7 NB3420-detected LAEs,
the simulations predict 4.3 $\pm$ 1.3 uncontaminated detections. For both samples, the simulation yields the
contamination-corrected average magnitude of the NB3420 detections. These values are used to compute the
sample-averaged NB3420$-V$ colors and flux-density ratios presented in Table \ref{tab_final_avg_flux_ratios} and
discussed in Section \ref{sec_res}. We note that the removal of contaminants must increase the non-ionizing UV to LyC
flux-density ratio for the ensembles containing the full sample of LBGs (or LAEs), as the total amount of LyC flux is
decreased when a contaminant is removed while the total amount of non-ionizing UV flux stays the same. For ensembles
that contain only objects with NB3420 detections, however, the flux-density ratio may not necessarily increase with
the removal of the contaminants. In these cases, when an NB3420 detection is identified as a contaminant, both its LyC
and non-ionizing UV emission are omitted from the sample when calculating the flux-density ratio as the object is no
longer considered to have a true LyC detection.

\subsection{Correction for IGM Absorption} \label{sssec_IGM_corr}

In addition to correcting for contamination by low-redshift galaxies, we must also correct the NB3420 photometry for
the absorption of LyC photons by neutral hydrogen in the IGM. In order to account for such attenuation, we ran a
second set of simulations to determine the mean IGM correction factor and its associated uncertainty. We note that, as
discussed in \citet{nestor11}, our simulations do not take into account the possible environmental effects due to the
presence of the $z\sim2.85$ protocluster and the proximity to the hyperluminous QSO (Q1549), and it is not clear
whether such effects would culminate in an increase or decrease of neutral hydrogen absorbers. We summarize the
methods employed below, and further details are described in \citet{nestor11} and \citet{nestor13}, the latter of
which describes the current version of the methodology.

\begin{figure}
\epsscale{1.0}
\plotone{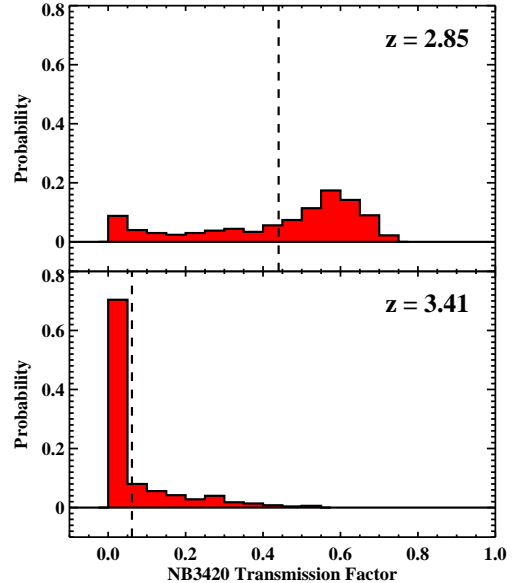}
\caption{
\small
The probability distribution of the LyC transmission factor through the NB3420 filter for an object at $z=2.85$ (top)
and an object at $z=3.41$ (bottom). The mean transmission at each redshift is indicated by the dashed lines. The top
panel represents the vast majority of LBGs and LAEs that lie near the redshift spike at $z=2.85$. The wide range of
possible transmission values at $z=2.85$ reflects the variation in IGM absorption at $z=2.85$ and the slight peak
around zero transmission corresponds to sightlines that encounter a Lyman limit system. The bottom panel corresponds
to the redshift of M23, an LBG with a redshift far removed from the spike, and represents the most extreme case of
attenuation. This large amount of attenuation is due to a combination of two effects. Not only is the sightline
through the IGM longer for the photons coming from $z=3.41$ (thus providing each photon with more time to encounter an
absorber), but the $z=3.41$ IGM has a higher fraction of neutral hydrogen $-$ and thus more absorbers $-$ than the
$z=2.85$ IGM.
\label{fig_IGM}
}
\end{figure}

First, we constructed 500 model sightlines simulating the distribution of HI absorbers in the IGM. For each model
sightline, absorbers were drawn randomly from their redshift and column density distributions \citep{rudie13},
spanning redshifts from $z=1.7$ to the redshift of each object of interest. The model sightlines for the LAEs were
created at the mean redshift of the LAEs ($z=2.85$) since the narrow width of the NB4670 filter implies that the LAE
targets lie within a small redshift range. Given that the LBGs span a wider range in redshift ($2.815 \leq z \leq
3.414$), we created a set of 500 model sightlines at the redshift of each LBG. We then calculated the mean
transmission of each model sightline in the LyC region. For the purposes of these simulations, the LyC region consists
of the fixed observed-frame bandpass of the NB3420 filter, taking into account the filter transmission profile. Figure
\ref{fig_IGM} shows the probability distributions of the LyC transmission factor for model sightlines at $z=2.85$
(representing typical LAE redshifts) and $z=3.41$ (corresponding to our highest-redshift LBG with the most extreme
case of attenuation).

The sample average transmission through the IGM ($\bar{t}_{sample}$) is equal to the mean transmission of the 500
$z=2.85$ sightlines for the LAEs and the mean transmission of all model sightlines for the LBGs at different
redshifts. The uncertainty in $\bar{t}_{sample}$ is estimated by first assuming an exponential distribution of
unattenuated LyC flux \citep[the parameters of this exponential function are fit to our data via a maximum likelihood
method, see][]{nestor13} and then creating 1000 realizations of our sample by randomly choosing for each galaxy an
unattenuated LyC flux from our exponential distribution and an attentuation factor from one of the simulated model
sightlines at the redshift of the galaxy. We set the uncertainty in $\bar{t}_{sample}$ equal to the standard deviation
of the 1000 simulated $\bar{t}_{sample}$ values. We find that $\bar{t}_{sample, LAE}=0.44 \pm 0.03$ and
$\bar{t}_{sample, LBG}=0.35 \pm 0.04$. For both the LBG and LAE samples, we multiply the contamination-corrected
non-ionizing to ionizing flux-density ratios by the transmission factor to obtain the IGM-and-contamination-corrected
values presented in Table \ref{tab_final_avg_flux_ratios}. Unlike the contamination correction discussed in Section
\ref{sssec_foreground_contam}, which decreases the average NB3420 flux, the IGM attenuation correction acts to
increase it.

\section{Results} \label{sec_res}

In order to study the amount of ionizing radiation escaping star-forming galaxies at high redshift, we have imaged a
large sample of $z \sim 2.85$ galaxies in a narrowband filter designed to probe LyC emission. As described in Section
\ref{ssec_det}, we have detected 4 out of 48 LBGs and 7 out of 90 LAEs in our NB3420 filter. After application of the
contamination and IGM corrections discussed in Sections \ref{sssec_foreground_contam} and \ref{sssec_IGM_corr}, the
average NB3420$-V$ properties of the galaxies in our sample provide information about the ratio of non-ionizing to
ionizing UV flux density and, with some assumptions, the LyC escape fraction.

\begin{figure*}
\epsscale{1.0}
\plotone{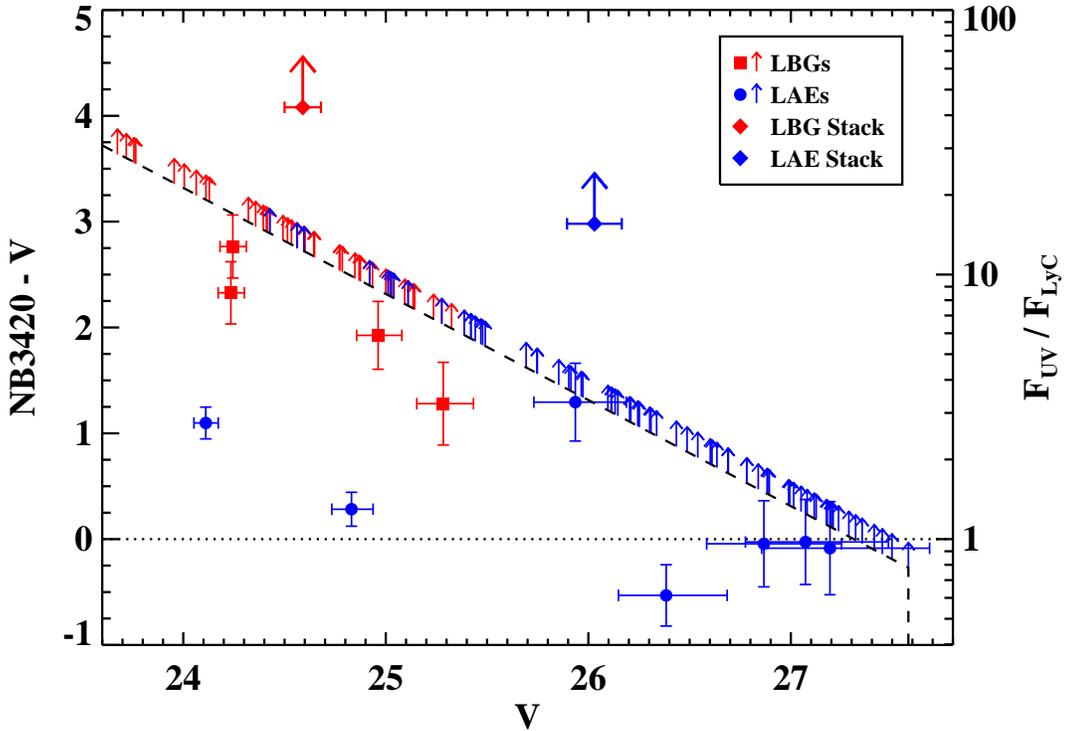}
\caption{
\small
Color-magnitude diagram of observed NB3420-$V$ vs. $V$ magnitude, with the equivalent values for $F_{UV}/F_{LyC}$
indicated on the right-hand axis. LBGs are indicated by red squares and LAEs by blue circles. LBGs (LAEs) without
NB3420 detections are plotted as 2$\sigma$ lower limits, indicated by the red (blue) arrows. The typical $V-$band
uncertainty associated with these lower limits is of similar size to the $V-$band uncertainty of data points, at a
given $V$ magnitude. The red (blue) diamond indicates the 3$\sigma$ limit in NB3420-$V$ color for the stack of all
LBGs (LAEs). All data represent observed values, uncorrected for foreground galaxy contamination and IGM absorption.
We note that an NB3420$-V$ color of zero ($F_{UV}/F_{LyC}=1$), corresponds to a flat spectrum with no Lyman break.
Such a blue spectrum cannot be reasonably explained by current stellar population synthesis models, and many of the
LAEs on this plot lie uncomfortably close to or below this limit.
\label{fig_cmd} 
}
\end{figure*}

\subsection{NB3420 minus $V$ Colors and Flux-Density Ratios} \label{sec_colors_fdrs}

In Figure \ref{fig_cmd}, we plot NB3420$-V$ color versus $V$ magnitude for LBGs and LAEs with NB3420 detections. The
colors of the LBG and LAE stacks are also plotted. All of the data presented in Figure \ref{fig_cmd} represent
observed values, uncorrected for foreground galaxy contamination and IGM absorption. Qualitatively, the
color-magnitude diagram (CMD) for objects in the HS1549 field agrees with the CMD for objects in the SSA22a field
\citep{nestor11}; both samples of galaxies exhibit narrowband minus continuum colors across a wide range of continuum
magnitudes that are extremely blue with respect to expectations from standard stellar population models. For example,
\citet{bc03} models with solar metallicity and a constant star-formation rate predict the intrinsic ratio of
non-ionizing to ionizing radiation for LAEs to be roughly 3 $-$ 6, which corresponds to an NB3420$-V$ color of 1.2 $-$
2. Out of the 7 LAEs with NB3420 detections, only one LAE (\emph{lae6662}) has a color redder than 1.2. Even more
surprising, especially given the small offsets ($\Delta_{UV,\,LyC} \leq 0\secpoint36=2.8$ kpc) of all LAEs except
\emph{lae6662}, 4 of the 6 remaining LAEs have NB3420$-V<0$, implying a complete absence of a Lyman break (i.e., a
flat spectrum). For typical LBG stellar populations \citep[e.g.,][]{shapley01,kornei10}, the intrinsic ratio is
predicted to be $>$6 (corresponding to an NB3420$-V$ color $>$2); however, two out of the four NB3420-detected LBGs
have NB3420$-V<2$. Thus, the HS1549 field provides the first confirmation that the blue non-ionizing to ionizing UV
colors observed in the SSA22a field are common among $z \sim 3$ galaxies, and not simply an unusual property of the
SSA22a field itself. Tables \ref{tab_LBG} and \ref{tab_LAE} display the raw non-ionizing to ionizing flux-density
ratios ($F_{UV}/F_{LyC}$) for individual LBGs and LAEs with NB3420 detections, and Table \ref{tab_stacks} shows the
same values for the stacks.

The average corrected and uncorrected NB3420$-V$ colors and flux-density ratios for the samples of LBGs and LAEs are
presented in Table \ref{tab_final_avg_flux_ratios}. Because photometry of the stacks of LBGs and LAEs without NB3420
detections does not yield any NB3420 signal, we determine average NB3420 and $V$ magnitudes by calculating the mean
flux of all the galaxies in each sample and assuming zero flux for undetected objects. These average magnitudes agree
with stacked photometry within the errors. As our $V$ and NB3420 filters probe rest-frame wavelengths of 1430\AA\ and
888\AA, respectively, we can convert our observed NB3420$-V$ colors directly into non-ionizing to ionizing
flux-density ratios (i.e., $F_{UV}/F_{LyC}$). We find uncorrected flux-density ratios of $(F_{UV}/F_{LyC})_{obs} = 92
\pm 49$ for LBGs and $(F_{UV}/F_{LyC})_{obs}= 11.2 \pm 5.4$ for LAEs. Considering only objects with NB3420 detections
in each sample, we find $(F_{UV}/F_{LyC})_{obs} = 7.2 \pm 2.2$ for LBGs and $(F_{UV}/F_{LyC})_{obs} = 1.6 \pm 0.9 $
for LAEs. In order to correct these colors and flux-density ratios for foreground contamination, we use adjusted
NB3420 and $V$ magnitudes derived from the contamination simulation described in Section
\ref{sssec_foreground_contam}. Correcting for IGM attenuation, however, only affects NB3420 magnitudes because
rest-frame $\sim$1500\AA\ flux is not affected by HI absportion in the IGM. Thus, for the IGM correction we simply
multiply the contamination-corrected flux-density ratio by the IGM transmission factor. With the application of these
two corrections to the colors and flux-density ratios, we find $(F_{UV}/F_{LyC})_{corr} = 82 \pm 45$ for LBGs and
$(F_{UV}/F_{LyC})_{corr} = 7.4 \pm 3.6$ for LAEs. For objects with NB3420 detections, we find corrected values of
$(F_{UV}/F_{LyC})_{corr} = 2.2 \pm 1.6$ for LBGs and $(F_{UV}/F_{LyC})_{corr} = 0.7 \pm 0.5$ for LAEs.

\begin{deluxetable*}{lcccccc}
\scriptsize
\tablewidth{0pt} 
\tablecaption{Average UV to LyC flux-density ratios. \label{tab_final_avg_flux_ratios}}
\tablehead{
\colhead{} & \colhead{} & \multicolumn{2}{c}{LBGs} & \colhead{} & \multicolumn{2}{c}{LAEs}\\
\cline{3-4}
\cline{6-7}
\colhead{Correction}  & \colhead{} & \colhead{$\left<\mathrm{NB3420}\right> - \left<V\right>$ \tablenotemark{a}} &
\colhead{$\left<F_{UV}/F_{LyC}\right>$\tablenotemark{b}} & \colhead{} & \colhead{$\left<\mathrm{NB3420}\right> -
\left<V\right>$\tablenotemark{a}} & \colhead{$\left<F_{UV}/F_{LyC}\right>$\tablenotemark{b}}
}
\startdata
\multicolumn{7}{c}{Full Ensembles} \\
\hline
none & & $4.91^{+0.47}_{-0.84}$ & $92 \pm 49$ & & $2.62^{+0.43}_{-0.72}$ & $11.2 \pm 5.4$ \\
contamination\tablenotemark{c} & & $5.93^{+0.47}_{-0.84}$ & $230 \pm 130$ & & $3.06^{+0.43}_{-0.72}$ & $16.8 \pm
8.1$\\
IGM + contamination\tablenotemark{d} & & $4.78^{+0.48}_{-0.88}$ & $82 \pm 45$ & & $2.17^{+0.43}_{-0.73}$ & $7.4 \pm
3.6$ \\
\hline
\multicolumn{7}{c}{Sources with NB3420 detections only} \\
\hline
none & & $2.15^{+0.29}_{-0.39}$ & $7.2 \pm 2.2$ & & $0.54^{+0.46}_{-0.80}$ & $1.6 \pm 0.9$\\
contamination\tablenotemark{c} & & $2.18^{+0.49}_{-0.91}$ & $7.5 \pm 4.2$ & & $0.50^{+0.56}_{-1.22}$ & $1.6 \pm 1.1$\\IGM + contamination\tablenotemark{d} & & $0.86^{+0.60}_{-1.44}$ & $2.2 \pm 1.6$ & & $-0.39^{+0.59}_{-1.36}$ & $0.7 \pm
0.5$
\enddata
\tablenotetext{a}{Color determined from average NB3240 and $V$-band fluxes. Uncertainties include photometric and
sample uncertainties.}
\tablenotetext{b}{Ratio and uncertainty in non-ionizing UV and LyC flux-densities inferred from
$\left<\mathrm{NB3420}\right> - \left<V\right>$ color.}
\tablenotetext{c}{Color and flux-density ratio after statistically correcting sample for foreground contamination of
NB3420 fluxes. For the ensembles containing all LBGs (or LAEs), the removal of contaminants must increase the
non-ionizing UV to LyC flux-density ratio as the total amount of LyC flux is decreased. For ensembles that contain
only objects with NB3420 detections, however, the flux-density ratio may not necessarily increase with the removal of
the contaminants.}
\tablenotetext{d}{Color and flux-density ratio after correcting sample for both foreground contamination and IGM
absorption of NB3420 fluxes.}
\end{deluxetable*}

It has so far proven difficult to reconcile the observed low non-ionizing to ionizing flux-density ratios with
intrinsic luminosity-density ratios predicted by current stellar-population synthesis models. Theoretical values of
the intrinsic luminosity-density ratio featured in previous works range from $(L_{UV}/L_{LyC})_{intr} = 3 - 6$
\citep{steidel01,shapley06,siana07}. Many of the values in Table \ref{tab_final_avg_flux_ratios} $-$ and especially
those that consider only the NB3420-detected galaxies $-$ do not fall within the theoretical range.  We also note that if we were to consider the flux-density ratio in the region defined by the NB3420 isophote (thus only including $V-$band light in the vicinity of the NB3420 emission, which more closely emulates a single stellar population), there would be even more tension between the measurements and the models.  \citet{nestor13}
examined a wider range of theoretical $(L_{UV}/L_{LyC})_{intr}$ values using stellar population synthesis models from
\citet{bc03} and the BPASS models from \citet{eldridgestanway09}, which include a more detailed treatment of massive
stellar binaries (specifically, Wolf-Rayet stars) and nebular emission. The models were used to describe galaxies with
an array of ages and metallicities, assuming a constant star-formation rate. The largest source of variation in
$(L_{UV}/L_{LyC})_{intr}$ is due to the age of the galaxy, as $(L_{UV}/L_{LyC})_{intr}$ increases quickly as the
stellar population evolves. The choice of model is also important, as the BPASS models predict ratios of
$(L_{UV}/L_{LyC})_{intr}$ a factor of $\sim$1.5 lower than the \citet{bc03} models for a given age and metallicity.
Changes in stellar metallicity cause smaller variations in $(L_{UV}/L_{LyC})_{intr}$ on the order of a few percent,
but the magnitude of these changes increases with the age of the stellar population. It is important to note that, if
the models are correct, the predicted value for $(L_{UV}/L_{LyC})_{intr}$ should serve as a lower limit to the
observed non-ionizing to ionizing flux-density ratio of the galaxy. The observed flux-density ratio will likely be
higher than this limit, as ionizing radiation may be absorbed by neutral hydrogen in both the interstellar medium
(ISM) of the galaxy and the IGM. Thus, for any given galaxy, the observed value of $F_{UV}/F_{LyC}$ should be greater
than or equal to $(L_{UV}/L_{LyC})_{intr}$.  We note that the two sets of stellar population synthesis models used in this analysis, as with all such models, are limited by the absence of direct observations in the LyC region of the O- and B-type stars that produce most of the ionizing radiation \citep{zastrow13}. Without direct observations to verify that the models are accurate, the fact that our LyC observations disagree with the models may indicate that there is something systematically incorrect with the model spectra at ultraviolet wavelengths.  Alternatively, the tension between the models and our observations may be due to the uncertainties involved in removing contaminants.  Any low-redshift contaminants left in our sample (e.g., an [OII]-emitter or foreground galaxy) would also produce colors different from those predicted for models of $z \sim 2.85$ galaxies.

Here, we consider our inferred values of $(F_{UV}/F_{LyC})_{corr}$ with respect to the model predictions of
$(L_{UV}/L_{LyC})_{intr}$. We first discuss the interpretation of $(F_{UV}/F_{LyC})_{corr}$ for samples of LBGs and
LAEs with NB3420 detections because these are the samples for which LyC emission is actually measured. We consider the
ensembles of NB3420-detected galaxies rather than individual galaxies because our statistical corrections for
foreground contamination and IGM absorption do not apply to individual galaxies.

For the sample of LBGs with NB3420 detections, $(F_{UV}/F_{LyC})_{corr} = 2.2 \pm 1.6$. This value of 2.2 implies the
unphysical ages of $\lesssim$ 10 Myr using the \citet{bc03} and BPASS models. Adding the 1$\sigma$ error to the LBG
flux-density ratio implies $(F_{UV}/F_{LyC})_{corr} = 3.8$, which is consistent with a 100 Myr galaxy using the BPASS
model. Typical LBGs are characterized by ages $>$100 Myr \citep[][although in Section \ref{ssec_IR} we show that this
average age may be somewhat over-estimated, as it is based only on LBG samples with near- or mid-infrared
detections]{kornei10,shapley05,shapley01} and not younger than $\sim$50 Myr \citep[given the typical LBG dynamical
timescale;][]{reddy12}. It is worth noting that using an average IGM correction for the NB3420-detections-only samples
probably constitutes an overcorrection, as NB3420-detected galaxies are likely to have clearer sightlines through the
IGM than the sample as a whole. Thus, the true value of $(F_{UV}/F_{LyC})_{corr}$ for LBGs is likely somewhere between
2.2 (corrected for contamination and IGM absorption) and 7.5 (corrected for contamination only). Values $\gtrsim$ 4.5
are consistent with galaxy ages $>$100 Myr using BPASS models, and values $\gtrsim$ 6.1 are consistent with the same
limit in age using \citet{bc03} models.

As for the LAE sample, the value of $(F_{UV}/F_{LyC})_{corr} = 0.7 \pm 0.5$ is inconsistent with all models presented
in \citet{nestor13}. For models with metallicity $Z=0.2\,Z_{\odot}$ and ages of 1, 10, and 100 Myr, respectively,
$(L_{UV}/L_{LyC})_{intr} =$ 1.33, 2.10, and 3.16 for BPASS models and $(L_{UV}/L_{LyC})_{intr} =$ 1.98, 3.59, and 6.17
for \citet{bc03} models. Only by considering values of the LAE flux-density ratio at $>1 \sigma$ from our measured
value can we reconcile our flux-density ratio with those predicted by young BPASS models with ages of several Myr.
Because the dynamical timescale argument (based on galaxy sizes and velocity dispersions) that limits LBG ages to
$>$50 Myr may not apply to LAEs, it is possible that the extremely low non-ionizing to ionizing flux-density ratios we
measure imply that LAEs are on average very young galaxies. Young ages for LAEs have been previously determined by SED
fits of stacked photometry \citep[e.g.,][]{gawiser07}. We also note that, once again, the correction for IGM
absorption may constitute an overcorrection, which would place the true value of $(F_{UV}/F_{LyC})_{corr}$ for LAEs
between 0.7 (corrected for contamination and IGM absorption) and 1.6 (corrected for contamination only).

Until now, we have only discussed the interpretation of $(F_{UV}/F_{LyC})_{corr}$ for samples with NB3420 detections.
We now consider the full samples of LBGs and LAEs, which include both objects with and without NB3420 detections. For
the full sample of LBGs, $(F_{UV}/F_{LyC})_{corr} = 82 \pm 45$. This number is significantly higher than the lower
limit required by stellar population synthesis models for typical LBGs with ages $>$100 Myr, namely
$(L_{UV}/L_{LyC})_{intr} > 4.5$ for BPASS models and $(L_{UV}/L_{LyC})_{intr} > 6.1$ for \citet{bc03} models. This
consistency with the models, however, arises not because all 48 LBGs have values of $F_{UV}/F_{LyC} \sim 82$, which
would not conflict with models, but rather because 4 LBGs have low values of $F_{UV}/F_{LyC}$ and the rest are
undetected in LyC. For the full sample of LAEs, $(F_{UV}/F_{LyC})_{corr} = 7.4 \pm 3.6$. As with the LBGs, these
values would not significantly conflict with models if they represented the typical LAE in the sample. However, the
LAE sample is also comprised of some objects with very strong LyC emission and some objects with no observed emission.
For both LBGs and LAEs, we must consider how values of $(F_{UV}/F_{LyC})_{corr}$ for individual galaxies compare with
stellar population synthesis models, as the ensemble-averaged value does not take into account variation within the
sample.

\begin{deluxetable*}{rcccc}
\tablewidth{0pt} 
\tabletypesize{\scriptsize}
\tablecaption{Contributions to the Ionizing Background. \label{tab_emissivity}}
\tablehead{
\colhead{} & \colhead{LF\tablenotemark{a}} & \colhead{$F_{UV}/F_{LyC}$\tablenotemark{b}} & \colhead{Magnitude
range\tablenotemark{c}} & \colhead{$\epsilon_{LyC}$\tablenotemark{d}}
}
\startdata 
(i)   & LBG & $82  \pm 45$  & $M_{AB} \leq -19.7$         & $1.4  \pm 0.8$ \\
(ii)  & LAE & $7.4 \pm 3.6$ & $-19.7 < M_{AB} \leq -17.7$ & $3.2  \pm 1.6$ \\
(iii) & LBG & $7.4  \pm 3.6 $ & $-19.7 < M_{AB} \leq -17.7$ & $13.6 \pm 6.7$ \\
(iv) & LBG & $82  \pm 45 $ & $M_{AB} \leq -17.7        $ & $2.6  \pm 1.5$ \\
 (v) & LAE & $7.4 \pm 3.6$ & $M_{AB} \leq -17.7        $ & $6.8  \pm 3.3$ \\
& Total (lum.-dep.)\tablenotemark{e} & \nodata & $M_{AB} \leq -17.7$  & $15.0  \pm 6.7$ \\
& Total (LAE-dep.)\tablenotemark{f} & \nodata & $M_{AB} \leq -17.7$  & $8.8  \pm 3.5$ 
\enddata 
\vspace{0.2cm}
\tablenotetext{a}{\mbox{Luminosity} function parameters are the same as those described in \citet{nestor13}.}
\tablenotetext{b}{Sample average flux-density ratio corrected for foreground contamination and IGM absorption, from
Table \ref{tab_final_avg_flux_ratios}.}
\tablenotetext{c}{Magnitude range over which the first moment of the luminosity function is determined. $M_{AB} =
-19.7$ and $-17.7$ correspond to $0.34L_{*}$ and $0.06L_{*}$, respectively.}
\tablenotetext{d}{Comoving specific emissivity of ionizing radiation in units of
$10^{24}$~ergs~s$^{-1}$~Hz$^{-1}$~Mpc$^{-3}$.}
\tablenotetext{e}{Totals for the luminosity-dependent model, determined by summing rows (i) and (iii).}
\tablenotetext{f}{Total for the LAE-dependent model, determined by summing $0.77 \times$ row (iv) and row (v).}
\end{deluxetable*}

\subsection{The LyC Escape Fraction}

Using the values of $(F_{UV}/F_{LyC})_{corr}$ for the full LBG and LAE samples presented in Table
\ref{tab_final_avg_flux_ratios}, we can determine the relative and absolute escape fractions for each sample. The
relative escape fraction, a measure of how the observed flux-density ratio $F_{UV}/F_{LyC}$ compares to the
theoretical ratio, is defined to be

\begin{equation}
f^{LyC}_{esc,\,rel} = \frac{\left(L_{UV}/L_{LyC}\right)_{intr}}{\left(F_{UV}/F_{LyC}\right)_{corr}},
\end{equation}

\noindent where $\left(L_{UV}/L_{LyC}\right)_{intr}$ is the intrinsic ratio of UV to LyC luminosity densities produced
in star-forming regions. The absolute escape fraction includes an additional term for the escape fraction of
non-ionizing UV photons ($f_{esc}^{UV}$) due to dust extinction:

\begin{equation}
f_{esc}^{LyC} = f^{LyC}_{esc,\,rel} \times f_{esc}^{UV}.
\end{equation}

\noindent Thus,

\begin{equation}
f_{esc}^{LyC} = \left(\frac{F_{UV}}{F_{LyC}}\right)_{corr}^{-1} \left(\frac{L_{UV}}{L_{LyC}}\right)_{intr}
\left(f_{esc}^{UV}\right)
\end{equation}

\noindent Both unknowns $f_{esc}^{UV}$ and $\left(L_{UV}/L_{LyC}\right)_{intr}$ are uncertain and likely to vary from
object to object. Following \citet{nestor13}, we estimate these values based on the observed ages and $E(B-V)$ values
of $z\sim3$ LBGs and LAEs. For the escape fraction of non-ionizing UV photons, we adopt $f_{esc, LBG}^{UV}=0.2$ and
$f_{esc, LAE}^{UV}=0.3$. For the intrinsic ratio of UV to LyC luminosity densities, we quote a range of values
bracketed by the BPASS and \citet{bc03} models. For LBGs, we use $10^{8}$ Myr, $Z=Z_{\odot}$ models to obtain
$\left(L_{UV}/L_{LyC}\right)_{intr,LBG} = 4.43 - 6.38$. For LAEs, which have been shown to be younger
\citep{gawiser07} and more metal-poor, we use $10^{6-7}$ Myr, $Z=0.2Z_{\odot}$ models to obtain
$\left(L_{UV}/L_{LyC}\right)_{intr,LAE} = 1.33 - 3.59$. With these assumptions, we derive relative LyC escape
fractions of $f_{esc,\,rel}^{LBG}=5-8$\% and $f_{esc,\,rel}^{LAE}=18-49$\%, and absolute escape fractions of $f_{esc}^{LBG}=1-2$\%
and $f_{esc}^{LAE}=5-15$\%. As our measured values of $F_{UV}/F_{LyC}$ have uncertainties of roughly fifty percent,
the uncertainty in $f_{esc}$ is at minimum fifty percent and likely higher due to uncertainties in our assumed values
of $f_{esc}^{UV}$ and $\left(L_{UV}/L_{LyC}\right)_{intr}$.
 
Our values of $f_{esc}$ are consistent with, though slightly lower than, those calculated by \citet{nestor13} for the
SSA22a field using very similar methods: $f_{esc,LBG}^{LyC}=5-7$\%, $f_{esc,LAE}^{LyC}=10-30$\%. Using IGM-corrected
values of $F_{UV}/F_{LyC}$ from the literature and our assumptions for $f_{esc}^{UV}$ and
$\left(L_{UV}/L_{LyC}\right)_{intr}$, we find additional values of $f_{esc}$ in the literature that range from
$19-27$\% \citep[considering 29 averaged $z=3.4$ LBG spectra]{steidel01}, $24-35$\% \citep[][for $z\sim3$ LBGs in the
SSA22a field]{iwata09}, and $4-6$\% \citep{shapley06}, although the two LBGs in \citet{shapley06} with putative LyC
emission were later shown to be a spurious detection \citep{nestor11,iwata09} and a foreground contaminant
\citep{nestor13} such that $f_{esc}$ calculated from this work would be consistent with zero. Using a slightly
different method to make a direct measurement of $f_{esc}$ for a sample of $3.4<z<4.5$ LBGs, \citet{vanzella10dec}
calculated $f_{esc}<5-20$\%. Finally, by measuring HI opacity along 32 GRB sightlines in the redshift range
$2.0<z<5.5$, \citet{fynbo09} determined $\langle f_{esc} \rangle = 0.02 \pm 0.02$ with a 95\% confidence level upper
limit of $\langle f_{esc} \rangle \leq 0.07$. This wide range of values highlights how difficult it has proven to
determine an accurate value for $f_{esc}$. These difficulties stem from many factors, including determination of
redshifts, spatial resolution, foreground contamination, and the fact that only $\sim$10\% of galaxies in the sample
will be detected in the LyC. In our study, only 12 out of 131 spectroscopically-confirmed galaxies were observed to
have NB3420 detections, and some of these are probably contaminated by foreground galaxies. Such a small sample size
of NB3420 detections, while an improvement upon many previous studies of spectroscopically-confirmed galaxies, limits
the precision of our $f_{esc}$ measurements. Until there exist uncontaminated samples of LyC-emitting galaxies an
order of magnitude larger, small sample sizes will be a constant source of uncertainty.

\subsection{The LyC Emissivity of Star-forming Galaxies} \label{ssec_lyc_emissivity}

We estimate the comoving specific emissivity of ionizing photons 
\begin{equation}
\epsilon_{LyC} = \left(\frac{F_{UV}}{F_{LyC}}\right)^{-1}_{corr} \, \int^{L_{max}}_{L_{min}} L \, \Phi \; dL
\end{equation}
\noindent following the assumptions of \citet{nestor13}. In estimating $\epsilon_{LyC}$, we assume that the UV
luminosity functions of LBGs and LAEs do not change significantly from $z=3.09$ to $z=2.85$ and the only change in the
value of $\int L$ $\Phi$ $dL$ is due to the integration bounds (i.e., $L_{min}$ and $L_{max}$). Comparing our observed
values for $(F_{UV}/F_{LyC})_{corr}$ to those of \citet{nestor13}, we obtain $(F_{UV}/F_{LyC})_{corr}^{HS1549} =
4.6(F_{UV}/F_{LyC})_{corr}^{SSA22a}$ for LBGs and $(F_{UV}/F_{LyC})_{corr}^{HS1549} =
2.0(F_{UV}/F_{LyC})_{corr}^{SSA22a}$ for LAEs with spectroscopic redshifts. \citet{nestor13} suggest two different
models for combining the LBG and LAE luminosity functions (see equations 4 and 5 in \citet{nestor13}) to calculate the
global ionizing emissivity. In the luminosity-dependent model, LAEs are assumed to represent galaxies with faint UV
continuum magnitudes ($0.06 L^{*} < L < 0.34 L^{*} $, corresponding to $25.5<V<27.5$) and LBGs represent brighter
galaxies ($L > 0.34 L^{*}$). However, the fact that our data show no change in average LAE NB3420$-V$ color across a
range in $V$ magnitude (Figure \ref{fig_cmd}) supports the idea that LAEs are a population of galaxies with properties
distinct from those of LBGs, and not simply faint LBG-analogs. This scenario is described by the LAE-dependent model,
in which LAEs are assumed to comprise 23\% of the LBG population \citep[see][]{nestor13}, galaxies identified both as
LBGs and LAEs are treated as LAEs, and the luminosity function is integrated over the full luminosity range
($0.06L^{*} < L < \infty$) for both LBGs and LAEs. Based on the values of $(F_{UV}/F_{LyC})_{corr}$ derived for the
HS1549 field, $\epsilon_{LyC}=15.0 \pm 6.7 \times 10^{24}$ ergs s$^{-1}$ Hz$^{-1}$ Mpc$^{-3}$ for the
luminosity-dependent model and $\epsilon_{LyC}=8.8 \pm 3.5 \times 10^{24}$ ergs s$^{-1}$ Hz$^{-1}$ Mpc$^{-3}$ for the
LAE-dependent model. The uncertainties in $\epsilon_{LyC}$ reflect only uncertainties in $(F_{UV}/F_{LyC})_{corr}$,
which dominate over uncertainties in the luminosity function. Our values of $\epsilon_{LyC}$ are roughly half of those
calculated by \citet{nestor13} for star-forming galaxies in the SSA22a field:
$\epsilon_{LyC}^{SSA22a}=32.2^{+12.0}_{-11.4} \times 10^{24}$ ergs s$^{-1}$ Hz$^{-1}$ Mpc$^{-3}$ for the
luminosity-dependent model and $\epsilon_{LyC}^{SSA22a}=16.8^{+6.9}_{-6.5} \times 10^{24}$ ergs s$^{-1}$ Hz$^{-1}$
Mpc$^{-3}$ for the LAE-dependent model. In Table \ref{tab_emissivity}, we summarize the contributions to
$\epsilon_{LyC}$ as determined from galaxies in the HS1549 field.

To place these values of $\epsilon_{LyC}$ for star-forming galaxies in context, we can compare to values of the total
ionizing emissivity ($\epsilon_{LyC}^{tot}$) derived from \lya-forest studies, which represent an upper bound on the
ionizing emissivity from star-forming galaxies. Using the formulation described in \citet{nestor11}, we define
$\epsilon_{LyC}^{tot}$ to be

\begin{equation}
\epsilon_{LyC}^{tot} =
\frac{\Gamma_{\mathrm{HI}}\,h\,(3-\alpha_{s})}{\sigma_{\mathrm{HI}}\,\lambda_{\mathrm{mfp}}\,(1+z)^{3}}
\end{equation}

\noindent where $\Gamma_{\mathrm{HI}}$ is the total hydrogen photoionization rate in the IGM at $z=2.85$ (measured by
\lya-forest studies), $h$ is Planck's constant, $\alpha_{s}$ is the power-law index of the UV spectral slope in the
LyC region ($f_{\nu}\propto\nu^{\alpha_{s}}$; we adopt $\alpha_{s}=-3$), $\sigma_{\mathrm{HI}} = 6.3 \times 10^{-18}$
cm$^{2}$ is the atomic hydrogen photoionization cross section, and $\lambda_{\mathrm{mfp}}=100$ Mpc is the ionizing
photon mean free path through the IGM at $z=2.85$ \citep[][]{rudie13,fauchergiguere08,songaila10}. Using
$\Gamma_{\mathrm{HI}} = 0.92 \times 10^{-12}$ s$^{-1}$ inferred from \citet{bolton07}, we derive the total ionizing
photon emissivity at $z=2.85$ to be $\epsilon_{LyC}^{tot} = 9.8 \pm 4.1 \times 10^{24}$ ergs s$^{-1}$ Hz$^{-1}$
Mpc$^{-3}$. Using $\Gamma_{\mathrm{HI}} = 0.53 \times 10^{-12}$ s$^{-1}$ inferred from \citet{fauchergiguere08}, we
derive $\epsilon_{LyC}^{tot} = 5.6 \pm 1.6 \times 10^{24}$ ergs s$^{-1}$ Hz$^{-1}$ Mpc$^{-3}$. Estimates of the
contribution of QSOs to the ionizing background at $z=2.85$ are lower, and range from $\epsilon_{LyC}^{QSO} \sim 1.5
\times 10^{24}$ ergs s$^{-1}$ Hz$^{-1}$ Mpc \citep{cowie09} to $\epsilon_{LyC}^{QSO} \sim 5.5 \times 10^{24}$ ergs
s$^{-1}$ Hz$^{-1}$ Mpc \citep{hopkins07}. Within their errors, the values of $\epsilon_{LyC}$ that we report for
star-forming galaxies in the HS1549 field are consistent with the inferred total ionizing emissivity from the
\lya-forest studies. This agreement between the total ionizing emissivity and the ionizing emissivity from
star-forming galaxies leaves little room for the contribution to the emissivity from QSOs or from fainter star-forming
galaxies not probed by our observations. Although the LyC emissivity we measure from star-forming galaxies may still
be an overestimation, it is in better agreement with $\epsilon_{LyC}^{tot}$ determined from \lya-forest studies than
the higher values determined in past work \citep[e.g.,][]{nestor13,nestor11,steidel01}. Even with the large
uncertainties in $\epsilon_{LyC}^{tot}$, $\epsilon_{LyC}^{QSO}$, and $\epsilon_{LyC}^{gal}$, our results may point to
star-forming galaxies providing the dominant contribution of ionizing photons at $z=2.85$.

\section{Properties of LyC-Emitting Galaxies} \label{sec_prop}

The measurements of the ionizing to non-ionizing flux-density ratios of the galaxies in our samples point to a large
spread in the amount of LyC emission escaping from galaxy to galaxy. While the majority of galaxies in our sample
($\sim$90\%) appear to have no leaking LyC radiation, the remaining 10\% exhibit very blue NB3420$-V$ colors,
indicating a high escape fraction. Two possible scenarios may describe our data. One possibility is that galaxies with
and without observed LyC emission have an intrinsically different physical property governing whether LyC radiation
escapes from or is absorbed by the ISM. A second possibility is that LyC emission escapes from all star-forming
galaxies, but only over a small solid angle where neutral hydrogren has been cleared away (e.g., by stellar winds or
supernovae). In this section, we explore the differential properties of galaxies with and without observed LyC
radiation, with the goal of distinguishing between these two scenarios. Specifically, we examine the rest-frame
near-infrared properties of both LBGs and LAEs and the \lya\ equivalent widths of the LAEs. The ultimate goal is to be
able to identify galaxies associated with strong LyC emission by some other galactic property and search for analogs
to such galaxies at higher redshifts, in regimes where the IGM is opaque to LyC photons.

\subsection{Rest-frame Near-Infrared Properties of LyC-emitting Galaxies} \label{ssec_IR}

In addition to the optical and near-UV data in the HS1549 field, there exists imaging in several infrared bands. A
small fraction of our objects lie in the footprint of the $HST$/WFC3 $F160W$ image (rest-frame 4150\AA) and the
\emph{Spitzer}/IRAC Channels 1 and 3 images (\emph{Ch1}, 3.6 $\mu$m; \emph{Ch3}, 5.8 $\mu$m; \emph{Spitzer} program
G03, PI: Steidel) corresponding respectively to rest-frame 0.9$\mu$m and 1.5$\mu$m. The entire field has been imaged
by Palomar/WIRC in J and K (rest-frame 3250\AA\ and 5700\AA, respectively) and \emph{Spitzer}/IRAC Channels 2 and 4
(\emph{Ch2}, 4.5 $\mu$m; \emph{Ch4}, 8 $\mu$m) corresponding respectively to rest-frame 1.2$\mu$m and 2.1$\mu$m. As
the \emph{Ch2} image is the deepest of these images, we focus our subsequent analysis on \emph{Ch2}. In the deepest
part of the mosiacs, the \emph{Ch2} image has an exposure time of 18500 seconds and a 3$\sigma$ limiting magnitude of
24.0 (AB). The typical \emph{Ch2} IRAC PSF FWHM ($\sim$2\secpoint5) is significantly larger than the seeing in the
LRIS optical imaging. All of the LBG and LAE targets lie within the footprint of the \emph{Ch2} image, and the
majority lie within the deepest, central regions. We performed PSF-fitting photometry of the LBGs and LAEs in the
\emph{Ch2} image using procedures described in \citet{reddy06dec}. Results of the photometry indicate that 24 out of
the 48 LBGs are detected in \emph{Ch2} at the 3$\sigma$ level, while there are \emph{Ch2} detections for only 7 out of
the 90 LAEs (2 of the 7 LAEs are also LBGs). The median limits in \emph{Ch2} magnitude and $V-$\emph{Ch2} color for
the nondetections are \emph{Ch2}$>$24.0 and $V-$\emph{Ch2}$<$0.7 for LBGs, and \emph{Ch2}$>$24.0 and
$V-$\emph{Ch2}$<$2.1 for LAEs.

In order to understand the typical properties of LBG stellar populations in the HS1549 field, we obtained SED fits for
each LBG with at least one infrared datapoint. While all 48 of the LBGs in our sample have $U$, $G$, and $\cal{R}$
photometry, only 33 of these also have detections in one or more infrared bands ($J$, $K$, $F160W$, and/or IRAC
\emph{Ch1}$-$\emph{Ch4}). For these 33 objects, we derived SED fits based on the \citet{bc03} models using a Chabrier
initial mass function (IMF). Following \citet{reddy12}, we adopted constant star-formation rate models, the
\citet{calzetti00} dust attenuation curve, and 50 Myr lower limits in age (representing the typical LBG dynamical
timescale). The ages of the galaxies range from 50 to 2200 Myr, with a median age of 202 Myr. Roughly one third of the
sample was assigned the lowest allowed age value (50 Myr). The star-formation rates (SFRs) range from 5 to 214
M$_{\odot}$ yr$^{-1}$, with a median SFR of 43 M$_{\odot}$ yr$^{-1}$. The stellar masses range from log(M/M$_{\odot}$)
= 9 to log(M/M$_{\odot}$) = 11 with a median of log(M/M$_{\odot}$) = 9.8. Finally, the values of $E(B-V)$ range from 0
to 0.31, with a median value of 0.2. These values agree well with typical values of LBG properties quoted in the
literature \citep[e.g.,][]{kornei10}. The 15 LBGs without IRAC photometry ($\sim$30\% of our sample) for which SED
fits could not be calculated are most likely undetected in IRAC bands because of their lower stellar mass and dust
content. Thus, the inclusion of these objects would likely change the average distributions of SED fit parameters. A
similar bias exists in the average LBG properties reported by \citet{kornei10}; of the 321 $z\sim3$ LBGs in their
sample, roughly 25\% of these LBGs did not have photometric detections redward of the Balmer break, and were therefore
not modeled with SED fits. Out of the 33 LBGs in our sample that were modeled with SED fits, only one (MD34) has an
NB3420 detection. The SED fit to MD34 produced the following values: age = 50 Myr (the minimum allowed), SFR = 137
M$_{\odot}$ yr$^{-1}$, log(M/M$_{\odot}$) = 9.8, $E(B-V)$ = 0.28. While it is not possible to draw conclusions about
the global population of LBGs emitting LyC based on only one object, we note that MD34 has been assigned the lowest
possible age allowed by our modeling and the second largest SFR of the 33 modeled galaxies.

\begin{figure*}
\epsscale{1.0}
\plotone{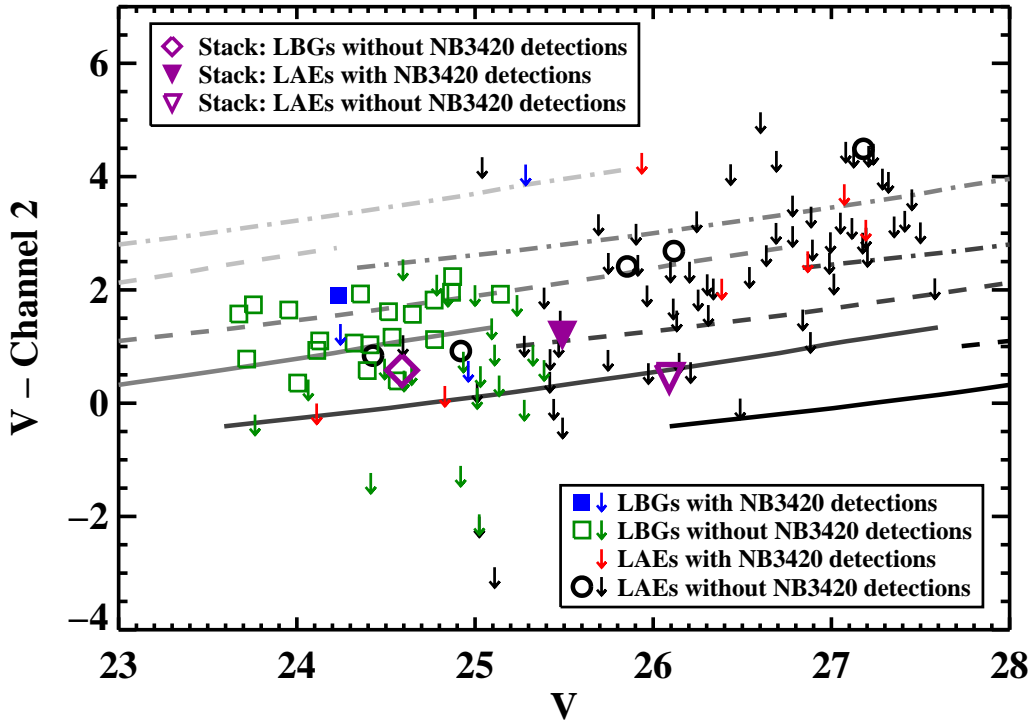}
\caption{ 
\small
Color-magnitude diagram plotting $V-$\emph{Ch2} color against $V$ magnitude for LBGs and LAEs with detections in $V$.
Individual LBGs are plotted using filled blue squares for objects with NB3420 detections and open green squares for
objects without NB3420 detections. LAEs without NB3420 detections are plotted with open black circles, and the two LAE
detections brightest in $V$ are also LBGs. Objects undetected in \emph{Ch2} are plotted as 3$\sigma$ upper limits in
$V-$\emph{Ch2} color following the color scheme for the detected objects. All LAEs with NB3420 detections are
undetected in \emph{Ch2}; the upper limits on these objects are indicated by red arrows. The open purple diamond
represents the color of the LBG stack of NB3420 non-detections. The filled (open) purple inverted triangle represents
the upper limit in $V-$\emph{Ch2} color for the stack of LAEs with NB3420 detections (non-detections). Gray curves
indicate lines of constant stellar mass, with $M_{stellar}=10^{8}\,\mbox{M}_{\odot}$, $10^{9}\,\mbox{M}_{\odot}$,
$10^{10}\,\mbox{M}_{\odot}$, and $10^{11}\,\mbox{M}_{\odot}$ shown in black, dark gray, medium gray, and light gray
respectively. Values of $E(B-V)$ are indicated by solid, dashed, and dot-dashed lines, corresponding to $E(B-V)$= 0.0,
0.15, and 0.3, respectively. The curves of constant mass are produced by a \citet{bc03} SED fit to a $z=2.85$ galaxy
using a Chabrier IMF. The length of each curve corresponds to ages ranging from 50 Myr to 2300 Myr (the age of the
universe at $z=2.85$).
\label{fig_irac_colors}
}
\end{figure*}

Because only a fraction of the LBGs and LAEs have sufficient infrared photometry to obtain SED fits, we use
$V-$\emph{Ch2} colors and limits to constrain the range of possible stellar populations of our full sample of
galaxies. For galaxies at $z\sim2.85$, the $V$ and \emph{Ch2} filters lie on either side of the Balmer Break and thus
are a direct probe of the stellar mass-to-light ratio of the galaxy. Thus, we can convert $V-$\emph{Ch2} colors to
stellar masses by assuming a range of values for $E(B-V)$ and using $V-$band imaging to estimate galaxy luminosty. We
present a $V-$\emph{Ch2} vs. $V$ color-magnitude diagram in Figure \ref{fig_irac_colors}. Photometry for individual
LBGs and LAEs is plotted for objects with and without NB3420 detections along with curves of constant stellar mass for
various assumptions of $E(B-V)$. In the sample of LBGs, there is only one galaxy (MD34) with both a \emph{Ch2}
detection and an NB3420 detection; MD34 does not differ significantly from the other \emph{Ch2}-detected LBGs in its
$V-$\emph{Ch2} vs. $V$ properties. In the sample of LAEs, there only five objects with both \emph{Ch2} and $V$
detections, none of which have NB3420 detections. While we cannot make any strong statements comparing the
$V-$\emph{Ch2} properties of LAEs with and without NB3420 detections, we can note that the limits of LAEs with NB3420
detections do not distinguish them from LAEs without NB3420 detections. Deeper \emph{Ch2} data is necessary to make
any stronger inferences. The preliminary results from this analysis seem to show no strong differences between the
stellar populations of galaxies with and without NB3420 detections, implying that the detection of LyC emission from a
small portion of galaxies stems from the effect of varying observer perspective with respect to the geometry of the
ISM of each galaxy, rather than intrinsic physical differences between galaxies. According to such a scenario,
galaxies are described by a constant escape fraction that appears to vary between objects based on the perspective of
the observer. Another important feature of Figure \ref{fig_irac_colors} is that the curves of constant stellar mass
indicate that the $V-$\emph{Ch2} limits of nearly all the LAEs are consistent with M$_{stellar} < 10^{10}$ M$_{\odot}$
galaxies with $E(B-V)$ $\lesssim$ 0.3. A notable outlier is \emph{lae1843}, the LAE with $V-$\emph{Ch2} $>$ 4.
\emph{Lae1843} is not located near an obvious contaminant in any of our images, but there is still the possibility of
a very red foreground galaxy not visible in the shorter-wavelength images or a foreground galaxy in those images that
cannot be distinguished from \emph{lae1843} at the current spatial resolution.

Because the \emph{Ch2} data are not deep enough to characterize the individual rest-frame near-infrared properties of
all of our target galaxies, we also performed photometry on stacked images of galaxies with and without LyC detections
for the LBG and LAE samples. The stacks were created using methods similar to those described in Appendix D of
\citet{reddy12}, with the requirement that objects included in the stack not be blended with a nearby neighbor in the
\emph{Ch2} image. $V-$\emph{Ch2} colors of the LBG and LAE stacks are plotted in Figure \ref{fig_irac_colors}. We use
stellar population synthesis models to estimate masses from the stacked photometry; while this method is not
necessarily equivalent to reporting the average properties of individual SED fits, it provides rough insights into the
relative properties of different subsamples in the absence of SED fits for every object in our sample.

Unfortunately, it is not possible to draw definitive conclusions from the LBG stacks. While the stack of LBGs without
NB3420 detections yields a \emph{Ch2} magnitude of $24.01 \pm 0.10$ and $V-$ \emph{Ch2} $=0.58$, we were unable to
make a useful stack for the LBGs with NB3420 detections because two of the four objects are blended with bright
neighbors in the \emph{Ch2} image. Of the two LBGs with NB3420 detections and unblended \emph{Ch2} photometry, MD34
has a magnitude of $22.34 \pm 0.12$ and MD5 has a 3$\sigma$ lower limit of $>$24.22. While the \emph{Ch2} photometry
of MD34 and MD5 is consistent with that of the LBGs undetected in NB3420, the small sample size precludes us from
making any comparisons between objects with and without LyC detections. We can also compare the LBG properties
determined by $V-$\emph{Ch2} color to those determined by the SED fits. The $V-$\emph{Ch2} color of the stack of LBGs
without NB3420 detections implies log(M/M$_{\odot}$) = 9.60 (9.10), assuming $E(B-V)$ = 0.0 (0.1). Higher values of
$E(B-V)$ imply younger galactic ages, and assumptions of $E(B-V)$ $>$ 0.1 do not result in meaningful fits to the
stacked color because they imply ages $<$50 Myr. The assumption of $E(B-V)$ = 0.1 requires an age of $\sim60$ Myr,
significantly younger than the median age derived for the sample of LBGs with SED fits (202 Myr). However, LBGs have
been shown to have typical reddening values of $E(B-V)$ $\sim 0.17$ \citep{kornei10}, and assuming such a value
further decreases the average age implied by the stacked photometry. This apparent discrepancy (i.e., the fact that
the median age implied by the stacks is lower than that implied by the SED fits) arises because the stacks include all
LBGs regardless of whether or not they are detected in \emph{Ch2}, while the SED fits only include the older or more
massive LBGs with IRAC photometry redwards of the Balmer break. We can further demonstrate this discrepancy by
contrasting the $V-$ \emph{Ch2} color of the stack of all LBGs without LyC detections ($V-$ \emph{Ch2} $=0.58$) with
the average $V-$ \emph{Ch2} color of the objects in that stack that have \emph{Ch2} detections ($V-$ \emph{Ch2}
$=1.33$). Because the stack includes all LBGs regardless of whether or not they have infrared data, the $V-$
\emph{Ch2} color of the stack is much bluer than the $V-$ \emph{Ch2} color of the subset of objects within the stack
that have \emph{Ch2} detections.

\begin{figure*}
\epsscale{1.0}
\plotone{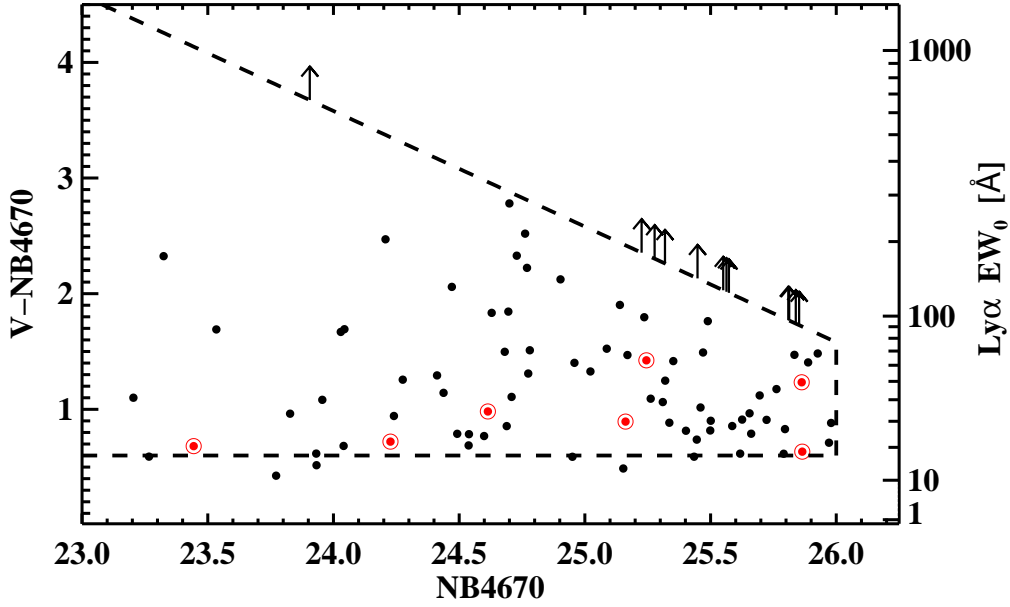}
\caption{
\small
Color-magnitude diagram plotting $V-$NB4670 vs. NB4670 for LAEs. The rest-frame \lya\ equivalent width corresponding
to a given $V-$NB4670 color is indicated on the right-hand axis. Galaxies without NB3420 detections are shown in black
and galaxies with NB3420 detections are shown as red with a circle around each point. Lower limits correspond to
objects undetected in $V$ and follow the same color scheme as the detections. The dashed diagonal line indicates the
observational limit on $V-$NB4670 color, which is determined by our $V-$band magnitude limit of 27.58. The vertical
dashed line indicates the cutoff of $m_{4670}<26.0$ applied in order to ensure the robustness of our sample (see
Section \ref{ssec_samp}). The horizontal dashed line corresponds to the color cut of $V-\mbox{NB4670}>0.6$ used for
LAE selection. A few LAEs fall below this line because they were selected as LAEs on the basis of earlier, shallower
photometry, and were already spectroscopically confirmed to lie at $z=2.85$.
\label{fig_lya_eqw}
}
\end{figure*}

In the case of the stacks of LAEs with and without NB3420 detections, neither stack was detected in \emph{Ch2}; the
stacks reached 3$\sigma$ limiting magnitudes of $>$24.30 and $>$25.64, respectively, corresponding to $V-$\emph{Ch2}
$<1.19$ and $V-$\emph{Ch2} $<0.45$. While these limits on LAE $V-$\emph{Ch2} color do not give any constraints on the
differential properties between LAEs with and without NB3420 detections, they do indicate that LAEs are preferentially
low-mass galaxies with small values of $E(B-V)$. Converting the limits on LAE $V-$\emph{Ch2} color into limits on
stellar mass results in log(M/M$_{\odot}$) $<$ 9.72 (9.38) for LAEs with NB3420 detections and log(M/M$_{\odot}$) $<$
8.87 (8.35) for LAEs without NB3420 detections, assuming $E(B-V)$ = 0.0 (0.1). The $E(B-V)$ = 0.1 fit to the deeper
LAE stack (LAEs without NB3420 detections) corresponds to an age slightly less than 50 Myr. While such a young age is
unphysical for LBGs based on dynamical timescale arguments, LAEs may be more compact systems where such young ages are
feasible; such young ages have been determined for LAEs in previous work \citep{gawiser07}.

In summary, the analysis of the infrared data does not yield firm results differentiating between the two proposed
models of LyC escape presented at the beginning of this section: that objects with and without LyC detections are
either intrinsically different, or intrinsically similar with anisotropic emission of LyC radiation. Although the
individual galaxies and stacks plotted in Figure \ref{fig_irac_colors} do not indicate strong differences in the
$V-$\emph{Ch2} properties of galaxies with and without LyC emission, the fact that only one galaxy (MD34) with an
NB3420 detection is actually detected in \emph{Ch2} severely limits our interpretation of the \emph{Ch2} data. At the
same time, we find that MD34 (the only object with a LyC detection modeled by an SED fit) has been assigned the lowest
possible age allowed by our modeling and the second largest SFR of the 33 modeled LBGs. A larger sample of LBGs with
both LyC detections and IRAC photometry is needed to test whether or not LBGs with LyC detections are preferentially
young. In terms of the overall sample properties, the $V-$\emph{Ch2} color of our LBG stacks (which include all LBGs
in the sample) indicate an age for LBGs that is younger than the median age of LBGs modeled with SED fits, suggesting
a possible bias in previous LBG stellar population studies \citep[e.g.,][]{shapley01,kornei10} limited to objects with
near- or mid-infrared detections. Finally, the $V-$\emph{Ch2} colors and limits we observe for the LAEs in our sample
indicate that they are preferentially low-mass galaxies with small values of $E(B-V)$, consistent with previous
results in $z\sim3$ LAE stellar population studies. The two obstacles hindering our understanding of the rest-frame
near-infrared properties of the galaxies in our sample are the low \emph{Ch2} detection rate of both LBGs and LAEs and
the lack of spatial resolution, resulting in the blending of several galaxies with bright nearby neighbors. Deeper and
higher resolution data enabling robust SED fits for all galaxies in our sample would constitute a much more powerful
tool to distinguish between the proposed models of LyC escape.

\subsection{The relationship between LyC emission and \lya\ emission} \label{ssec_lya_eqw}

In addition to exploring the differential rest-frame UV $-$ near-infrared colors of galaxies with and without LyC
emission, we studied their relative \lya\ properties. Figure \ref{fig_lya_eqw} shows the $V-$NB4670 vs. NB4670
color-magnitude diagram for LAEs with and without NB3420 detections (shown in red and black, respectively), indicating
the rest-frame \lya\ equivalent width (EW) on the right-hand axis. Although the LAEs with NB3420 detections have lower
measured values of \lya\ EW on average than LAEs without NB3420 detections, they span a range in NB4670 magnitude
equivalent to that of the full LAE sample. In order to account for the fact that for several LAEs the measured EWs are
only lower limits, we used the package ASURV (``$A$stronomy $SURV$ival $A$nalysis'') Rev 1.2 \citep{asurv1,asurv2} to
calculate the \lya\ EW mean and standard error of LAEs with and without NB3420 detections using survival analysis.
This analysis shows that the mean \lya\ EW for objects with NB3420 detections is $34.0 \pm 5.9$ \AA\ and the mean
\lya\ EW for objects without NB3420 detections is $80.1 \pm 8.3$ \AA. These results confirm that galaxies with NB3420
detections tend to have lower \lya\ EWs, on average, than galaxies without NB3420 detections, a correlation that has
been previously reported in the SSA22a field by \citet{nestor11}.

In testing the connection between \lya\ and LyC emission, we must also consider changes in LyC emission when \lya\ EW
is the independent variable. To determine if $(F_{UV}/F_{LyC})_{corr}$ is positively correlated with \lya\ EW, we
calculated $(F_{UV}/F_{LyC})_{corr}$, the non-ionizing to ionizing flux-density ratio corrected for contamination and
IGM absorption, for LAEs in two bins of \lya\ EW separated by the mean LAE \lya\ EW (46 \AA). For LAEs with smaller
and larger EWs, respectively, we find $(F_{UV}/F_{LyC})_{corr}^{\mathrm{EW}\leq46\mathrm{\AA}} = 4.9 \pm 2.6$ and
$(F_{UV}/F_{LyC})_{corr}^{\mathrm{EW}>46\mathrm{\AA}} = 33 \pm 23$; LAEs with lower EWs have smaller values of
$(F_{UV}/F_{LyC})_{corr}$ (indicating stronger LyC emission) and LAEs with larger EWs have larger values of
$(F_{UV}/F_{LyC})_{corr}$. These two values are consistent with the results of the survival analysis: LAEs with LyC
detections are characterized by a smaller average \lya\ EW than those without LyC detections.

One explanation of the observed anticorrelation between \lya\ EW and LyC emission strength consists of the fact that
\lya\ photons are simply reprocessed LyC photons. As a LyC photon propagates through the ISM of a galaxy, one
possibility is that the photon escapes without being absorbed by a neutral hydrogen atom, and LyC emission is observed
from the galaxy. A second possibility is that the photon encounters a neutral hydrogen atom and is absorbed, ionizing
the atom. When the free electron recombines with the hydrogen ion, the emission cascade spectrum will often culminate
in the release of a \lya\ photon. Thus, for a fixed production rate of LyC photons, galaxies with larger \lya\ EWs
should tend to exhibit less LyC emission.

The fact that the stellar populations of galaxies with and without LyC detections do not appear to differ
significantly (see Section \ref{ssec_IR}) suggests that detecting LyC in a given galaxy depends on the orientation of
the galaxy with respect to the observer. If the hypothesis that LyC detectability depends on observer orientation is
correct, then the anticorrelation between observed LyC and \lya\ emission implies that \lya\ EW must also depend on
orientation, and in such a way to support the anticorrelation. Results from several recent galaxy simulations (of both
isolated galaxies and those set in cosmological context) suggest that the observed \lya\ EW is strongly dependent on
the orientation of the observer \citep{verhamme12,barnes11,yajima12}. Specifically, these simulations predict
orientation effects for disk galaxies: a disk galaxy, when viewed face-on, will have a much greater observed \lya\ EW
than the same galaxy viewed edge-on. This orientation effect stems from the fact that \lya\ photons are resonantly
scattered; for a \lya\ photon, the direction of final escape from the galaxy is nearly independent of its original
trajectory. As trajectories perpendicular to the plane of the galaxy present the paths of least opacity, more \lya\
photons will escape perpendicular to the galactic plane. It is still unclear how the spatial redistribution of \lya\
photons relates to that of LyC photons, and how these distributions are affected by the clumpy morphologies of
high-redshift galaxies. If, for example, \lya\ and LyC photons are both more likely to escape along lines of sight
with lower hydrogen column densities, a positive correlation between observed \lya\ and LyC emission might be expected
$-$ contrary to the anticorrelation observed in the LAE populations of the HS1549 and SSA22a fields. In any case, it
would be extremely useful for simulations to examine the effect that galaxy orientation produces on the relationship
between observed \lya\ and LyC emission. It would also be useful for such simulations to consider morphologies that
more closely represent those of high-redshift galaxies \citep[i.e., clumpy and irregular morphologies;][]{law07}. If
orientation effects \emph{cannot} explain the observed trend that smaller observed \lya\ EWs in LAEs coincide with
more LyC emission, then the difference in \lya\ EWs must indicate some intrinsic difference between galaxies
exhibiting strong or weak LyC emission that can be traced by \lya\ emission. Along these lines, we note that the mean
rest-frame \lya\ EW for LAEs without NB3420 detections (80~\AA) is close to that expected by the ``Case B" assumption
for 100\% absorption of LyC photons in a dust-free environment \citep[$\sim$100~\AA;][]{steidel11}. In the presence of
dust, the resonant scattering of \lya\ photons would tend to decrease the expected ``Case B" \lya\ EW. If \lya\
orientation effects do not correlate with LyC in the manner described above, then the larger \lya\ EWs in our sample
of LAEs without NB3420 detections indicates a higher rate of LyC absorption within the ISM of these objects.

Also relevant for understanding the connection between \lya\ and LyC emission, we find a lower average non-ionizing to ionizing UV flux-density ratio among LAEs (which have a median \lya\ EW of 46 \AA) than among LBGs (which have a much lower median \lya\ EW of 5 \AA).  Considering LAEs in the same magnitude range as LBGs ($V<25.4$), we derive
$(F_{UV}/F_{LyC})_{corr}^{LAE,\,V<25.4} = 4.1 \pm 2.7$ which is significantly lower than the value for LBGs,
$(F_{UV}/F_{LyC})_{corr}^{LBG} = 82 \pm 45$. The relative values of $(F_{UV}/F_{LyC})_{corr}$
for LAEs and LBGs apparently suggest a \emph{positive} correlation between \lya\ EW and LyC emission, in the opposite
sense of the trend traced by the LyC vs. \lya\ emission properties of LAEs. However, LBGs have been shown to differ
from LAEs in several galactic properties, including dust content and gas/dust covering fraction. As dust destroys UV
photons, and \lya\ photons travel particularly long path lengths before they escape their galaxy due to resonant
scattering, \lya\ photons suffer greater dust attenuation than other UV photons. Thus, while the amount of dust in a
galaxy will not affect the creation of \lya\ photons from LyC photons, it will affect the observed \lya\ EW.
Observational studies have confirmed that LBGs with larger \lya\ EWs have smaller values of $E(B-V)$
\citep[e.g.,][]{atek09,shapley03} and simulations have reproduced the same result \citep[e.g.,][]{yajima12,dayal09}.
Also, LBGs with larger \lya\ EWs have weaker interstellar absorption lines, indicating lower gas covering fraction
\citep{shapley03}. Thus, it is possible that the low median \lya\ EW observed in LBGs compared to LAEs is due to
increased dust attenuation \citep[LBGs have median values of $E(B-V)$ that are typically higher than those of
LAEs;][]{gronwall07,blanc11} and larger gas covering fraction. In summary, the observed strengths of \lya\ and LyC
emission may be anti-correlated (as seen in the LAE sample) except in the cases of higher interstellar extinction
where \lya\ photons are systematically destroyed (as in the case of the LBGs).

\section{Summary} \label{sec_summary}

We have presented an analysis of the $z \sim 3$ LyC properties of the HS1549 field. Multiple LyC studies of
star-forming galaxies have been conducted in the SSA22a field
\citep[e.g.,][]{inoue05,shapley06,iwata09,nestor11,nestor13} and HS1549 therefore provides an important independent
set of $z \sim 3$ LyC measurements. The HS1549 field contains a galaxy protocluster at $z=2.85$, and our narrowband
(NB3420) imaging has targeted the LyC spectral region of 131 spectroscopically confirmed galaxies at $z \geq 2.82$ (49
LBGs and 91 LAEs, 9 of which are constituents of both samples). We have detected 12 galaxies in NB3420 and implemented
simulations to correct the average NB3420 magnitudes of subsamples of these galaxies for both foreground galaxy
contamination and absorption from neutral hydrogren in the IGM. In addition to more than 19 hours of imaging in the
NB3420 filter, we have analyzed ancillary narrowband \lya\ imaging and broadband imaging in the rest-frame
non-ionizing UV, optical, and near-infrared in order to determine non-ionizing to ionizing UV flux-density ratios and
place limits on other galactic properties. Our main conclusions are as follows:

\begin{enumerate}

\item We find 4 LBGs and 7 LAEs with NB3420 detections, along with one additional object (MD12/\emph{lae3540}) that we have removed from both samples. The NB3420 detections of the LBGs range in magnitude from
$26.56 \leq m_{3420} \leq 27.01$ and those of the LAEs range from $25.11 \leq m_{3420} \leq 27.23$. Our contamination
simulations predict that 1.5 $\pm$ 1.0 LBG NB3420 detections and 4.3 $\pm$ 1.3 LAE NB3420 detections are
uncontaminated by foreground galaxies.

\item The ratio of non-ionizing to ionizing UV radiation of the LBGs and LAEs is traced by the observed NB3420$-V$
color. The colors of our targets vary widely, with NB3420-detected galaxies at a given $V$ magnitude having NB3420
magnitudes up to a factor of ten brighter than the NB3420 limits of undetected galaxies; i.e., a small percentage of
galaxies are detected strongly in NB3420 and the rest remain undetected. These results confirm those presented in
\citet{nestor11} for galaxies in the SSA22a field, and lend further support to the hypothesis that LyC radiation
escapes galaxies through cleared holes in the ISM and can only be viewed by the observer with fortuitous orientation.
The large variation in individual galaxy NB3420 properties makes it necessary to consider properties of the full
ensemble of LBGs and LAEs in order to determine the average LyC properties of $z\sim3$ galaxies.

\item One of the principal difficulties in performing accurate LyC photometry is the presence of low-redshift
foreground galaxies that introduce contaminating light into the NB3420 filter. At the spatial resolution of $HST$,
galaxies at $z \sim 3$ tend to exhibit clumpy morphology. Separating a $z \sim 3$ clump from a foreground contaminant
is impossible unless the individual redshifts of each clump are known. While simulations of the amount of foreground
contamination can help correct the observed non-ionizing to ionizing UV flux-density ratios, it is very difficult to
study the properties of LyC-emitting galaxies when the simulations predict that roughly half of the NB3420-detected
sample may be contaminated. To obtain spectroscopic or photometric redshifts at the high spatial resolution necessary
to distinguish each clump requires imaging from a high-resolution, space-based observing facility such as $HST$
coupled with deep, high-resolution spectroscopy from the ground.

\item Reconciling the low observed non-ionizing to ionizing UV flux-density ratios ($F_{UV}/F_{LyC}$) with intrinsic
values predicted by stellar population synthesis models has been a challenge in all $z\sim3$ LyC studies. As in the
SSA22a field, the observed flux-density ratios of many of the LBGs and LAEs with putative LyC detections are too low
to be reasonably explained by current models. There are possible ways of solving the problem; for example, one could
make the assumption of no IGM absorption along the line of sight for galaxies with observed LyC detections. However,
this condition is unlikely to hold for all galaxies with putative LyC detections. As of now, the discrepancy between
the LyC observations and models remains unresolved.
     
\item Using estimates of the UV escape fraction from observational studies of $z\sim3$ galaxies and a range of
estimated values for the theoretical ratio of intrinsic UV to LyC luminosities given by the \citet{bc03} and BPASS
\citep{eldridgestanway09} models, we derive relative LyC escape fractions of $f_{esc,\,rel}^{LBG}=5-8$\% and
$f_{esc,\,rel}^{LAE}=18-49$\%, and absolute escape fractions of $f_{esc}^{LBG}=1-2$\% and $f_{esc}^{LAE}=5-15$\%. The
uncertainties on the escape fractions we calculate are on the order of 50\%. We also determine a comoving specific
ionizing emissivity of LyC photons ($\epsilon_{LyC}$) in the range of 8.8 $-$ 15.0 $\times 10^{24}$ ergs s$^{-1}$
Hz$^{-1}$ Mpc$^{-3}$. This range of emissivity values was determined by using two different models to combine the LBG
and LAE samples (as discussed in Section \ref{ssec_lyc_emissivity}); the luminosity-dependent model yields
$\epsilon_{LyC}=15.0 \pm 6.7 \times 10^{24}$ ergs s$^{-1}$ Hz$^{-1}$ Mpc$^{-3}$ and the LAE-dependent model yields
$\epsilon_{LyC}=8.8 \pm 3.5 \times 10^{24}$ ergs s$^{-1}$ Hz$^{-1}$ Mpc$^{-3}$. These values of $\epsilon_{LyC}$ for
star-forming galaxies are consistent with the total ionizing emissivity inferred from the \lya-forest studies of
\citet{bolton07} and \citet{fauchergiguere08}. The rough agreement between the total ionizing emissivity and the
ionizing emissivity from star-forming galaxies leaves little room for the contribution to the emissivity from QSOs or
from fainter star-forming galaxies not probed by our observations. While the LyC emissivity we measure from
star-forming galaxies may still be an overestimation, it is in better agreement with the total LyC emissivity
determined from \lya-forest studies than the higher values determined in past work.

\item Examining the rest-frame near-infrared properties of galaxies in our sample based on IRAC \emph{Ch2} imaging, we
find no significant difference in the \emph{Ch2} or $V-$\emph{Ch2} properties of objects with and without NB3420
detections. This agreement indicates that objects with and without LyC detections are drawn from a populations of
star-forming galaxies with similar distributions of stellar mass, age, $E(B-V)$, and star-formation rate. We caution,
however, that our interpretation is limited by the fact that only one galaxy (MD34) with an NB3420 detection is
detected in \emph{Ch2}. While the $V-$\emph{Ch2} vs. $V$ properties of this object do not distinguish it from LBGs
without NB3420 detections, in the SED modeling, it has been assigned the lowest possible age allowed and the second
largest SFR of the 33 modeled LBGs. A much larger sample of LBGs with both LyC detections and IRAC photometry is
needed to test whether LBGs with LyC detections are preferentially young, or whether LBGs with and without LyC
detections come from similar populations.

\item While we do not find significant differences in the stellar populations of objects with and without LyC
detection, we do find that LAEs with LyC detections have smaller \lya\ equivalent widths on average than those
without. A comparison of $(F_{UV}/F_{LyC})_{corr}$ for LAEs in two bins of equivalent width yields
$(F_{UV}/F_{LyC})_{corr}^{\mathrm{EW}\leq46\mathrm{\AA}} = 4.9 \pm 2.6$ and
$(F_{UV}/F_{LyC})_{corr}^{\mathrm{EW}>46\mathrm{\AA}} = 33 \pm 23$. These results imply an inverse relationship
between the amount of \lya\ and LyC emission observed from LAEs. One possibility is that this relationship stems from
the fact that \lya\ photons are reprocessed LyC photons, so an overall increase in \lya\ photons must correlate with a
decrease in LyC photons. The details of LyC and \lya\ radiative transfer through the ISM of clumpy, high-redshift
galaxies, however, are still unclear, as are the effects of varying the observer's orientation while observing \lya\
and LyC emission. We also note that despite this apparent anti-correlation between the strength of \lya\ and LyC
emission, LBGs (which have lower median \lya\ equivalent widths than LAEs) display much weaker LyC emission than LAEs
in the same magnitude range. This trend may be explained by the higher dust content in LBGs systematically destroying
both LyC photons and resonantly-scattered \lya\ photons.

\end{enumerate}

Future progress in this field is contingent upon amassing a large and well-defined sample of LyC-emitting star-forming
galaxies and eradicating the possibility that LyC measurements are contaminated by foreground interlopers. A detailed,
multi-wavelength analysis of such a sample would reveal definitively whether LyC-emitting objects possess specific
properties that facilitate LyC escape or whether LyC emission escapes from all galaxies through randomly-oriented
paths cleared through the ISM. With the goals of removing contamination and obtaining multiwavelength photometry, we
are currently pursuing follow-up observations of our NB3420-detected galaxies in the HS1549 field with $HST$/WFC3.

\medskip

We thank Kevin Hainline for his helpful discussions regarding our photometric simulations, Katherine Kornei for advice
on the spectroscopic data reduction, and the anonymous referee for constructive suggestions. R.E.M., A.E.S., and
D.B.N. acknowledge support from the David and Lucile Packard Foundation. C.C.S. acknowledges additional support from
NSF grant AST-0908805 and GO-11638.01 from the Space Telescope Science Institute. We wish to extend special thanks to
those of Hawaiian ancestry on whose sacred mountain we are privileged to be guests. Without their generous
hospitality, most of the observations presented herein would not have been possible.

\appendix
\section{GNBs and Faint LAEs}
\begin{figure*}
\epsscale{0.65}
\plotone{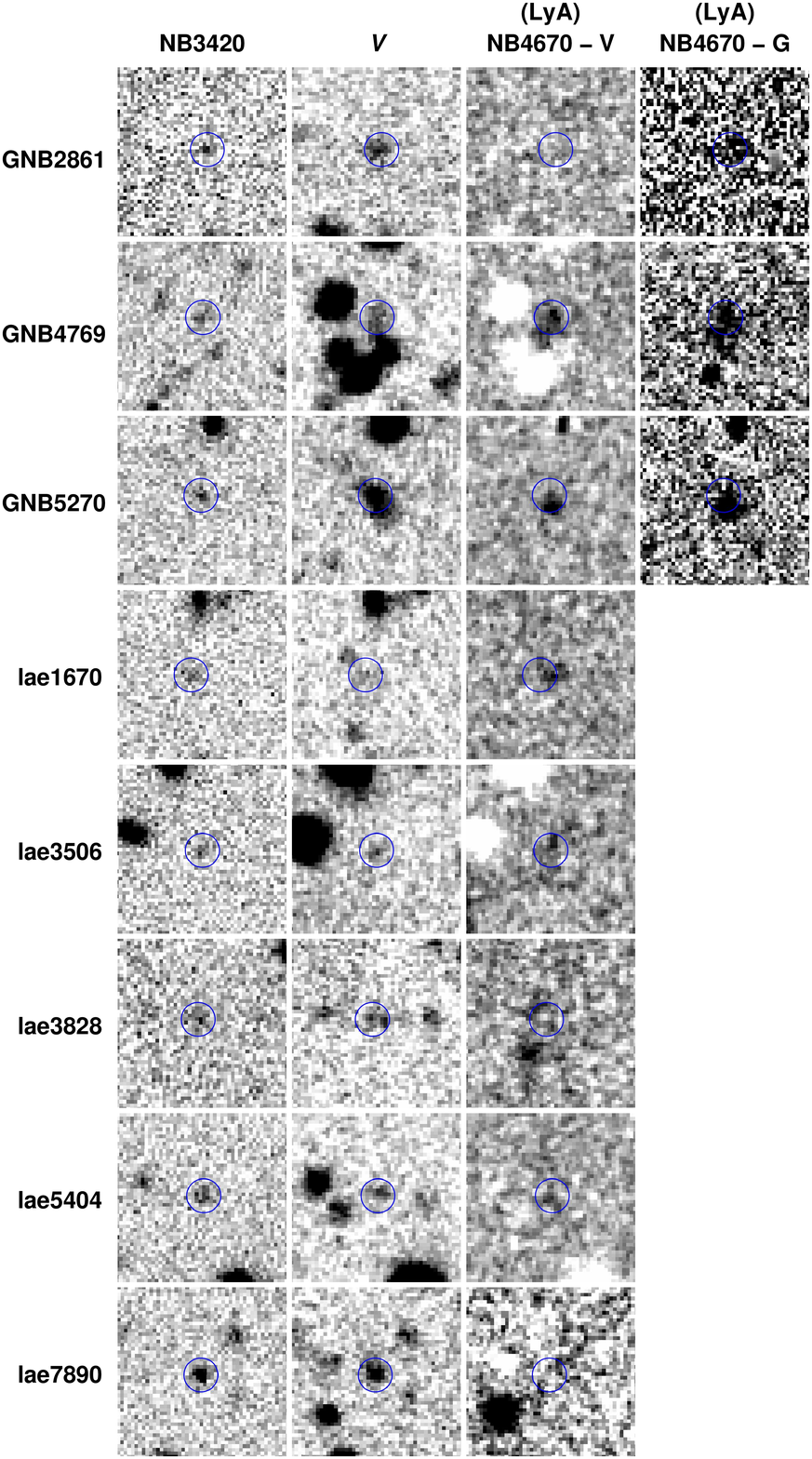}
\caption{
\small
10\secnopoint\ $\times$ 10\secnopoint\ postage stamp images of the three GNBs and five faint LAEs with NB3420
detections. Objects are displayed in the following bands: NB3420 (indicating the LyC), $V$ (indicating the
non-ionizing UV continuum), NB4670$-V$ (indicating \lya\ emission), and NB4670$-G$ (also indicating \lya\ emission;
GNBs only). All postage stamps are centered on the NB4670 centroid and blue circles (1\secnopoint\ radius) indicate
the centroid of the NB3420 emission. All postage stamps follow the conventional orientation, with north up and east to
the left.
\label{fig_GNB_faint_LAE} }
\end{figure*}

Here we present photometry (Table \ref{tab_GNB_faint_LAE}) and postage stamp images (Figure \ref{fig_GNB_faint_LAE})
for eight additional \lya-emitting galaxies with NB3420 detections for which we have obtained spectroscopic redshifts
(see Section \ref{ssec_spec}). This sample comprises three GNBs and five faint LAEs ($m_{4670} > 26$). As we do not
have a complete and unbiased sample of these objects, we do not include them in the main LAE sample.

\section{Photometric LAE Sample}

\begin{figure*}
\epsscale{1.0}
\plotone{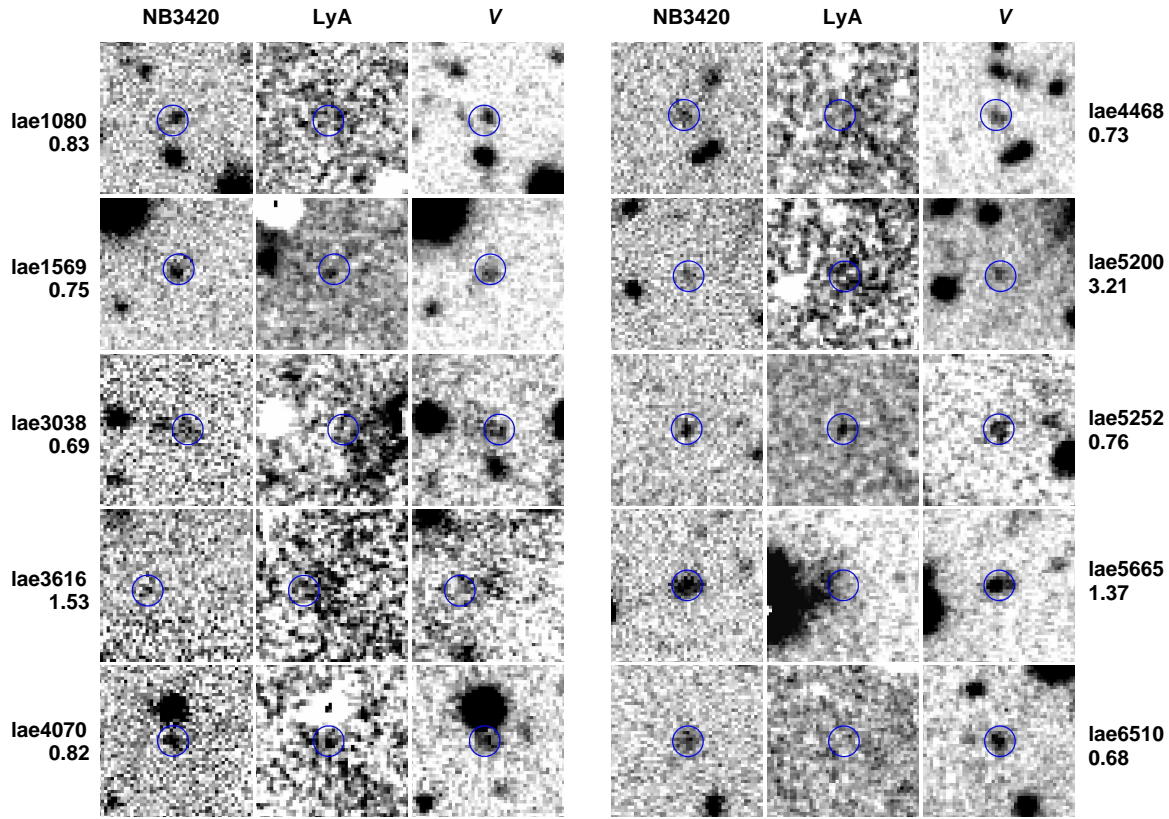}
\caption{
\small
10\secnopoint\ $\times$ 10\secnopoint\ postage stamp images of the 10 LAE photometric candidates with NB3420
detections. Objects are displayed in the following bands: NB3420 (indicating the LyC), NB4670$-V$ (indicating \lya\
emission), and $V$ (indicating the non-ionizing UV continuum). The $V-$NB4670 color of each LAE is indicated below the
object name. All postage stamps are centered on the $V$ centroid and blue circles (1\secnopoint\ radius) indicate the
centroid of the NB3420 emission. All postage stamps follow the conventional orientation, with north up and east to the
left. We note that LAEs with more diffuse \lya\ emission may be difficult to distinguish in the NB4670$-V$ image, even
though their $V-$NB4670 colors identify them as LAEs; for such objects, we have increased the stretch of the
NB4670$-V$ image to make the diffuse emission more easily visible.
\label{fig_LAE_phot} }
\end{figure*}

Here we present photometry of 33 LAE photometric candidates without spectroscopically confirmed redshifts (see Section
\ref{ssec_samp}). Ten of these LAEs have NB3420 detections. Table \ref{tab_LAE_phot} contains photometric information
for all 33 LAE photometric candidates, and Figure \ref{fig_LAE_phot} displays postage stamp images for the ten
candidates with NB3420 detections. While some of these NB3420 detections may be true LyC detections, the high rate of
NB3420 detections in this sample may be due to the increased probability of contamination for objects without
spectroscopic redshifts. For this reason, we do not include these objects in the LAE analysis.

\begin{deluxetable*}{lccccccccc}
\tablewidth{0pt}
\tabletypesize{\scriptsize}
\tablecaption{Photometry for GNBs and LAEs with $m_{4670} > 26.0$ \label{tab_GNB_faint_LAE}}
\tablehead{
\colhead{ID} & \colhead{RA}\tablenotemark{a} & \colhead{Dec}\tablenotemark{a} & \colhead{z} & \colhead{NB4670} &
\colhead{$V$} & \colhead{NB3420} & \colhead{$\Delta_{UV,\,LyC}$\tablenotemark{b}} &
\colhead{$\Delta_{Ly\alpha,\,LyC}$\tablenotemark{c}} & \colhead{$\frac{F_{UV}}{F_{LyC}}_{obs}$\tablenotemark{d}} \\
 & \colhead{(J2000)} & \colhead{(J2000)} & & & & & & &
}
\startdata
GNB2861 & 15:51:53.603 & 19:11:36.53 & 2.844 & 26.80 & 26.08 & 26.86 & 0.28 & 0.33 & 2.1 $\pm$ 1.0 \\
GNB4769 & 15:51:59.873 & 19:08:42.73 & 2.849 & 25.30 & 26.54 & 27.33 & 0.34 & 0.52 & 2.1 $\pm$ 0.9 \\
GNB5270 & 15:51:57.364 & 19:09:53.38 & 2.847 & 24.84 & 25.14 & 26.43 & 0.40 & 0.30 & 3.3 $\pm$ 1.0 \\
lae1670 & 15:51:45.121 & 19:10:15.34 & 2.846 & 26.18 & 27.27 & 26.45 & 0.25 & 0.65 & 0.5 $\pm$ 0.1 \\
lae3506 & 15:51:52.242 & 19:11:41.01 & 2.841 & 26.71 & 27.58 & 27.09 & 0.11 & 0.08 & 0.6 $\pm$ 0.3 \\
lae3828 & 15:51:53.228 & 19:13:08.50 & 2.892 & 26.64 & 26.85 & 26.27 & 0.28 & 0.33 & 0.6 $\pm$ 0.3 \\
lae5404 & 15:52:00.008 & 19:08:54.42 & 2.816 & 26.60 & 27.16 & 26.53 & 0.21 & 0.15 & 0.6 $\pm$ 0.3 \\
lae7890 & 15:52:01.943 & 19:12:42.47 & 2.850 & 26.02 & 26.13 & 25.91 & 0.23 & 0.17 & 0.8 $\pm$ 0.3 \\
\enddata
\tablenotetext{a}{GNB and LAE coordinates are based on NB4670 centroids.}
\tablenotetext{b}{Spatial offset between the centroids of $V$ and NB3420 emission.}
\tablenotetext{c}{Spatial offset between the centroids of NB4760 and NB3420 emission.}
\tablenotetext{d}{Observed ratio and uncertainty in the ratio of non-ionizing UV to LyC flux-densities, inferred from
the NB3420$-V$ color. This value has not been corrected for either contamination by foreground sources or IGM
absorption.}
\end{deluxetable*}

\begin{deluxetable*}{lcccccccc}
\tablewidth{0pt}
\tabletypesize{\scriptsize}
\tablecaption{Photometry for LAE Photometric Candidates \label{tab_LAE_phot}}
\tablehead{
\colhead{ID} & \colhead{RA}\tablenotemark{a} & \colhead{Dec}\tablenotemark{a} & \colhead{NB4670} & \colhead{$V$} &
\colhead{NB3420} & \colhead{$\Delta_{UV,\,LyC}$\tablenotemark{b}} &
\colhead{$\Delta_{Ly\alpha,\,LyC}$\tablenotemark{c}} & \colhead{$\frac{F_{UV}}{F_{LyC}}_{obs}$\tablenotemark{d}} \\
 & \colhead{(J2000)} & \colhead{(J2000)} & & & & & &
}
\startdata
lae759 & 15:51:41.039 & 19:09:57.61 & 25.54 & 26.79 & $>$27.30 & \nodata & \nodata & $>$1.6 \\  
lae803 & 15:51:41.598 & 19:10:19.54 & 25.68 & 27.02 & $>$27.30 & \nodata & \nodata & $>$1.3 \\  
lae1043 & 15:51:42.623 & 19:13:00.70 & 26.00 & 27.07 & $>$27.30 & \nodata & \nodata & $>$1.2 \\  
lae1058 & 15:51:42.791 & 19:11:07.95 & 26.00 & 27.14 & $>$27.30 & \nodata & \nodata & $>$1.2 \\  
lae1080 & 15:51:42.918 & 19:12:33.53 & 25.92 & 26.37 & 26.03 & 0\secpoint3 & 0\secpoint4 & 0.7 $\pm$ 0.2 \\
lae1569 & 15:51:44.416 & 19:08:43.92 & 25.16 & 25.84 & 25.51 & 0\secpoint3 & 0\secpoint4 & 0.7 $\pm$ 0.2 \\
lae1840 & 15:51:45.875 & 19:11:55.76 & 25.98 & 27.19 & $>$27.30 & \nodata & \nodata & $>$1.1 \\  
lae1883 & 15:51:45.893 & 19:12:23.30 & 25.49 & 26.28 & $>$27.30 & \nodata & \nodata & $>$2.6 \\  
lae2278 & 15:51:47.324 & 19:12:01.54 & 25.77 & $>$27.58 & $>$27.30 & \nodata & \nodata & \nodata \\  
lae2369 & 15:51:47.863 & 19:11:24.63 & 25.93 & $>$27.58 & $>$27.30 & \nodata & \nodata & \nodata \\  
lae2431 & 15:51:48.048 & 19:09:04.50 & 25.97 & $>$27.58 & $>$27.30 & \nodata & \nodata & \nodata \\  
lae2482 & 15:51:48.129 & 19:09:48.31 & 25.29 & $>$27.58 & $>$27.30 & \nodata & \nodata & \nodata \\  
lae2664 & 15:51:49.090 & 19:11:00.51 & 25.52 & $>$27.58 & $>$27.30 & \nodata & \nodata & \nodata \\  
lae2666 & 15:51:49.161 & 19:13:17.59 & 25.80 & 26.89 & $>$27.30 & \nodata & \nodata & $>$1.4 \\  
lae3038 & 15:51:50.091 & 19:09:02.19 & 25.74 & 26.25 & 26.44 & 0\secpoint7 & 0\secpoint2 & 1.2 $\pm$ 0.4 \\
lae3348 & 15:51:51.561 & 19:11:20.54 & 25.90 & 26.98 & $>$27.30 & \nodata & \nodata & $>$1.3 \\  
lae3365 & 15:51:51.691 & 19:11:20.50 & 25.90 & 26.78 & $>$27.30 & \nodata & \nodata & $>$1.6 \\  
lae3616 & 15:51:52.218 & 19:08:21.79 & 25.10 & 26.96 & 27.17 & 1\secpoint9 & 1\secpoint4 & 1.2 $\pm$ 0.6 \\
lae4070 & 15:51:54.054 & 19:10:26.76 & 25.93 & 27.47 & 25.88 & 0\secpoint2 & 0\secpoint1 & 0.2 $\pm$ 0.1 \\
lae4079 & 15:51:54.023 & 19:10:05.98 & 25.19 & 25.71 & $>$27.30 & \nodata & \nodata & $>$4.3 \\  
lae4468 & 15:51:55.367 & 19:12:57.55 & 25.82 & 26.46 & 25.97 & 0\secpoint6 & 0\secpoint4 & 0.6 $\pm$ 0.2 \\
lae5157 & 15:52:01.016 & 19:08:50.34 & 26.00 & $>$27.58 & $>$27.30 & \nodata & \nodata & \nodata \\  
lae5200 & 15:52:01.083 & 19:11:25.97 & 25.83 & 27.19 & 26.70 & 0\secpoint2 & 0\secpoint1 & 0.6 $\pm$ 0.3 \\
lae5252 & 15:52:00.519 & 19:08:31.67 & 26.00 & 26.38 & 25.72 & 0\secpoint0 & 0\secpoint1 & 0.5 $\pm$ 0.2 \\
lae5446 & 15:51:58.774 & 19:12:46.40 & 25.43 & 26.01 & $>$27.30 & \nodata & \nodata & $>$3.3 \\  
lae5661 & 15:51:59.092 & 19:12:39.74 & 25.89 & 26.31 & $>$27.30 & \nodata & \nodata & $>$2.5 \\  
lae5665 & 15:51:57.738 & 19:12:27.59 & 23.96 & 25.31 & 24.78 & 0\secpoint0 & 0\secpoint7 & 0.6 $\pm$ 0.1 \\
lae6041 & 15:51:57.864 & 19:11:14.43 & 25.56 & 25.58 & $>$27.30 & \nodata & \nodata & $>$4.9 \\  
lae6436 & 15:52:07.285 & 19:11:54.07 & 25.50 & 26.30 & $>$27.30 & \nodata & \nodata & $>$2.5 \\  
lae6510 & 15:52:06.976 & 19:12:03.51 & 25.80 & 25.94 & 25.41 & 0\secpoint1 & 0\secpoint1 & 0.6 $\pm$ 0.2 \\
lae7110 & 15:52:02.523 & 19:12:07.71 & 25.98 & 26.96 & $>$27.30 & \nodata & \nodata & $>$1.4 \\  
lae7247 & 15:52:04.452 & 19:10:48.30 & 25.96 & 26.83 & $>$27.30 & \nodata & \nodata & $>$1.5 \\  
lae7642 & 15:52:03.478 & 19:12:58.32 & 25.26 & 26.20 & $>$27.30 & \nodata & \nodata & $>$2.7  
\enddata
\tablenotetext{a}{Coordinates for LAEs are based on NB4670 centroids.}
\tablenotetext{b}{Spatial offset between the centroids of $V$ and NB3420 emission.}
\tablenotetext{c}{Spatial offset between the centroids of NB4760 and NB3420 emission.}
\tablenotetext{d}{Observed ratio and uncertainty in the ratio of non-ionizing UV to LyC flux-densities, inferred from
the NB3420$-V$ color. This value has not been corrected for either contamination by foreground sources or IGM
absorption.}
\end{deluxetable*}

\bibliographystyle{apj}

\end{document}